\documentclass[12pt,twoside,a4paper]{article}
\pdfoutput=1

\usepackage{amsmath, amsthm, amsfonts, amssymb}
\usepackage{graphicx}
\usepackage{multirow}
\usepackage{float}
\usepackage{color}
\usepackage{cite}
\usepackage[width=\textwidth,font=small,labelfont=bf,
labelsep=endash]{caption}
\usepackage{titlesec}

\titleformat{\section}
  {\normalfont\fontsize{13}{16}\bfseries}{\thesection}{1em}{}
\titleformat{\subsection}
  {\normalfont\fontsize{11}{14}\bfseries}{\thesubsection}{1em}{}

\normalsize
\newcommand\tr            {\mathrm{Tr}}
\newcommand\csch            {\mathrm{csch}}
\newcommand\sech            {\mathrm{sech}}
\newcommand\arccosh        {\mathrm{arccosh}}

\newcommand\doi[2]        {\href{http://dx.doi.org/#1}{#2}}

\renewcommand{\theequation}{\arabic{section}.\arabic{equation}}
\makeatletter
\newcommand{\vast}{\bBigg@{4}}
\newcommand{\Vast}{\bBigg@{5}}
\makeatother

\begin{document}

\title{\textbf{An Inverse Mass Expansion for the Mutual Information in Free Scalar QFT at Finite Temperature}}
\author{Dimitrios Katsinis$^{1,2}$ and Georgios Pastras$^2$}
\date{\small $^1$Department of Physics, National and Kapodistrian University of Athens,\\University Campus, Zografou, Athens 15784, Greece\\
$^2$NCSR ``Demokritos'', Institute of Nuclear and Particle Physics,\\Aghia Paraskevi 15310, Attiki, Greece\linebreak \vspace{8pt}
\texttt{dkatsinis@phys.uoa.gr, pastras@inp.demokritos.gr}}

\vskip .5cm

\maketitle

\begin{abstract}
We study the entanglement entropy and the mutual information in coupled harmonic systems at finite temperature. Interestingly, we find that the mutual information does not vanish at infinite temperature, but it rather reaches a specific finite value, which can be attributed to classical correlations solely. We further obtain high and low temperature expansions for both quantities. Then, we extend the analysis performed in the seminal paper by Srednicki \cite{Srednicki:1993im} for free real scalar field theories in Minkowski space-time in $3+1$ dimensions at a thermal state. We find that the mutual information obeys an area law, similar to that obeyed by the entanglement entropy at vanishing temperature. The coefficient of this area law does not vanish at infinite temperature. Then, we calculate this coefficient perturbatively in an $1/\mu$ expansion, where $\mu$ is the mass of the scalar field. Finally, we study the high and low temperature behaviour of the area law term.
\end{abstract}

\newpage

\tableofcontents

\newpage

\setcounter{equation}{0}
\section{Introduction}
\label{sec:intro}

Quantum entanglement is a fundamental phenomenon without classical analogue, which plays an important role in quantum physics. When a complex quantum system lies in an entangled state, even if this is a pure state, there is no answer to the question ``what is the state that describes the subsystem A?''. However, the latter can be described by a density matrix, the \emph{reduced density matrix}, which can be derived from the density matrix of the composite system, via the tracing out of the degrees of freedom of the subsystem $A^C$, which is complementary to $A$,
\begin{equation}
\rho_A = \tr_{A^C} \rho .
\end{equation}
Under the assumption that the quantum composite system lies in a pure state, at the limit the subsystems $A$ and $A^C$ become disentangled, the reduced density matrix $\rho_A$ corresponds to a pure state, and, thus, the question ``what is the state of the subsystem A?'' acquires an answer. Therefore, it is natural to claim that entanglement is encoded in the spectrum of the reduced density matrix, and the most natural measure of entanglement is Shannon entropy applied to the latter, i.e.,
\begin{equation}
S_A := - \tr \rho_A \ln \rho_A ,
\end{equation}
which is the so called entanglement entropy.

Entanglement is a property that depends on the specific separation of the composite system to the pair of complementary subsystems $A$ and $A^C$. Naturally, one would require from a measure of entanglement to obey the property
\begin{equation}
S_A = S_{A^C} ,
\end{equation}
which can indeed be shown to hold, when the composite system lies in a pure state.

In a seminal paper \cite{Srednicki:1993im}, Srednicki showed that entanglement entropy has a particularly interesting property in free massless scalar quantum field theory: Assuming that the system lies at its ground state, and separating the degrees of freedom to two subsystems, one containing the degrees of freedom inside a given sphere of radius $R$, and another being its complementary, it was shown that entanglement entropy is proportional to the area of the sphere. This property is somehow expected from the physics of entanglement: As already mentioned, entanglement characterizes the separation of the composite system to two subsystems and not the subsystems themselves. Thus, the entanglement entropy cannot depend on the properties of any of the two subsystems (such as the volume of subsystem $A$), but on those of their only common feature, i.e. their boundary. Nevertheless, this finding is highly intriguing, since it resembles the area law of the black hole entropy. This similarity motivates the investigation of whether the black hole entropy can be attributed completely or partially to quantum entanglement. Bombelli et.al. \cite{Bombelli:1986rw} also motivated by the similarity to black hole physics calculated the entanglement entropy for a scalar field in the background of a Schwarzschild black hole resulting in similar conclusions.

However, the entanglement entropy is a good measure for entanglement, or more generally of correlations between the subsystems, only when the composite system lies in a pure state. If this is not the case, the entanglement entropy will inherit contributions that originate from the classical entropy of the composite system, and, thus, they do not characterize the entanglement between the two subsystems. In general, when the composite system lies in a mixed state,
\begin{equation}
S_A \neq S_{A^C} .
\end{equation}
In field theory, the above argument implies that when the composite system lies in a thermal state, the entanglement entropy will have contributions originating from the thermal entropy of the composite system, and, thus, will be proportional to the volume of the subsystem.

Entanglement in field theory at finite temperature has been studied mainly in the context of two-dimensional conformal field theory \cite{Calabrese:2004eu,Cardy:2014jwa,Chen:2014unl,Calabrese:2014yza} with the use of the replica trick \cite{Callan:1994py,Holzhey:1994we}. Much fewer works focus on gapped systems \cite{Herzog:2012bw} or to higher dimensional theories \cite{Herzog:2014fra,Herzog:2014tfa,Herzog:2015cxa}. In more recent years, entanglement in thermal states has also been studied through the holographic duality. The issue has been posted in the original works that established the Ryu-Takayanagi conjecture \cite{Ryu:2006bv,Ryu:2006ef}, where the problem of the non-symmetry property of the entanglement entropy is resolved by the existence of more than one minimal surfaces, due to the presence of the black hole, which are homologous to complementary boundary region. This study has been extended in several works (see e.g. \cite{Fischler:2012ca}). Most of these focus mainly on the geometry of the BTZ black hole \cite{Fischler:2012uv,Hubeny:2013gta,Kundu:2016dyk,Erdmenger:2017pfh}, which is also relevant to two-dimensional CFTs, as this is the only black hole geometry where minimal surfaces can be expressed analytically. Entanglement in harmonic lattice systems at finite temperature has been studied in \cite{Cramer:2005mx}. However, there is not much attention to the study of entanglement in field theory at finite temperature via the techniques originally used in \cite{Srednicki:1993im}.

When the composite system lies in a mixed state, a better measure of the correlation between the two subsystems is the mutual information,
\begin{equation}
I \left( A , A^C \right) := S_A + S_{A^C} - S_{A \cup A^C} ,
\label{eq:intro_def_mutual}
\end{equation}
which has the symmetric property by construction. Following the arguments above, the mutual information should characterize the separation of the composite system to two subsystems and, thus, in field theory it should depend only on the properties of the entangling surface, even at mixed, e.g. thermal, states. It has been shown that in lattice spin systems the mutual information obeys an area law bound \cite{Wolf:2007tdq}. It was recently shown \cite{thermal_short} that appropriate generalization of the techniques of \cite{Srednicki:1993im} can be used to calculate the mutual information in free scalar field theory at finite temperature and indeed it is proportional to the area of the entangling surface.

In \cite{Katsinis:2017qzh}, the authors developed a perturbative approach in order to study the area law of the entanglement entropy in scalar field theory at its ground state, analytically, bypassing the numerical part of the original calculation in \cite{Srednicki:1993im}. In this paper, we extend this method, in order to calculate perturbatively the entanglement entropy and the mutual information in free scalar field theory at a thermal state. In section \ref{sec:qm_2}, we study the system of two harmonically coupled oscillators at finite temperature. In section \ref{sec:qm_many} we generalize to a coupled harmonic system with an arbitrary number of degrees of freedom at a thermal state. In section \ref{sec:chain} we develop the hopping expansion for chains of coupled oscillators, i.e. systems where only neighbouring oscillators are coupled. In section \ref{sec:qft} we use the results of the previous sections, in order to study the entanglement entropy and the mutual information in free scalar field theory in 3+1 dimensions. Finally, in section \ref{sec:discussion} we discuss our results. There are also several appendices containing more details of the related algebra.

\setcounter{equation}{0}
\section{A Pair of Coupled Harmonic Oscillators}
\label{sec:qm_2}
In order to study entanglement entropy and mutual information in free scalar field theory at finite temperature, we first study systems of coupled harmonic oscillators with a finite number of degrees of freedom. The simplest such system, which is the subject of this section, is a system of two coupled harmonic oscillators at finite temperature. The analysis closely follows the original treatment presented in \cite{Srednicki:1993im}, in the sense that it is performed in coordinate representation and presents several technical similarities. A short account of this analysis recently appeared in \cite{thermal_short}.

\subsection{A Single Harmonic Oscillator at Finite Temperature}
\label{subsec:one_osc}

First, we would like to recall some formulas related to the problem of a single harmonic oscillator at finite temperature in coordinate representation \cite{Feynman}, which will be useful in the following. Without loss of generality, we consider the mass of the harmonic oscillator to be equal to one, i.e. the Hamiltonian of the system is
\begin{equation}
H = \frac{1}{2}{p^2} + \frac{1}{2}{\omega ^2}{x^2} .
\end{equation}
In coordinate representation, the energy eigenstates and the corresponding eigenvalues of the harmonic oscillator are
\begin{equation}
{\psi _n}\left( x \right) = \frac{1}{{\sqrt {{2^n}n!} }}\sqrt[4]{{\frac{\omega }{\pi }}}{e^{ - \frac{{\omega {x^2}}}{2}}}{H_n}\left( {\sqrt \omega  x} \right) , \quad {E_n} = \omega \left( {n + \frac{1}{2}} \right) ,
\label{eq:ho_eigenstates}
\end{equation}
where ${H_n}$ is the Hermite polynomial of order $n$. The equation \eqref{eq:ho_eigenstates} trivially implies that the density matrix describing a quantum harmonic oscillator at finite temperature $T$ is given by
\begin{equation}
\rho \left( {x,x'} \right) = \sum\limits_{n = 0}^\infty  {2\sinh \frac{\omega }{{2T}}{e^{ - \frac{\omega }{T}\left( {n + \frac{1}{2}} \right)}}\frac{1}{{{2^n}n!}}\sqrt {\frac{\omega }{\pi }} {e^{ - \frac{{\omega \left( {{x^2} + x{'^2}} \right)}}{2}}}{H_n}\left( {\sqrt \omega  x} \right){H_n}\left( {\sqrt \omega  x'} \right)} .
\label{eq:ho_density_1}
\end{equation}
As a consequence of Mehler's formula,
\begin{equation}
\sum\limits_{n = 0}^\infty  {\frac{{{H_n}\left( x \right){H_n}\left( y \right)}}{{n!}}{{\left( {\frac{w}{2}} \right)}^n}}  = \frac{1}{{\sqrt {1 - {w^2}} }}{e^{\frac{{2xyw - \left( {{x^2} + {y^2}} \right){w^2}}}{{1 - {w^2}}}}} ,
\end{equation}
the density matrix \eqref{eq:ho_density_1} can be written in a simpler form, namely
\begin{equation}
\rho \left( {x,x'} \right) = \sqrt {\frac{\omega }{\pi }\left( {a + b} \right)} {e^{ - \frac{{a\left( {{x^2} + x{'^2}} \right)}}{2}}}{e^{ - bxx'}} ,
\label{eq:ho_density_2}
\end{equation}
where we defined the quantities $a$ and $b$ as
\begin{equation}
a \equiv \omega \coth \frac{\omega }{T},\quad b \equiv  - \omega \csch \frac{\omega }{T} .
\end{equation}

Finally, it is a matter of simple algebra to show that the thermal entropy of the single quantum harmonic oscillator at temperature $T$ equals
\begin{equation}
{S_{\textrm{th}}} =  - \ln \left( {1 - {e^{ - \frac{\omega }{T}}}} \right) + \frac{\omega }{T}\frac{1}{{{e^{\frac{\omega }{T}}} - 1}} .
\label{eq:N_thermal_one_thermal}
\end{equation}
Expanding the above equation at high temperatures yields
\begin{equation}
{S_{\textrm{th}}} = \ln \frac{T}{\omega } + 1 + \frac{{{\omega ^2}}}{{24}}\frac{1}{{{T^2}}} - \frac{{{\omega ^4}}}{{960}}\frac{1}{{{T^4}}} + \mathcal{O}\left( {\frac{1}{{{T^6}}}} \right) ,
\label{eq:N_thermal_highT}
\end{equation}
whereas expanding it at low temperature yields
\begin{equation}
{S_{\textrm{th}}} \simeq \left( {\frac{\omega }{T} + 1} \right){e^{ - \frac{\omega }{T}}} +  \ldots
\end{equation}

\subsection{Two Coupled Harmonic Oscillators}
\label{subsec:two_osc}

Now, let us consider a system of two coupled oscillators at finite temperature. The oscillator described by coordinate $x$ and canonical momentum $p$ is constituting the subsystem $A$, whereas the other oscillator, which obviously coincides with subsystem $A^C$, is described by coordinate $x^C$ and canonical momentum $p^C$. All oscillator masses are taken equal to one. The Hamiltonian of the system is
\begin{equation}
H = \frac{1}{2}\left[ {{p}^2 + \left({p^C}\right)^2 + {k_0}\left( {{x}^2 + \left({x^C}\right)^2} \right) + {k_1}{{\left( {{x^C} - {x}} \right)}^2}} \right] .
\label{eq:hamiltonian_original_coordinates}
\end{equation}
When the Hamiltonian is written in terms of the canonical coordinates,
\begin{equation}
{x_ \pm } \equiv \frac{{{x^C} \pm {x}}}{{\sqrt 2 }},\quad {p_ \pm } \equiv \frac{{{p^C} \pm {p}}}{{\sqrt 2 }} ,
\label{eq:normal_coordinates}
\end{equation}
it assumes the form
\begin{equation}
H = \frac{1}{2}\left( {p_+ ^2 + p_- ^2 + {\omega _ + }^2x_+ ^2 + {\omega _ - }^2x_- ^2} \right) ,
\label{eq:hamiltonian_normal_coordinates}
\end{equation}
where $\omega_\pm$ are the eigenfrequencies of the normal modes, namely, $\omega_+ = \sqrt{k_0}$ and $\omega_- = \sqrt{k_0 + 2 k_1}$.

The Hamiltonian \eqref{eq:hamiltonian_normal_coordinates} describes two decoupled oscillators, corresponding to the two normal modes of the system. It follows that the density matrix that describes the composite system at finite temperature can be trivially written as the tensor product of the thermal density matrix \eqref{eq:ho_density_2}, for each of the two normal modes,
\begin{multline}
\rho \left( {{x_ + },{x_ + }',{x_ - },{x_ - }'} \right) = \rho \left( {{x_ + },{x_ + }'} \right) \otimes \rho \left( {{x_ - },{x_ - }'} \right)\\
= \frac{{\sqrt {\left( {{a_ + } + {b_ + }} \right)\left( {{a_ - } + {b_ - }} \right)} }}{\pi }{e^{ - \frac{{{a_ + }\left( {{x_ + }^2 + {x_ + }{'^2}} \right) + {a_ - }\left( {{x_ - }^2 + {x_ - }{'^2}} \right)}}{2}}}{e^{ - {b_ + }{x_ + }{x_ + }'}}{e^{ - {b_ - }{x_ - }{x_ - }'}},
\end{multline}
where
\begin{equation}
a_\pm \equiv \omega_\pm \coth \frac{\omega_\pm }{T},\quad b_\pm \equiv  - \omega_\pm \csch \frac{\omega_\pm }{T} .
\end{equation}
In order to find the reduced density matrix of the subsystem $A$, this density matrix has to be expressed in terms of the original coordinates $x$ and $x^C$ prior to tracing out the $A^C$ degrees of freedom,
\begin{multline}
\rho \left( {{x},{x}',{x^C},{x^C}'} \right) = \frac{{\sqrt {\left( {{a_ + } + {b_ + }} \right)\left( {{a_ - } + {b_ - }} \right)} }}{\pi } \\
\times {e^{ - \frac{{{a_ + }\left( {{{\left( {{x} + {x^C}} \right)}^2} + {{\left( {{x}' + {x^C}'} \right)}^2}} \right) + {a_ - }\left( {{{\left( {{x^C} - {x}} \right)}^2} + {{\left( {{x^C}' - {x}'} \right)}^2}} \right)}}{4}}}\\
\times {e^{ - \frac{{{b_ + }\left( {{x} + {x^C}} \right)\left( {{x}' + {x^C}'} \right)}}{2}}}{e^{ - \frac{{{b_ - }\left( {{x^C} - {x}} \right)\left( {{x^C}' - {x}'} \right)}}{2}}} .
\end{multline}

We proceed to trace out the degree of freedom of the subsystem $A^C$, integrating out $x^C$. After some simple algebra we find
\begin{equation}
\rho \left( {{x},{x}'} \right) = \int {d{x^C}\rho \left( {{x},{x}',{x^C},{x^C}} \right)} = \sqrt {\frac{{\gamma  - \beta }}{\pi }} {e^{ - \frac{{\left( {{x}^2 + {x}{'^2}} \right)\gamma }}{2}}}{e^{{x}{x}'\beta }} ,
\label{eq:two_osc_reduced}
\end{equation}
where
\begin{equation}
\gamma  - \beta  = 2\frac{{\left( {{a_ + } + {b_ + }} \right)\left( {{a_ - } + {b_ - }} \right)}}{{ {{a_ + } + {a_ - } + {b_ + } + {b_ - }} }} , \quad \gamma + \beta = \frac{1}{2} \left( {{a_ + } + {a_ - } - {b_ + } - {b_ - }} \right) .
\end{equation}

Similarly to the ground state case analysis \cite{Srednicki:1993im}, one can show that the functions
\begin{equation}
{f_n}\left( x \right) = {H_n}\left( {\sqrt \alpha  x} \right){e^{ - \frac{{\alpha {x^2}}}{2}}} ,
\label{eq:two_eigenfunctions}
\end{equation}
where
\begin{equation}
\alpha \equiv \sqrt {{\gamma ^2} - {\beta ^2}}  = \sqrt {\frac{{\left( {{a_ + } + {b_ + }} \right)\left( {{a_ - } + {b_ - }} \right)\left( {{a_ + } + {a_ - } - {b_ + } - {b_ - }} \right)}}{{{a_ + } + {a_ - } + {b_ + } + {b_ - }}}} ,
\end{equation}
are the eigenfunctions of the reduced density matrix. The respective eigenvalues are
\begin{equation}
{p_n} = \left( {1 - \frac{\beta }{{\gamma  + \alpha }}} \right){\left( {\frac{\beta }{{\gamma  + \alpha }}} \right)^n} \equiv \left( {1 - \xi } \right){\xi ^n} ,
\end{equation}
where
\begin{equation}
\xi \equiv \frac{\beta }{{\gamma  + \alpha }} = \frac{{\sqrt {\frac{{\gamma  + \beta }}{{\gamma  - \beta }}} - 1}}{{\sqrt {\frac{{\gamma  + \beta }}{{\gamma  - \beta }}} + 1}} .
\end{equation}
This can be expressed in terms of the physical quantities of the problem, i.e. the eigenfrequancies of the normal modes and the temperature,
\begin{equation}
\xi  = \frac{{\frac{1}{2} \left( {\frac{1}{\omega _ + }\coth \frac{{{\omega _ + }}}{{2T}} + \frac{1}{\omega _ - }\coth \frac{{{\omega _ - }}}{{2T}}} \right) ^{\frac{1}{2}} \left( {{\omega _ + }\coth \frac{{{\omega _ + }}}{{2T}} + {\omega _ - }\coth \frac{{{\omega _ - }}}{{2T}}} \right)^{\frac{1}{2}} - 1}}{{\frac{1}{2} \left( {\frac{1}{\omega _ + }\coth \frac{{{\omega _ + }}}{{2T}} + \frac{1}{\omega _ - }\coth \frac{{{\omega _ - }}}{{2T}}} \right)^{\frac{1}{2}} \left( {{\omega _ + }\coth \frac{{{\omega _ + }}}{{2T}} + {\omega _ - }\coth \frac{{{\omega _ - }}}{{2T}}} \right)^{\frac{1}{2}} + 1}} .
\end{equation}
Then, it is straightforward to calculate the entanglement entropy, which equals
\begin{equation}
S_A =  - \ln \left( {1 - \xi } \right) - \frac{\xi }{{1 - \xi }}\ln \xi .
\label{eq:two_osc_SA}
\end{equation}

As we argued in the introduction, the entanglement entropy is not a very good measure for the quantum entanglement when the overall system lies at a mixed state, like the scenario under consideration. In general, it contains a contribution from the thermal entropy of the overall system. Indeed, the entanglement entropy does not vanish at the limit $k_1 \to 0$ as one would expect from a good measure of quantum entanglement. It rather tends to the thermal entropy of a single oscillator with eigenfrequency $\sqrt{k_0}$ at temperature $T$. In the case of the two coupled oscillators that we study here, it holds that $S_{A^C} = S_A$, due to the symmetry of the system. Therefore, the mutual information is given by,
\begin{equation}
I\left( {A:{A^C}} \right) = 2 {S_A} - S_{\textrm{th}} ,
\end{equation}
where $S_A$ is given by \eqref{eq:two_osc_SA} and $S_{\textrm{th}}$ is obviously given by the sum of two versions of equation \eqref{eq:N_thermal_one_thermal}, one for each normal mode.

\subsection{Similarity to a Single Harmonic Oscillator}

One may observe that the reduced density matrix \eqref{eq:two_osc_reduced} is identical to the thermal density matrix of a single harmonic oscillator \eqref{eq:ho_density_2}, after some appropriate identifications. More specifically, there is no experiment that can be performed to the one of the two coupled oscillators at finite temperature $T$ that can distinguish it from a single \emph{effective} harmonic oscillator with eigenfrequency equal to
\begin{equation}
\omega_{\textrm{eff}} = \alpha
\label{eq:effective_omega}
\end{equation}
at an effective temperature equal to
\begin{equation}
T_{\textrm{eff}} =  - \frac{\alpha }{{\ln \xi }} .
\label{eq:effective_T}
\end{equation}
The latter is always higher than the physical temperature $T$.

This identification obeys some obvious consistency checks. For example, at the limit $k_1 \to 0$, the two oscillators become decoupled, each having eigenfrequency equal to $\sqrt{k_0}$. It follows that at this limit, the system is separable, i.e. $\rho = \rho_1 \otimes \rho_2$, and, thus, the reduced density matrix should be identical to $\rho_1$, i.e. the thermal density matrix of a single harmonic oscillator with eigenfrequency $\sqrt{k_0}$ at temperature $T$. Indeed, expanding $\omega_{\textrm{eff}}$ and $T_{\textrm{eff}}$ around $k_1 = 0$ yields
\begin{align}
{\omega _{\textrm{eff}}} &= \sqrt {{k_0}}  + \frac{1}{{2\sqrt {{k_0}} }}{k_1} - \left( {\frac{3}{{8\sqrt {k_0^3} }} + \frac{{\csch\frac{{\sqrt {{k_0}} }}{T}}}{{4{k_0}T}}} \right)k_1^2 + \mathcal{O}\left( {k_1^3} \right) ,\\
{T_{\textrm{eff}}} &= T + \frac{1}{{8\sqrt {k_0^5} }}\left( { - \sqrt {{k_0}} T + {T^2}\sinh \frac{{\sqrt {{k_0}} }}{T} + {k_0}\tanh \frac{{\sqrt {{k_0}} }}{{2T}}} \right)k_1^2 + \mathcal{O}\left( {k_1^3} \right) .
\end{align}

Similarly, at the limit $T \to 0$, one finds the following
\begin{align}
{\omega _{\textrm{eff}}} &= \omega _{\textrm{eff}}^0\left[ {1 + \frac{{{\omega _ - } - {\omega _ + }}}{{{\omega _ - } + {\omega _ + }}}\left( {{e^{ - \frac{{{\omega _ - }}}{T}}} - {e^{ - \frac{{{\omega _ + }}}{T}}}} \right) + \dots} \right] , \\
{T_{\textrm{eff}}} &= T_{\textrm{eff}}^0 \left[ {1 + \frac{{{\omega _ - } - {\omega _ + }}}{{{\omega _ - } + {\omega _ + }}}\left( {{e^{ - \frac{{{\omega _ - }}}{T}}} - {e^{ - \frac{{{\omega _ + }}}{T}}}} \right) + \frac{{ 4 \left( {\omega _ - } + {\omega _ + } \right) T_{\textrm{eff}}^0 }}{{{{\left( {{\omega _ - } - {\omega _ + }} \right)}^2}}}\left( {{e^{ - \frac{{{\omega _ - }}}{T}}} + {e^{ - \frac{{{\omega _ + }}}{T}}}} \right) + \dots} \right] , 
\end{align}
where
\begin{equation}
\omega _{\textrm{eff}}^0 = \sqrt {{\omega _ + }{\omega _ - }} , \quad T_{\textrm{eff}}^0 = - \frac{{\omega _{\textrm{eff}}^0}}{{\ln {\xi ^0}}} , \quad {\xi ^0} = {\left( {\frac{{\sqrt {{\omega _ - }}  - \sqrt {{\omega _ + }} }}{{\sqrt {{\omega _ - }}  + \sqrt {{\omega _ + }} }}} \right)^2} .
\end{equation}
Therefore, we recover correctly the ground state result \cite{Srednicki:1993im}. At low temperatures the corrections to the zero-temperature values of $\omega _{\textrm{eff}}$ and $T _{\textrm{eff}}$ are exponentially suppressed and tend to reduce the eigenfrequency of the effective oscillator, whereas they tend to increase its temperature. This expansion is an asymptotic expansion and not a usual Taylor series. This is due to the fact that the involved functions are not analytic at $T=0$. The results are expressed at first order in the exponentials ${e^{ - \frac{{{\omega _ \pm }}}{T}}}$, but one has to be careful with this kind of expansion; for example, depending on the values of $\omega_\pm$, the second order term in the exponential of $\omega_+$ may be a more significant contribution that the first order term in the exponential of $\omega_-$.

In a similar manner at high temperatures we find
\begin{align}
{\omega _{{\rm{eff}}}} &= \sqrt {\frac{{2\omega _ + ^2\omega _ - ^2}}{{\omega _ + ^2 + \omega _ - ^2}}} \left[ {1 + \frac{1}{{48}}\frac{{{{\left( {\omega _ + ^2 - \omega _ - ^2} \right)}^2}}}{{\omega _ + ^2 + \omega _ - ^2}}\frac{1}{{{T^2}}} + \mathcal{O}\left( {\frac{1}{{{T^4}}}} \right)} \right] , \\
{T_{{\rm{eff}}}} &= T \left[ {1 + \frac{1}{{24}}\frac{{{{\left( {\omega _ + ^2 - \omega _ - ^2} \right)}^2}}}{{\omega _ + ^2 + \omega _ - ^2}}\frac{1}{{{T^2}}} + \mathcal{O}\left( {\frac{1}{{{T^4}}}} \right)} \right] .
\end{align}
This implies that at high temperatures, the eigenfrequency of the effective oscillator tends to a finite given value,
\begin{equation}
{\omega_{{\rm{eff}}}^\infty} = \sqrt {\frac{{2\omega _ + ^2\omega _ - ^2}}{{\omega _ + ^2 + \omega _ - ^2}}} ,
\label{eq:effective_omega_infinity}
\end{equation}
whereas the effective temperature is dominated by the physical temperature of the composite system.

A very interesting question that can be posted is whether the fact that the subsystem $A$ can be described by an effective thermal reduced density matrix can be attributed to the eigenstate thermalization hypothesis \cite{Srednicki_ETH}. Naturally, this should not be expected, since the system under consideration is integrable.

When we consider either a thermal state or the ground state for the overall system, its density matrix is time independent. This implies that the same holds for the reduced density matrix, which describes the considered subsystem. However, the subsystem is an open system, and, thus, a time-independent state, has to be a non-trivial state that describes a system in equilibrium with its environment (not necessarily thermal).

This behaviour becomes clearer in the case of many harmonic oscillators that we are about to study in next section. There, we will analyse a system of $N$ coupled oscillators, considering as subsystem $A$ an arbitrary subset comprising of $n$ oscillators. Although we are not going to discuss on the similarity of the reduced density matrix to the density matrix of a harmonic system of $n$ oscillators at an appropriate state, the entanglement entropy is identical to the sum of the thermal entropies of $n$ effective oscillators, each lying at a different temperature. This is consistent with the picture of a harmonic system with $n$ degrees of freedom, where each normal mode has been heated to a different temperature. Since, the normal modes of a harmonic system do not interact, this is an equilibrium, time-independent state, which nevertheless is not thermal. It follows that the reduced system is not thermalized; actually, it is as far as possible from a thermalized state, as imposed by its integrability. 

In the case of the two coupled oscillators, the considered subsystem contains a single degree of freedom, and thus, such a state is a thermal one. Thus, the fact that the reduced density matrix appears to be thermal is not a consequence of thermalization, but rather a technical coincidence due to the specific selection of the state of the overall system and the number of the degrees of freedom.

\subsection{High and Low Temperature Expansions}

At temperatures much higher than the system eigenfrequencies, the entanglement entropy and mutual information have asymptotic expansions of the form
\begin{multline}
{S_A} = \frac{1}{2}\ln \frac{{\left( {{k_0} + {k_1}} \right){T^2}}}{{{k_0}\left( {{k_0} + 2{k_1}} \right)}} + 1 + \frac{{{k_0} + {k_1}}}{{24{T^2}}}\\
 + \frac{{3k_0^4 + 12k_0^3{k_1} + 20k_0^2k_1^2 + 16{k_0}k_1^3 + 9k_1^4}}{{2880{{\left( {{k_0} + {k_1}} \right)}^2}{T^4}}} + \mathcal{O}\left( {\frac{1}{{{T^6}}}} \right)
\label{eq:2_SEE_highT_expansion}
\end{multline}
and
\begin{equation}
I\left( {A:{A^C}} \right) = \frac{1}{2}\ln \frac{{{{\left( {{k_0} + {k_1}} \right)}^2}}}{{{k_0}\left( {{k_0} + 2{k_1}} \right)}} + \frac{{k_1^2\left( {{k_0} - {k_1}} \right)\left( {{k_0} + 3{k_1}} \right)}}{{1440{{\left( {{k_0} + {k_1}} \right)}^2}{T^4}}} + \mathcal{O}\left( {\frac{1}{{{T^6}}}} \right) ,
\label{eq:2_I_highT_expansion}
\end{equation}
respectively. Notice that the coefficients of the high temperature expansion of the mutual information do vanish when the oscillators are decoupled, i.e. $k_1 \to 0$, as expected. Furthermore, the coefficient of the $1 / T^2$ term in the mutual information vanishes, which turns out to be a more general feature, as we will show in next section.

Finally, it is evident that the mutual information does not vanish at infinite temperature, but rather it tends to the value
\begin{equation}
I^\infty = \frac{1}{2}\ln \frac{{{{\left( {{k_0} + {k_1}} \right)}^2}}}{{{k_0}\left( {{k_0} + 2{k_1}} \right)}} = 2 \ln \frac{\omega _{\textrm{eff}}^0}{\omega _{\textrm{eff}}^\infty}.
\label{eq:2_mutual_infinity}
\end{equation}
It is well known that in qubit systems, the mutual information vanishes at infinite temperature. It is natural to wonder what is the underlying reason for this difference between qubits and oscillators. The answer to this seeming inconsistency is related to the dimensionality of the Hilbert space of our problem. In all qubit systems, the related Hilbert spaces are finite dimensional. Trivially, at the infinite temperature limit, the density matrices of the composite system tends to
\begin{equation}
\mathop {\lim }\limits_{T \to \infty } \rho = \frac{1}{{\dim {\mathcal{H}_{A \cup {A^C}}}}}{I_{\dim {\mathcal{H}_{A \cup {A^C}}}}}.
\end{equation}
This is a separable density matrix, implying trivially that
\begin{equation}
\mathop {\lim }\limits_{T \to \infty } {\rho _A} = \frac{1}{{\dim {\mathcal{H}_A}}}{I_{\dim {\mathcal{H}_A}}},\quad \mathop {\lim }\limits_{T \to \infty } {\rho _{{A^C}}} = \frac{1}{{\dim {\mathcal{H}_{{A^C}}}}}{I_{\dim {\mathcal{H}_{{A^C}}}}} .
\end{equation}

It follows that the entanglement entropies tend to
\begin{equation}
\mathop {\lim }\limits_{T \to \infty } {S_A} = \ln \dim {\mathcal{H}_A},\quad \mathop {\lim }\limits_{T \to \infty } {S_{{A^C}}} = \ln \dim {\mathcal{H}_{{A^C}}} ,
\end{equation}
whereas the thermal entropy tends to
\begin{equation}
\mathop {\lim }\limits_{T \to \infty } {S_{A \cup {A^C}}} = \ln {\dim {\mathcal{H}_{A \cup {A^C}}}} .
\end{equation}
The above imply that the mutual information vanishes at infinite temperature,
\begin{equation}
\mathop {\lim }\limits_{T \to \infty } I\left( {A:{A^C}} \right) = 0 .
\end{equation}
However, in our case the corresponding Hilbert spaces are infinite dimensional and the above arguments cannot be applied equally well. Both entanglement entropies ${S_A}$ and ${S_{{A^C}}}$ diverge at infinite temperature as $\ln T$. This divergence is cancelled in the mutual information, via the same mechanism that enforces the mutual information to vanish in qubit systems; however, there is a finite remnant.

In general, the mutual information measures both classical and quantum correlations. So, another natural question concerns the origin of this mutual information remnant at infinite temperature. The mutual information $I^\infty$ coincides with the mutual information that one can calculate via a classical analysis, as shown in the appendix \ref{sec:classical_mutual} (see also \cite{Cramer:2005mx}). Therefore, this infinite temperature remnant should be attributed solely to classical correlations. As intuitively expected, at infinite temperature the classical fluctuations completely dominate and yield the quantum fluctuations irrelevant.

This is also in line to the fact that another measure of quantum entanglement, the entanglement negativity, also vanishes at infinite temperature. Actually, the negativity vanishes above a finite critical temperature, as shown in appendix \ref{sec:negativity}, a phenomenon widely known as sudden death of entanglement. However, this does not necessarily imply that there is really such a finite temperature phase transition in the system of coupled oscillators. The absence of negativity is not a proof of lack of entanglement in infinite dimensional Hilbert spaces, as in finite dimensional ones \cite{Peres_PH,Horodecki_PH}. This issue requires further investigation. 

At low temperatures, the entanglement entropy tends to the zero temperature result, plus exponentially suppressed corrections
\begin{equation}
{S_A} = S_A^0 + \frac{{{\omega _ - } + {\omega _ + }}}{{4T_{\mathrm{eff}}^0}}\left( {{e^{ - \frac{{{\omega _ - }}}{T}}} + {e^{ - \frac{{{\omega _ + }}}{T}}}} \right) +\dots .
\label{eq:2_SEE_smallT_expansion}
\end{equation}
Similarly, the mutual information is equal to
\begin{multline}
I\left( {A:{A^C}} \right) = 2S_A^0 + \left( {\frac{{{\omega _ - } + {\omega _ + }}}{{2T_{\mathrm{eff}}^0}} - \frac{{{\omega _ - }}}{T} - 1} \right){e^{ - \frac{{{\omega _ - }}}{T}}} \\
+ \left( {\frac{{{\omega _ - } + {\omega _ + }}}{{2T_{\mathrm{eff}}^0}} - \frac{{{\omega _ + }}}{T} - 1} \right){e^{ - \frac{{{\omega _ + }}}{T}}} +\dots .
\label{eq:2_I_smallT_expansion}
\end{multline}

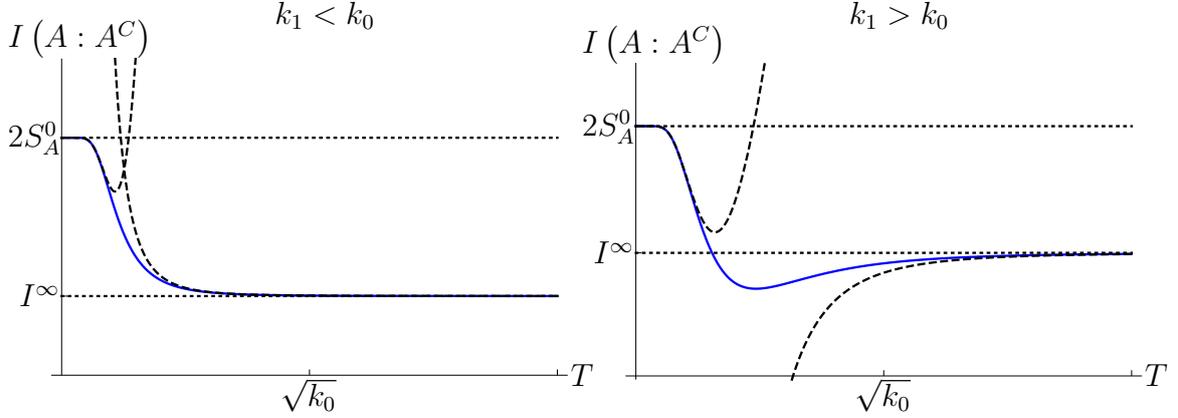
\begin{figure}[ht]
\centering
\begin{picture}(100,38)
\put(3.5,2.5){\includegraphics[width = 0.45\textwidth]{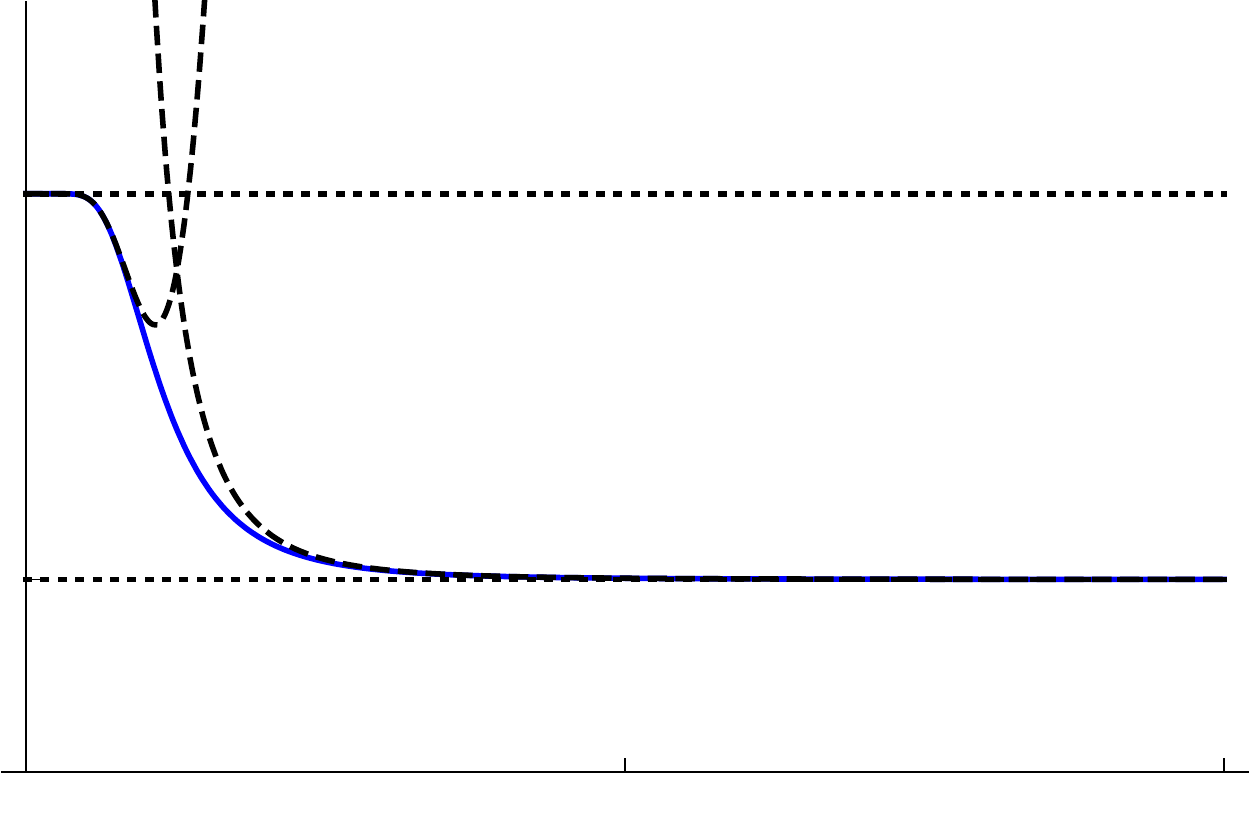}}
\put(53.5,2.5){\includegraphics[width = 0.45\textwidth]{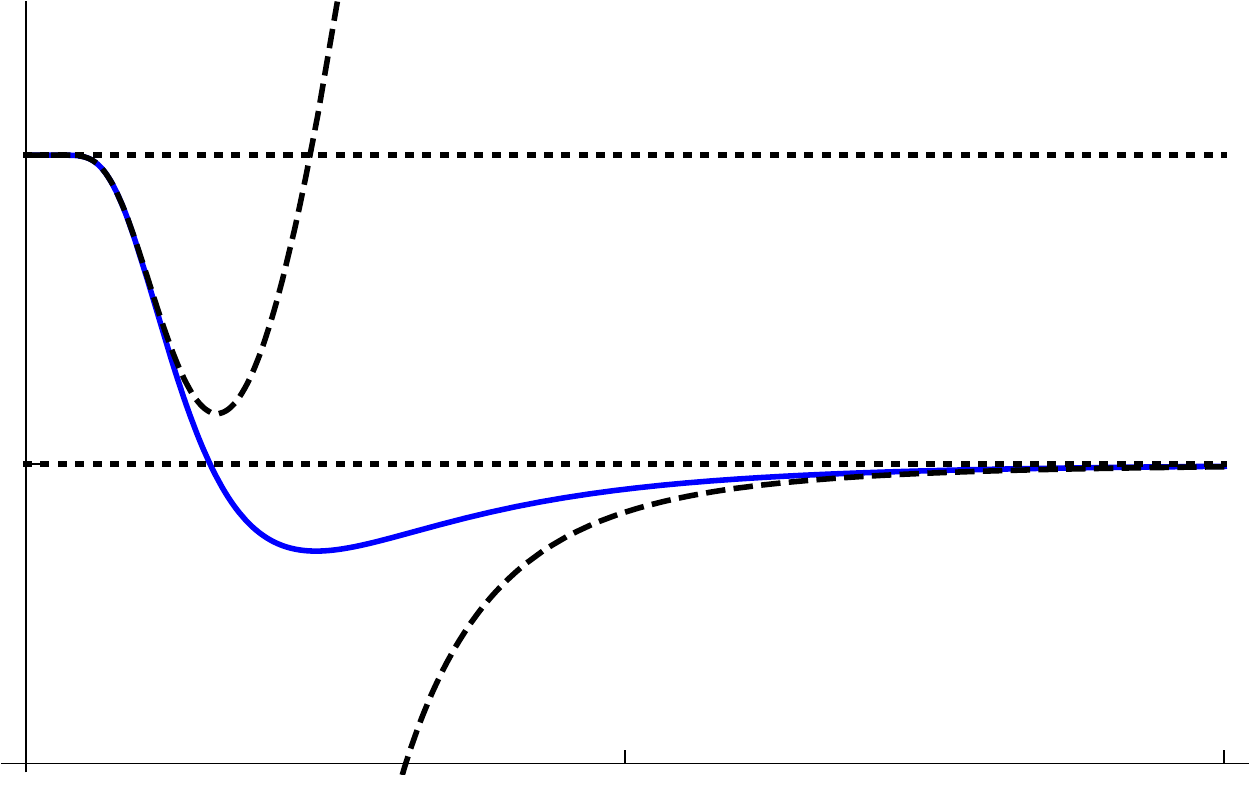}}
\put(23,35){$k_1 < k_0$}
\put(73,35){$k_1 > k_0$}
\put(-0.25,33){$I\left( {A:{A^C}} \right)$}
\put(49.75,32.5){$I\left( {A:{A^C}} \right)$}
\put(48.75,3.25){$T$}
\put(98.75,3.25){$T$}
\put(-0.25,24.25){$2S_A^0$}
\put(49.75,25.25){$2S_A^0$}
\put(0.75,10.25){$I^\infty$}
\put(50.75,14){$I^\infty$}
\put(23.5,1.5){$\sqrt{k_0}$}
\put(73.5,1.5){$\sqrt{k_0}$}
\end{picture}
\caption{The mutual information as function of the temperature. The dashed lines are the low and high temperature expansions of the mutual information, whereas the dotted lines are the asymptotic values for $T \to 0$ and $T \to \infty$.}
\label{fig:2osc_mi}
\end{figure}

As shown in figure \ref{fig:2osc_mi}, where the mutual information is plotted as a function of the temperature, the mutual information may be a monotonous function of the temperature or not. This depends on the relevant magnitude of the couplings $k_0$ and $k_1$, which determines the sign of the coefficient of the $1/T^4$ term in the high temperature expansion.

\setcounter{equation}{0}
\section{Arbitrary Number of Harmonic Oscillators}
\label{sec:qm_many}

\subsection{Entanglement Entropy and Mutual Information}
Building on the results of Section \ref{sec:qm_2}, we proceed to study a system of $N$ coupled harmonic oscillators. In this analysis, the subsystem $A^C$ coincides with any subset of $n$ oscillators. Without loss of generality, all oscillators are considered having unit mass. The Hamiltonian is given by
\begin{equation}
H = \frac{1}{2}\sum\limits_{i = 1}^N {{p_i}^2}  + \frac{1}{2}\sum\limits_{i,j = 1}^N {{x_i}{K_{ij}}{x_j}} .
\end{equation}
The matrix $K$ is symmetric and all its eigenvalues are positive, since the above Hamiltonian should describe an oscillatory system around a stable equilibrium. Writing down the Hamiltonian in terms of the normal coordinates $y_i$, which are related to the initial coordinates $x_i$ via an orthogonal transformation $O$, yields
\begin{equation}
H = \frac{1}{2}\sum\limits_{i = 1}^N {{q_i}^2}  + \frac{1}{2}\sum\limits_{i = 1}^N {{\omega _i}^2{y_i}^2} ,
\end{equation}
where $\omega_i$ are the frequencies of the normal modes. In other words, the orthogonal transformation $O$ diagonalizes the matrix $K$, or
\begin{equation}
K = {O^T}{K_D}O ,
\end{equation}
where $\left( K_D \right)_{ij} = \omega_i^2 \delta_{ij}$.

We define the matrices
\begin{equation}
a = \sqrt K \coth\frac{{\sqrt K }}{T},\quad b =  - \sqrt K \csch\frac{{\sqrt K }}{T}.
\label{eq:N_ab_defs}
\end{equation}
These matrices can be related to the eigenfrequencies of the system as
\begin{equation}
a = {O^T}{a_D}O ,\quad b = {O^T}{b_D}O,
\end{equation}
where
\begin{equation}
\left( a_D \right)_{ij} = \omega_i \coth\frac{{\omega_i }}{T} \delta_{ij} \equiv a_i \delta_{ij}, \quad \left( b_D \right)_{ij} =  - \omega_i \csch\frac{{\omega_i }}{T} \delta_{ij} \equiv b_i \delta_{ij}.
\end{equation}

Since the oscillators corresponding to the normal modes are decoupled, the density matrix of the overall system can be written as the tensor product of the density matrices corresponding to each of the normal modes,
\begin{equation}
\begin{split}
\rho \left( {{\bf{y}},{\bf{y}}'} \right) &= \bigotimes\limits_{i = 1}^N \rho \left( {{y_i},{y_1}'} \right) \\
&= \prod\limits_{i = 1}^N {\sqrt {\frac{{{a_i} + {b_i}}}{\pi }} {e^{ - \frac{{{a_i}}}{2}\left( {y_i^2 + {y_i}{'^2}} \right) - {b_i}{y_i}{y_i}'}}} \\
&= \sqrt {\frac{{\det \left( {{a_D} + {b_D}} \right)}}{{{\pi ^N}}}} {e^{ - \frac{{{{\bf{y}}^T}{a_D}{\bf{y}} + {\bf{y}}{'^T}{a_D}{\bf{y}}'}}{2}}}{e^{ - {{\bf{y}}^T}{b_D}{\bf{y}}'}} .
\end{split}
\end{equation}
We express the density matrix in terms of the original $x$ coordinates, using the orthogonal transformation $O$,
\begin{equation}
\rho \left( {{\bf{x}},{\bf{x}}'} \right) = \sqrt {\frac{{\det \left( {a + b} \right)}}{{{\pi ^N}}}} {e^{ - \frac{{{{\bf{x}}^T}a{\bf{x}} + {\bf{x}}{'^T}a{\bf{x}}'}}{2}}}{e^{ - {{\bf{x}}^T}b{\bf{x}}'}} .
\label{eq:many_thermal_rho}
\end{equation}

In the following, we use the block form notation
\begin{equation}
{\bf{x}} = \left( {\begin{array}{*{20}{c}}
{{x^C}}\\
{{x}}
\end{array}} \right), \quad
\textrm{where}
\quad 
x^C = \left( {\begin{array}{*{20}{c}}
{{x_1}}\\
 \vdots \\
{{x_n}}
\end{array}} \right),\quad x = \left( {\begin{array}{*{20}{c}}
{{x_{n+1}}}\\
 \vdots \\
{{x_N}}
\end{array}} \right) .
 \label{eq:x_block}
\end{equation}
We will also write any symmetric matrix $M$ in block form, using the notation
\begin{equation}
M = \left( {\begin{array}{*{20}{c}}
{{M_A}}&{{M_B}}\\
{M_B^T}&{{M_C}}
\end{array}} \right) ,
\end{equation}
where $M_A$ is an $n\times n$ matrix, $M_C$ is an $\left( N - n \right) \times \left( N - n \right)$ matrix and finally $M_B$ is an $n\times \left( N - n \right)$ matrix. The indices $A$, $B$ and $C$ will always indicate the corresponding blocks of such matrices. Then, the density matrix $\rho \left( {{\bf{x}},{\bf{x}}'} \right)$ can be expressed as,
\begin{multline}
\rho \left( {{\bf{x}},{\bf{x}}'} \right) = \sqrt {\frac{{\det \left( {a + b} \right)}}{{{\pi ^N}}}} {e^{ - \frac{{{x^C}^T a_A {x^C} + 2{x^C}^T a_B {x} + {x}^T a_C {x} + {x^C}{'^T} a_A {x^C}' + 2{x^C}{'^T} a_B {x}' + {x}{'^T} a_C {x}'}}{2}}}\\
\times {e^{ - \left( {{x^C}^T b_A {x^C}' + {x^C}^T b_B {x}' + {x^C}{'^T} b_B{x} + {x}^T b_C{x}'} \right)}} .
\end{multline}

We proceed to trace out the first $n$ degrees of freedom to find the reduced density matrix for the remaining $N-n$ ones. Simple algebra with Gaussian integrals yields
\begin{equation}
\begin{split}
&{\rho}\left( {{x},{x}'} \right) = \int {dx^C {\rho}\left( {\left\{ {x},{x^C} \right\},\left\{ {x}',{x^C} \right\}} \right) }\\
&= \sqrt {\frac{{\det \left( {a + b} \right)}}{{{\pi ^N}}}} \int {\left( \prod\limits_{i = 1}^n {d{x_i}} \right) {e^{ - {x^C}^T\left( {a_A + b_A} \right){x^C} + {x^C}^T\left( {a_B + b_B} \right)\left( {{x} + {x}'} \right)}}} {{e^{ - \frac{{{x}^T a_C {x} + {x}{'^T} a_C {x}' + 2{x}^T b_C {x}'}}{2}}}} \\
&= \sqrt {\frac{{\det \left( {\gamma  - \beta } \right)}}{{{\pi ^{N - n}}}}} {e^{ - \frac{{{x}^T\gamma {x} + {x}{'^T}\gamma {x}'}}{2}}}{e^{{x}^T\beta {x}'}} ,
\end{split}
\end{equation}
where
\begin{align}
\gamma  &= a_C - \frac{1}{2}\left( {{a_B^T} + b_B^T} \right){\left( {a_A + b_A} \right)^{ - 1}}\left( {a_B + b_B} \right) , \label{eq:N_gamma_def}\\
\beta  &=  - b_C + \frac{1}{2}\left( {{a_B^T} + b_B^T} \right){\left( {a_A + b_A} \right)^{ - 1}}\left( {a_B + b_B} \right) . \label{eq:N_beta_def}
\end{align}
Similarly to the ground state case \cite{Srednicki:1993im}, one may find the spectrum of the reduced density matrix, via the explicit construction of its eigenfunctions. It reads
\begin{equation}
{p_{{n_{n + 1}}, \ldots ,{n_N}}} = \prod\limits_{i = n + 1}^N {\left( {1 - {\xi _i}} \right)\xi _i^{{n_i}}} ,\quad {n_i} \in \mathbb{Z} ,
\end{equation}
where the quantities $\xi_i$ are given by
\begin{equation}
{\xi _i} = \frac{{{\beta _{Di}}}}{{1 + \sqrt {1 - \beta _{Di}^2} }}
\label{eq:N_xi_of_beta}
\end{equation}
and $\beta _{Di}$ are the eigenvalues of the matrix $\gamma^{-1}\beta$.
It follows that the entanglement entropy is given by
\begin{equation}
S = \sum\limits_{j = n + 1}^N {\left( { - \ln \left( {1 - {\xi _j}} \right) - \frac{{{\xi _j}}}{{1 - {\xi _j}}}\ln {\xi _j}} \right)} .
\label{eq:N_S_of_xi}
\end{equation}
Notice that this formula is identical to the formula that would provide the thermal entropy of $n$ independent oscillators, each with eigenfrequency $\sqrt {1 - \beta _{Di}^2}$ and at temperature $- \sqrt {1 - \beta _{Di}^2} / \ln \xi_i$.

As a consistency check, let us consider the special case where the two subsystems are decoupled, i.e. $K_B=0$. It holds that
\begin{align}
a &= \left( {\begin{array}{*{20}{c}}
{\sqrt {{K_A}} \coth \frac{{\sqrt {{K_A}} }}{T}}&0\\
0&{\sqrt {{K_C}} \coth \frac{{\sqrt {{K_C}} }}{T}}
\end{array}} \right), \\
b &=  - \left( {\begin{array}{*{20}{c}}
{\sqrt {{K_A}} \csch \frac{{\sqrt {{K_A}} }}{T}}&0\\
0&{\sqrt {{K_C}} \csch \frac{{\sqrt {{K_C}} }}{T}}
\end{array}} \right).
\end{align}
In this case, it is straightforward that
\begin{align}
\gamma  &= {a_C} = \sqrt {{K_C}} \coth \frac{{\sqrt {{K_C}} }}{T} , \\
\beta  &=  - {b_C} = \sqrt {{K_C}} {\mathop{\rm csch}\nolimits} \frac{{\sqrt {{K_C}} }}{T} , \\
{\gamma ^{ - 1}}\beta  &= {\mathop{\rm sech}\nolimits} \frac{{\sqrt {{K_C}} }}{T} .
\end{align}
Therefore the eigenvalues $\beta_{Di}$ of the matrix ${\gamma ^{ - 1}}\beta$ can be expressed in terms of the eigenvalues of the matrix $K_C$, i.e. the eigenfrequencies $\omega _i$ of the decoupled subsystem $A$. Notice that the eigenfrequencies, as well as the thermal entropy of the subsystem $A$ are well defined in this limit, since the two subsystems are decoupled. The eigenvalues $\beta_{Di}$ read
\begin{equation}
{\beta _{Di}} = {\mathop{\rm sech}\nolimits} \frac{{{\omega _i}}}{T} .
\end{equation}
It follows that
\begin{equation}
{\xi _i} = \frac{{{\mathop{\rm sech}\nolimits} \frac{{{\omega _i}}}{T}}}{{1 + \sqrt {1 - {{{\mathop{\rm sech}\nolimits} }^2}\frac{{{\omega _i}}}{T}} }} = {e^{ - \frac{{{\omega _i}}}{T}}} .
\end{equation}
Comparing equations \eqref{eq:N_thermal_one_thermal} and \eqref{eq:N_S_of_xi}, we conclude that in the $K_B = 0$ case, the entanglement entropy is simply equal to the thermal entropy of the subsystem $A$. This is expected, since at this limit, the composite system density matrix is separable. This also implies that the mutual information vanishes at this limit.

\subsection{High and Low Temperature Expansions}

A high temperature expansion of the above result can be performed. The details are included in the appendix \ref{sec:highT}. The high temperature expansions of the entanglement entropy and mutual information are
\begin{multline}
S_A = - \frac{1}{2}\ln \det \frac{{K_C} - K_B^T{\left( {{K_A}} \right)^{ - 1}}{K_B}}{T^2} + N - n + \frac{1}{{24T^2}} \tr {K_C} \\
- \frac{1}{2880 T^4} \left\{ 3 \tr {\left( {{K^2}} \right)_C} + 4 \tr \left[ {{{\left( {K_B^T{{\left( {{K_A}} \right)}^{ - 1}}{K_B}} \right)}^2}} \right] - \tr \left( {K_B^T{K_B}} \right) \right\} \\
+ \mathcal{O} \left( \frac{1}{T^6} \right) 
\label{eq:N_EE_high_T}
\end{multline}
and
\begin{multline}
I\left( {A:{A^C}} \right) = - \frac{1}{2}\ln \det \left[ {I - {{\left( {{K_A}} \right)}^{ - 1}}{K_B}{{\left( {{K_C}} \right)}^{ - 1}}K_B^T} \right] + \frac{0}{T^2} \\
- \frac{1}{{720 T^4}}\left\{ {\tr \left[ {{{\left( {K_B^T{{\left( {{K_A}} \right)}^{ - 1}}{K_B}} \right)}^2}} \right] + \tr \left[ {{{\left( {{K_B}{{\left( {{K_C}} \right)}^{ - 1}}K_B^T} \right)}^2}} \right] - \frac{1}{2}\tr \left( {K_B^T{K_B}} \right)} \right\} \\
+ \mathcal{O} \left( \frac{1}{T^6} \right) ,
\label{eq:N_MI_high_T}
\end{multline}
respectively. Interestingly, \emph{the coefficient of $1/T^2$ in the high temperature expansion of the mutual information vanishes for any system}. It is trivial to show that in the case of the two oscillators, where the matrices of the above formula are simply numbers, namely, $K_A = K_C = k_0 + k_1$ and $K_B = - k_1$, the above formulae reproduce the expansions \eqref{eq:2_SEE_highT_expansion} and \eqref{eq:2_I_highT_expansion}. Furthermore, in the case where the two subsystems are decoupled, i.e. the matrix $K_B$ vanishes, the above formula implies that the first terms in the expansion of the mutual information are vanishing, as expected.

At low temperatures, the situation is a little less transparent. As in the case of the two oscillators, the involved functions are not analytical at $T=0$. Nevertheless, we may obtain an asymptotic expansion, approximating the hyperbolic functions with exponentials. It turns out that the matrix $\gamma^{-1} \beta$, whose eigenvalues determine the entanglement entropy is given in this expansion by
\begin{multline}
\left(\gamma^{-1}\beta\right) = \left(\gamma^{-1}\beta\right)^{(0)} + \left(1-\left(\gamma^{-1}\beta\right)^{(0)}\right)\left(\tilde{\Omega}_C-\tilde{\Omega}_B^T\Omega_A^{-1}\Omega_B\right)\left(1+\left(\gamma^{-1}\beta\right)^{(0)}\right)\\
+\left(\gamma^{-1}\right)^{(0)}\left(\Omega \tilde{\Omega}\right)_C\left(1-\left(\gamma^{-1}\beta\right)^{(0)}\right) + \dots,
\label{eq:N_gammabeta_low_T}
\end{multline}
where $\left(\gamma^{-1}\beta\right)^{(0)}$ is the matrix $\left(\gamma^{-1}\beta\right)$ at zero temperature and
\begin{equation}
\tilde{\Omega}=\mathrm{Exp}\left(-\Omega/T\right) , \quad \Omega = \sqrt{K} .
\end{equation}
The details of this calculation are included in the appendix \ref{sec:lowT}. It is not possible to obtain a generic expression for the low temperature expansion of entanglement entropy or the mutual information in this limit. However, the equation \eqref{eq:N_gammabeta_low_T} implies that at low temperatures the corrections to the zero temperature result are exponentially suppressed as $\exp \left( - \omega_i / T \right)$, where $\omega_i$ are the eigenfrequencies of the overall system. In the case of the two oscillators, it can be shown that the above formula correctly reproduces the results \eqref{eq:2_SEE_smallT_expansion} and \eqref{eq:2_I_smallT_expansion}.

\setcounter{equation}{0}
\section{Chains of Oscillators}
\label{sec:chain}

In this section, we consider systems of coupled oscillators, with the specific property that only adjacent degrees of freedom are coupled. In other words, we consider a coupling matrix $K$ of the form
\begin{equation}
{K_{ij}} =  {{k_i}{\delta _{ij}} + \left( {{l_i}{\delta _{i,j + 1}} + {l_j}{\delta _{i + 1,j}}} \right)} .
\end{equation}
We will refer to such systems as ``chains of oscillators''. This class of harmonic systems, apart from their own interest, will be essential in the study of the free scalar quantum field theory in next section.

\subsection{A Hopping Expansion}

Assuming that the diagonal elements of the matrix $K$ are much larger than the off-diagonal ones, one may follow the approach of a hopping expansion, in the spirit of \cite{Katsinis:2017qzh}, in order to calculate the entanglement entropy and the mutual information for this class of systems perturbatively. One may define
\begin{equation}
{K_{ij}} \equiv \frac{1}{\varepsilon }{K_{ij}^{\left( 0 \right)}} + {K_{ij}^{\left( 1 \right)}} ,
\end{equation}
where
\begin{equation}
{K_{ij}^{\left( 0 \right)}} = \varepsilon {{k_i}{\delta _{ij}}} , \quad {K_{ij}^{\left( 1 \right)}} = {{l_i}{\delta _{i,j + 1}} + {l_j}{\delta _{i + 1,j}}}
\end{equation}
and perform an expansion in $\varepsilon$ (or equivalently in $l / k$).

In the following, we will adopt a particular notation for the matrix elements of all the involved matrices. The subscript denotes the line of the element, when it lies on top of the main diagonal and its column, when it lies below the main diagonal. The superscript denotes the diagonal (i.e. superscript $0$ implies that the element lies in the main diagonal, superscript $1$ implies that it lies in the first superdiagonal, superscript $-1$ implies that it lies in the first subdiagonal and so on). In other words $M_{i,j} \equiv M_{\min\left(i,j\right)}^{j-i}$. Obviously for symmetric matrices $M$ it holds that $M_i^j = M_i^{-j}$ and we will not post the results for both. Finally, the second superscript, which will appear into parentheses, denotes the order of the element in the $\varepsilon$ expansion.

Furthermore, for simplicity we define the functions
\begin{align}
{f_1}\left( x \right) &= \sqrt x \coth \sqrt x , \\
{f_2}\left( x \right) &=  - \sqrt x \csch\sqrt x , \\
{f_3}\left( x \right) &= {f_1}\left( x \right) + {f_2}\left( x \right) = \sqrt x \tanh \left( \sqrt x / 2 \right) , \\
{f_4}\left( x \right) &=  - {f_2}\left( x \right)/{f_1}\left( x \right) = {{\sech \sqrt x }},
\end{align}
which will appear throughout the calculations of this section.

Expanding the matrix ${\gamma ^{ - 1}}\beta$ in $\varepsilon$,
\begin{equation}
{\gamma ^{ - 1}}\beta  = {\left( {{\gamma ^{ - 1}}\beta } \right)^{\left( 0 \right)}} + \varepsilon {\left( {{\gamma ^{ - 1}}\beta } \right)^{\left( 1 \right)}} + {\varepsilon ^2}{\left( {{\gamma ^{ - 1}}\beta } \right)^{\left( 2 \right)}} + \mathcal{O}\left( {{\varepsilon ^3}} \right) ,
\end{equation}
one can show that the zeroth and first order terms are given by
\begin{equation}
\left( {{\gamma ^{ - 1}}\beta } \right)_i^{0\left( 0 \right)} = {f_4}\left( {\frac{{{k_{n + i}}}}{{{T^2}}}} \right)
\label{eq:chain_gb0}
\end{equation}
and
\begin{equation}
\left( {{\gamma ^{ - 1}}\beta } \right)_i^{ \pm 1 \left( 1 \right)} = \frac{{{l_{n + i}}}}{{{k_{n + i}} - {k_{n + i + 1}}}}\left( {{f_4}\left( {\frac{{{k_{n + i}}}}{{{T^2}}}} \right) - {f_4}\left( {\frac{{{k_{n + i + 1}}}}{{{T^2}}}} \right)} \right) 
\label{eq:chain_gb1}
\end{equation}
and all other elements are vanishing. The second order result is given by a little more complicated expressions. We provide here only its diagonal part, which is crucial for the following
\begin{multline}
\left( {{\gamma ^{ - 1}}\beta } \right)_i^{0\left( 2 \right)} = \frac{{l_{n + i - 1}^2}}{{{k_{n + i - 1}} - {k_{n + i}}}}\left( {\frac{{{f_4}\left( {\frac{{{k_{n + i - 1}}}}{{{T^2}}}} \right) - {f_4}\left( {\frac{{{k_{n + i}}}}{{{T^2}}}} \right)}}{{{k_{n + i - 1}} - {k_{n + i}}}} + \frac{1}{{2{T^2}}}\frac{{{f_4}\left( {\frac{{{k_{n + i}}}}{{{T^2}}}} \right)}}{{{f_1}\left( {\frac{{{k_{n + i}}}}{{{T^2}}}} \right)}}} \right)\\
 - \left( {1 - {\delta _{i,N - n}}} \right)\frac{{l_{n + i}^2}}{{{k_{n + i}} - {k_{n + i + 1}}}}\left( {\frac{{{f_4}\left( {\frac{{{k_{n + i}}}}{{{T^2}}}} \right) - {f_4}\left( {\frac{{{k_{n + i + 1}}}}{{{T^2}}}} \right)}}{{{k_{n + i}} - {k_{n + i + 1}}}} + \frac{1}{{2{T^2}}}\frac{{{f_4}\left( {\frac{{{k_{n + i}}}}{{{T^2}}}} \right)}}{{{f_1}\left( {\frac{{{k_{n + i}}}}{{{T^2}}}} \right)}}} \right)\\
+ {\delta _{i1}}\frac{{l_n^2}}{{{{\left( {{k_n} - {k_{n + 1}}} \right)}^2}}}\vast[ { \frac{{{f_1}\left( {\frac{{{k_{n + 1}}}}{{{T^2}}}} \right) - {f_2}\left( {\frac{{{k_{n + 1}}}}{{{T^2}}}} \right)}}{{2f_1^2\left( {\frac{{{k_{n + 1}}}}{{{T^2}}}} \right)}}\frac{{{{\left( {{f_3}\left( {\frac{{{k_n}}}{{{T^2}}}} \right) - {f_3}\left( {\frac{{{k_{n + 1}}}}{{{T^2}}}} \right)} \right)}^2}}}{{{f_3}\left( {\frac{{{k_n}}}{{{T^2}}}} \right)}}} \\
+ \frac{{{f_2}\left( {\frac{{{k_{n + 1}}}}{{{T^2}}}} \right)}}{{{f_1}\left( {\frac{{{k_{n + 1}}}}{{{T^2}}}} \right)}}\frac{{{{\left( {{f_1}\left( {\frac{{{k_n}}}{{{T^2}}}} \right) - {f_1}\left( {\frac{{{k_{n + 1}}}}{{{T^2}}}} \right)} \right)}^2}}}{{{f_1}\left( {\frac{{{k_n}}}{{{T^2}}}} \right){f_1}\left( {\frac{{{k_{n + 1}}}}{{{T^2}}}} \right)}} 
{ - \frac{{\left( {{f_1}\left( {\frac{{{k_n}}}{{{T^2}}}} \right) - {f_1}\left( {\frac{{{k_{n + 1}}}}{{{T^2}}}} \right)} \right)\left( {{f_2}\left( {\frac{{{k_n}}}{{{T^2}}}} \right) - {f_2}\left( {\frac{{{k_{n + 1}}}}{{{T^2}}}} \right)} \right)}}{{{f_1}\left( {\frac{{{k_n}}}{{{T^2}}}} \right){f_1}\left( {\frac{{{k_{n + 1}}}}{{{T^2}}}} \right)}}}
 \vast] . \\
\label{eq:chain_gb2}
\end{multline}
There is a special contribution in the very first element, which originates from the $\left( {{a_B^T} + b_B^T} \right){\left( {a_A + b_A} \right)^{ - 1}}\left( {a_B + b_B} \right)$ term of the $\gamma$ and $\beta$ matrices. This is going to play an important role in the following. More details are provided in the appendix \ref{subsec:app_hopping_matrix}.

The eigenvalues of the matrix $\gamma^{-1} \beta$ have to be perturbatively calculated in the $\varepsilon$ expansion. The problem is more difficult than the zero temperature problem \cite{Katsinis:2017qzh}; In this case, the elements of the matrix $\gamma^{-1} \beta$ obey an hierarchy in both its directions, i.e. the leading contribution to the element $\left( \gamma^{-1} \beta \right)_{ij}$ is of order $i+j$. This hierarchy is inherited to the eigenvalues, setting their perturbative calculation a simple task. However, in the case of finite temperature, the thermal contributions have changed this structure; The leading contribution to the element $\left( \gamma^{-1} \beta \right)_{ij}$ is of order $\left| i - j \right|$. It follows that a more systematic approach is required.

In order to obtain the expressions \eqref{eq:chain_gb0}, \eqref{eq:chain_gb1} and \eqref{eq:chain_gb2}, we only needed to demand that the diagonal elements of the matrix $K$ are larger than the non-diagonal ones. However, this does not suffice for the perturbative specification of the eigenvalues of the matrix $\gamma^{-1} \beta$. In order to clarify this, we post a simple, indicative example: Assume the Hamiltonian
\begin{equation}
H = \left( {\begin{array}{*{20}{c}}
{{h_1}}&g\\
g&{{h_2}}
\end{array}} \right) ,
\end{equation}
where the diagonal elements are much larger than the off-diagonal ones. In order to calculate its eigenvalues perturbatively, naively one would consider the diagonal part of this Hamiltonian as an exactly solvable unperturbed Hamiltonian and the off-diagonal elements as a perturbation. However, this is not necessarily a good approach. This is evident in this two by two example, since the problem is simple enough to find its answer analytically,
\begin{equation}
\lambda  = \frac{{{h_1} + {h_2}}}{2} \pm \sqrt {{{\left( {\frac{{{h_1} - {h_2}}}{2}} \right)}^2} + {g^2}} .
\end{equation}
Following this approach is equivalent to Taylor expanding the above eigenvalues with respect to the parameter $g$. However, this expansion does not converge whenever $g > \frac{{{h_1} - {h_2}}}{2}$. In this case, one should perform a Taylor expansion in ${h_1} - {h_2}$, which implies that another setup for the perturbative calculation of the eigenvalues should have been considered. The unperturbed Hamiltonian should be considered proportional to the identity matrix. Then, there are two perturbations: one that consists of the non-diagonal part of the Hamiltonian and a manifestly smaller one, which is diagonal and proportional to the difference of the two diagonal elements. Now the unperturbed problem is degenerate and the basic eigenvectors are determined by the large perturbation.

Thus, the appropriate structure of the perturbation theory depends on the ratio of the diagonal elements to the \emph{difference} of the diagonal ones. The assumption we have made for the matrix $K$ does not determine this ratio. It follows that there are two distinct approaches in determining the eigenvalues of $\gamma^{-1} \beta$, which we will call ``non-degenerate'' and ``degenerate'' perturbation theory. They are presented in appendices \ref{subsec:app_hopping_nondeg} and \ref{subsec:app_hopping_deg}, respectively.

When the diagonal elements have differences of the same order of magnitude as themselves, the non-degenerate perturbation theory applies and it yields
\begin{multline}
\beta _{Di} = {f_4}\left( {\frac{{{k_{n + i}}}}{{{T^2}}}} \right) \\
+ \frac{1}{{2{T^2}}}\frac{{{f_4}\left( {\frac{{{k_{n + 1}}}}{{{T^2}}}} \right)}}{{{f_1}\left( {\frac{{{k_{n + 1}}}}{{{T^2}}}} \right)}}\left( {\frac{{l_{n + i - 1}^2}}{{{k_{n + i - 1}} - {k_{n + i}}}} - \frac{{l_{n + i}^2}}{{{k_{n + i}} - {k_{n + i - 1}}}}\left( {1 - {\delta _{i,N - n}}} \right)} \right)\\
 + {\delta _{i1}}\frac{{l_n^2}}{{{{\left( {{k_n} - {k_{n + 1}}} \right)}^2}}}\frac{1}{{{f_1}\left( {\frac{{{k_{n + 1}}}}{{{T^2}}}} \right)}} \vast[ {{f_1}\left( {\frac{{{k_n}}}{{{T^2}}}} \right)\left( {{f_4}\left( {\frac{{{k_n}}}{{{T^2}}}} \right) - {f_4}\left( {\frac{{{k_{n + 1}}}}{{{T^2}}}} \right)} \right)} \\
{ + \left( {1 + {f_4}\left( {\frac{{{k_{n + 1}}}}{{{T^2}}}} \right)} \right)\frac{{{{\left( {{f_3}\left( {\frac{{{k_n}}}{{{T^2}}}} \right) - {f_3}\left( {\frac{{{k_{n + 1}}}}{{{T^2}}}} \right)} \right)}^2}}}{{2{f_3}\left( {\frac{{{k_n}}}{{{T^2}}}} \right)}}} \vast] + \mathcal{O} \left( l^4 \right) .
\end{multline}
The unique second order contribution to $\left( \gamma^{-1} \beta \right)_{11}$ has affected a single eigenvalue at this order. This is similar to the zero temperature case; however, the other eigenvalues do not vanish. These eigenvalues imply that the entanglement entropy for subsystem $A$ equals
\begin{multline}
S_A = \sum\limits_{i = 1}^{N - n} {\frac{{\sqrt {{k_{n + i}}} }}{T}\frac{{{e^{ - \frac{{\sqrt {{k_{n + i}}} }}{T}}}}}{{1 - {e^{ - \frac{{\sqrt {{k_{n + i}}} }}{T}}}}} - \ln \left( {1 - {e^{ - \frac{{\sqrt {{k_{n + i}}} }}{T}}}} \right)} \\
 + \frac{1}{{4{T^2}}}\sum\limits_{i = 1}^{N - n - 1} {\frac{{l_{n + i}^2}}{{{k_{n + i}} - {k_{n + i + 1}}}}} \left( {\frac{{{f_4}\left( {\frac{{{k_{n + i + 1}}}}{{{T^2}}}} \right)}}{{1 - {f_4}\left( {\frac{{{k_{n + i + 1}}}}{{{T^2}}}} \right)}} - \frac{{{f_4}\left( {\frac{{{k_{n + i}}}}{{{T^2}}}} \right)}}{{1 - {f_4}\left( {\frac{{{k_{n + i}}}}{{{T^2}}}} \right)}}} \right)\\
 + \frac{1}{2}\frac{{l_n^2}}{{{k_n} - {k_{n + 1}}}}\frac{1}{{1 - {f_4}\left( {\frac{{{k_{n + 1}}}}{{{T^2}}}} \right)}} \vast[ {\frac{1}{{2{T^2}}}{f_4}\left( {\frac{{{k_{n + 1}}}}{{{T^2}}}} \right)} \\
{ + \frac{1}{{{k_n} - {k_{n + 1}}}} \vast( {\left( {1 + {f_4}\left( {\frac{{{k_{n + 1}}}}{{{T^2}}}} \right)} \right){{\frac{{\left( {{f_3}\left( {\frac{{{k_n}}}{{{T^2}}}} \right) - {f_3}\left( {\frac{{{k_{n + 1}}}}{{{T^2}}}} \right)} \right)}}{{2{f_3}\left( {\frac{{{k_n}}}{{{T^2}}}} \right)}}}^2} } }  \\
 { + {f_1}\left( {\frac{{{k_n}}}{{{T^2}}}} \right)\left( {{f_4}\left( {\frac{{{k_n}}}{{{T^2}}}} \right) - {f_4}\left( {\frac{{{k_{n + 1}}}}{{{T^2}}}} \right)} \right)} \vast) \vast] .
\end{multline}
The first two lines of the the above expression contain the contributions from the generic eigenvalues. The rest originates from the special eigenvalue $\beta_{D1}$. The entanglement entropy $S_{A^C}$ has a similar structure.

The contributions to the entanglement entropy from all the generic eigenvalues are identical to those of the thermal entropy, and, thus, at this order in $l / k$, the mutual information receives contributions only from the two special eigenvalues,  one from each subsystem. It is equal to
\begin{equation}
I = \frac{{l_n^2}}{{4{T^2}\left( {{k_n} - {k_{n + 1}}} \right)}}\left( {\frac{1}{{{f_3}\left( {\frac{{{k_{n + 1}}}}{{{T^2}}}} \right)}} - \frac{1}{{{f_3}\left( {\frac{{{k_n}}}{{{T^2}}}} \right)}}} \right) + \mathcal{O}\left( {{l^3}} \right) .
\label{eq:chain_I_non_degenerate}
\end{equation}
Expanding for high temperatures the above result yields
\begin{equation}
I = \frac{{l_n^2}}{{2{k_n}{k_{n + 1}}}} + \frac{{l_n^2}}{{1440{T^4}}} + \mathcal{O}\left( {\frac{1}{{{T^6}}}} \right) ,
\label{eq:chain_I_non_degenerate_high_T}
\end{equation}
which coincides with the $l/k$ expansion of the high temperature formula for the generic oscillatory system \eqref{eq:N_MI_high_T}. 

In the case the differences of the diagonal elements are smaller than the non-diagonal ones, one should apply degenerate perturbation theory. We will focus on a subclass of this kind of problems that emerges from the discretization of $1+1$ dimensional field theory, namely the case where the matrix $K$ is of the form
\begin{equation}
k_i = k , \quad l_i = l .
\end{equation}

It is a matter of algebra (see appendix \ref{subsec:app_hopping_deg}) to show that the matrix ${{\gamma ^{ - 1}}\beta }$ can be perturbatively calculated as
\begin{align}
\left( {{\gamma ^{ - 1}}\beta } \right)_i^{0\left( 0 \right)} &= {f_4}\left( {\frac{k}{{{T^2}}}} \right) , \\
\left( {{\gamma ^{ - 1}}\beta } \right)_i^{1\left( 1 \right)} &= \frac{l}{{{T^2}}}{f_4}'\left( {\frac{k}{{{T^2}}}} \right) , \\
\left( {{\gamma ^{ - 1}}\beta } \right)_i^{2\left( 2 \right)} &= \frac{{{l^2}}}{{2{T^4}}}{f_4}''\left( {\frac{k}{{{T^2}}}} \right) , \\
\left( {{\gamma ^{ - 1}}\beta } \right)_i^{0\left( 2 \right)} &= \frac{{{l^2}}}{{2{T^4}}}\left( {{f_4}''\left( {\frac{k}{{{T^2}}}} \right)\left( {2 - {\delta _{i,1}} - {\delta _{i,N - n}}} \right) + {\beta _1}{\delta _{i,1}}} \right) ,
\end{align}
where
\begin{multline}
{\beta _1} = \frac{1}{{{{\left( {{f_1}\left( {\frac{k}{{{T^2}}}} \right)} \right)}^2}}}\left[ {\left( {{f_1}\left( {\frac{k}{{{T^2}}}} \right) - {f_2}\left( {\frac{k}{{{T^2}}}} \right)} \right) \frac{{{{\left[ {{f_3}'\left( {\frac{k}{{{T^2}}}} \right)} \right]}^2}}}{{{f_3}\left( {\frac{k}{{{T^2}}}} \right)}} } \right.\\
\left. \phantom{\frac{{{{\left[ {{f_3}'\left( {\frac{k}{{{T^2}}}} \right)} \right]}^2}}}{{{f_3}\left( {\frac{k}{{{T^2}}}} \right)}}} { - \left( {{f_1}\left( {\frac{k}{{{T^2}}}} \right){f_2}''\left( {\frac{k}{{{T^2}}}} \right) - {f_1}''\left( {\frac{k}{{{T^2}}}} \right){f_2}\left( {\frac{k}{{{T^2}}}} \right)} \right)} \right] .
\end{multline}

The above imply that the eigenvalues at zeroth order are
\begin{equation}
\beta _D^{j\left( 0 \right)} = {f_4}\left( {\frac{k}{{{T^2}}}} \right) .
\end{equation}
As expected, they are all equal, and, thus, they do not determine the eigenvectors. At first order the matrix $\gamma^{-1} \beta$ is proportional to the matrix ${{\delta _{i + 1,j}} + {\delta _{i,j + 1}}}$. The determination of its eigenvectors is a simple problem. The normalized eigenvectors $v^j$ are
\begin{equation}
v_i^j = \sqrt {\frac{2}{{N + 1}}} \sin \frac{{ij\pi }}{{N + 1}}
\end{equation}
and the eigenvalues of the matrix $\gamma^{-1} \beta$ at first order equal
\begin{equation}
\beta _D^{j\left( 1 \right)} = \frac{{2l}}{{{T^2}}}{f_4}'\left( {\frac{k}{{{T^2}}}} \right)\cos \frac{{j\pi }}{{N - n + 1}}.
\end{equation}
Now we may apply degenerate perturbation theory to determine the eigenvalues at second order. They equal
\begin{equation}
\beta _D^{j\left( 2 \right)} = \left\langle {{v^j}} \right|{\left( {{\gamma ^{ - 1}}\beta } \right)^{\left( 2 \right)}}\left| {{v^j}} \right\rangle .
\end{equation}
It is a matter of algebra to show that
\begin{equation}
\beta _D^{j\left( 2 \right)} = \frac{{{l^2}}}{{{T^4}}}\left( {2{f_4}''\left( {\frac{k}{{{T^2}}}} \right){{\cos }^2}\frac{{j\pi }}{{N - n + 1}} + \frac{{{\beta _1}}}{{N - n + 1}}{{\sin }^2}\frac{{j\pi }}{{N - n + 1}}} \right) .
\end{equation}
The above eigenvalues imply that the entanglement entropy equals
\begin{multline}
{S_A} = \left( {N - n} \right)\left[ {\frac{{\sqrt k }}{T}\frac{{{e^{ - \frac{{\sqrt k }}{T}}}}}{{1 - {e^{ - \frac{{\sqrt k }}{T}}}}} - \ln \left( {1 - {e^{ - \frac{{\sqrt k }}{T}}}} \right)} \right]\\
 + \frac{{{l^2}}}{{32{k^{\frac{3}{2}}}{T^3}}}\left[ {\sqrt k T \csch^2 \frac{{\sqrt k }}{{2T}} + \coth \frac{{\sqrt k }}{{2T}}\left( {2{T^2} + k\left( {2\left( {N - n} \right) - 1} \right) \csch^2 \frac{{\sqrt k }}{{2T}}} \right)} \right] \\
+ \mathcal{O}\left( {{l^3}} \right) .
\end{multline}
Interestingly enough, a similar cancellation between the contributions from all eigenvalues, but two, one from each subsystem, occurs in the calculation of mutual information in this case too. One can show that at this order 
\begin{equation}
I = \frac{{{l^2}}}{{16{k^{\frac{3}{2}}}{T^2}}} \csch^2 \frac{{\sqrt k }}{{2T}}\left( {\sqrt k  + T\sinh \frac{{\sqrt k }}{T}} \right) + \mathcal{O}\left( {{l^3}} \right) .
\end{equation}

The above formula may look quite dissimilar to the formula \eqref{eq:chain_I_non_degenerate} that we found in the case of the non-degenerate perturbation theory. However, it is exactly the smooth limit of the latter as $k_i \to k$ and $l_i \to l$, i.e.
\begin{equation}
I =  - \frac{{{l^2}}}{{4{T^2}}}\frac{d}{{dk}}\left( {\frac{1}{{{f_3}\left( {\frac{k}{{{T^2}}}} \right)}}} \right) + \mathcal{O}\left( {{l^3}} \right) .
\label{eq:chain_I_degenerate}
\end{equation}
The non-degenerate and degenerate perturbation theories resulted in different results for the entanglement entropy, but in the \emph{same result for the mutual information}. This hints that the mutual information is determined by an underlying matrix object, which has the same double hierarchy as the matrix $\gamma^{-1} \beta$ at zero temperature, and, thus, at this order in the $l/k$ expansion it has only two non-vanishing elements. This is not unexpected, since the symmetry property of the mutual information enforces the latter to depend only on the entangling surface (in this case the point that separates the two subsystems) and not the subsystems. Whether the two approaches provide different results at higher orders is an issue that requires further investigation. At leading order, the difference of the two approaches, is restricted to the thermal contributions to the entanglement entropy, thus, irrelevant to our interests.

The formula \eqref{eq:chain_I_degenerate} also has a high temperature expansion of the form
\begin{equation}
I = \frac{{l^2}}{{2{k^2}}} + \frac{{l^2}}{{1440{T^4}}} + \mathcal{O}\left( {\frac{1}{{{T^6}}}} \right) ,
\label{eq:chain_I_degenerate_high_T}
\end{equation}
which coincides with the $l/k$-expansion of the high temperature formula \eqref{eq:N_MI_high_T}.

\subsection{Low Temperature Expansion}

In the previous section, we managed to find an $l/k$ expansion for the mutual information in the case of a chain of oscillators. Although there is an ambiguity at the process of the perturbative calculation of the eigenvalues of the matrix $\gamma^{-1} \beta$, as long as the mutual information is considered, this ambiguity disappears, at least at this order in the perturbation theory.

We also showed that the expressions agree with the expected form for the high temperature expansion of the mutual information. However, as we will see in the next subsection with the study of two indicative example chains of oscillators, at low temperatures, the expressions we obtained with the $l/k$ expansion fail to approximate successfully the actual mutual information. The underlying reason for this is the fact that at low temperatures, most eigenvalues tend to zero (at least at this order in the perturbation theory). As a result, the expansive formulae for the calculation of the contribution of an eigenvalue to the entanglement entropy are not correct, since they reach a singular point. The case of low temperatures should be dealt separately, making the appropriate adaptations of the relevant formulae. This is performed in the appendix \ref{sec:Small_T_chain}. It turns out that the low temperature expansion of the mutual information is given by
\begin{multline}
I =  - \log \left( {\frac{{\beta _n^{\left( 0 \right)}}}{2}} \right)\left( {1 + 2\beta _n^{\left( 0 \right)}} \right) + \left( n \to n + 1 \right) \\
 + \left[ { - \log \left( {\frac{{\beta _n^{\left( 0 \right)}}}{2}} \right)\left( {1 + \beta _n^{\left( 0 \right)}} \right) - \left( {1 + \frac{{\sqrt {{k_n}} }}{T}\left( {1 + \frac{{k_n^{\left( 2 \right)}}}{{2k_n^{\left( 0 \right)}}}} + \mathcal{O}\left( {{l^3}} \right) \right)} \right)} \right] \\
\times \exp\left[ { - \frac{{\sqrt {{k_n}} }}{T}\left( {1 + \frac{{k_n^{\left( 2 \right)}}}{{2k_n^{\left( 0 \right)}}}} + \mathcal{O}\left( {{l^3}} \right) \right)} \right] + \left( n \to n + 1 \right) + \ldots ,
\label{eq:chain_low_T}
\end{multline}
where $\beta _n^{\left( 0 \right)}$ is the non-vanishing eigenvalue of the matrix $\gamma^{-1} \beta$ at zero temperature, which at this order in the $l/k$ expansion reads
\begin{equation}
\beta _{n}^{\left( 0 \right)} = \frac{{l_n^2}}{{2\sqrt {{k_n}} \sqrt {{k_{n + 1}}} {{\left( {\sqrt {{k_n}}  + \sqrt {{k_{n + 1}}} } \right)}^2}}}
\end{equation}
and $k_i^{\left( 2 \right)}$ is the second order correction of the eigenvalues of the matrix $K$ in a non-degenerate perturbation theory approach, namely
\begin{equation}
k_i^{\left( 2 \right)} =  - \left( {\frac{{l_{i - 1}^2}}{{{k_{i - 1}} - {k_i}}} - \frac{{l_i^2}}{{{k_i} - {k_{i + 1}}}}} \right) .
\end{equation}
The first line of the equation \eqref{eq:chain_low_T} is trivially twice the zero temperature entanglement entropy. The second line is the thermal correction to the mutual information at low temperatures, which clearly is exponentially suppressed.

\subsection{Two Characteristic Examples}

\begin{figure}[p]
\centering
\begin{picture}(100,110)
\put(3.5,74.5){\includegraphics[width = 0.45\textwidth]{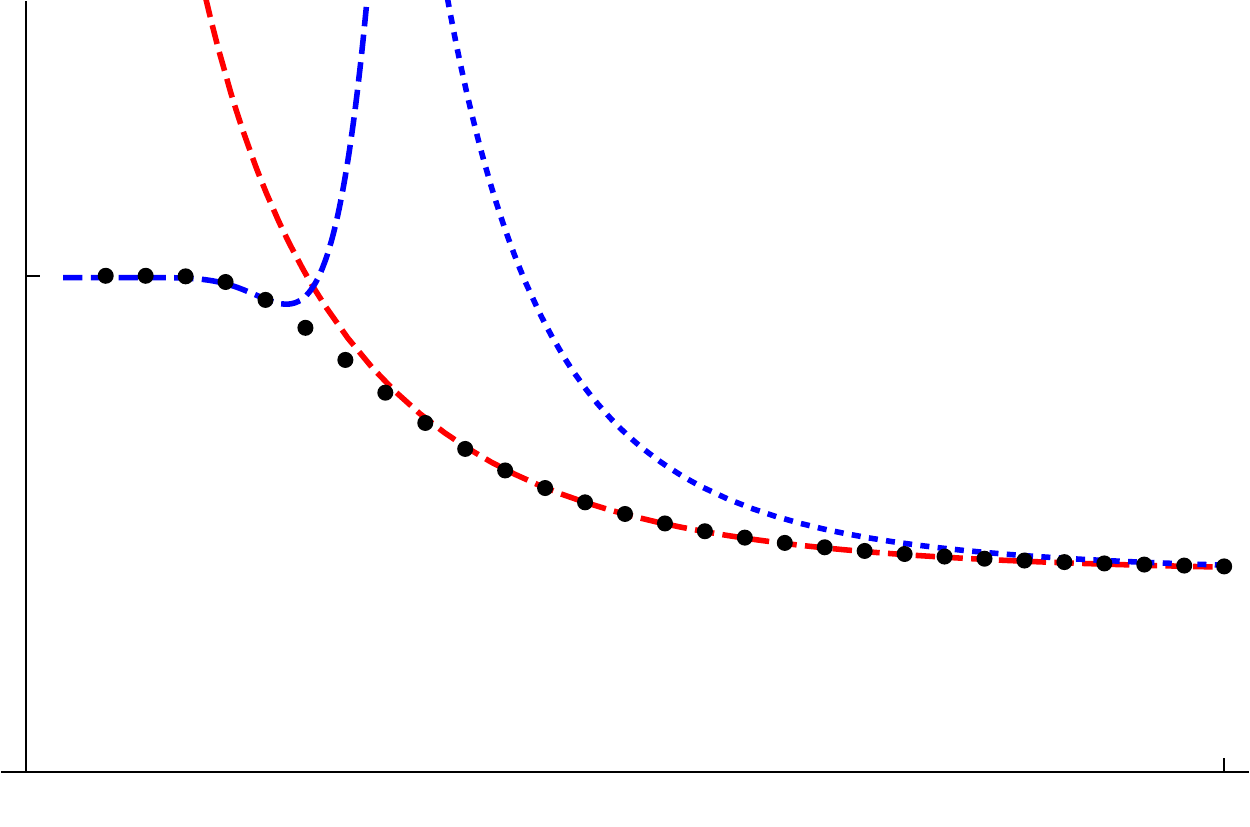}}
\put(53.5,74.5){\includegraphics[width = 0.45\textwidth]{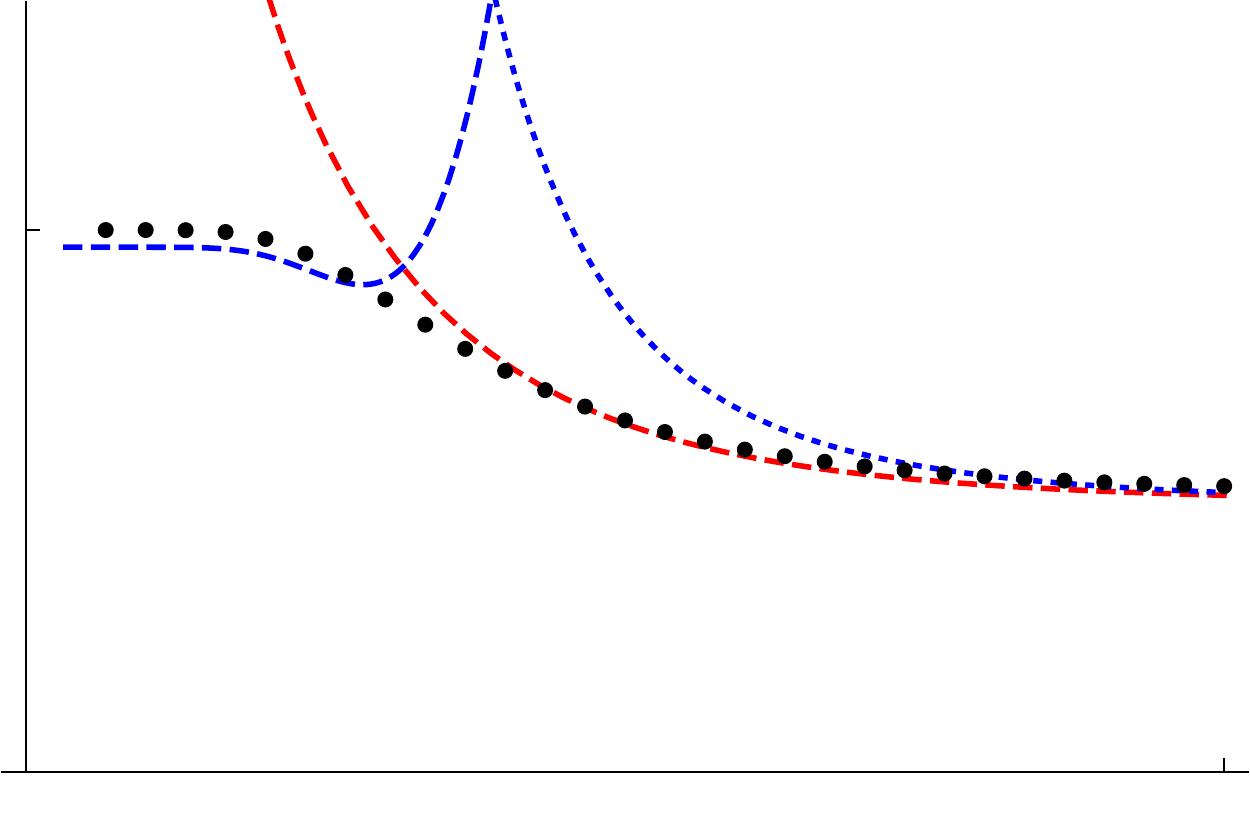}}
\put(3.5,38.5){\includegraphics[width = 0.45\textwidth]{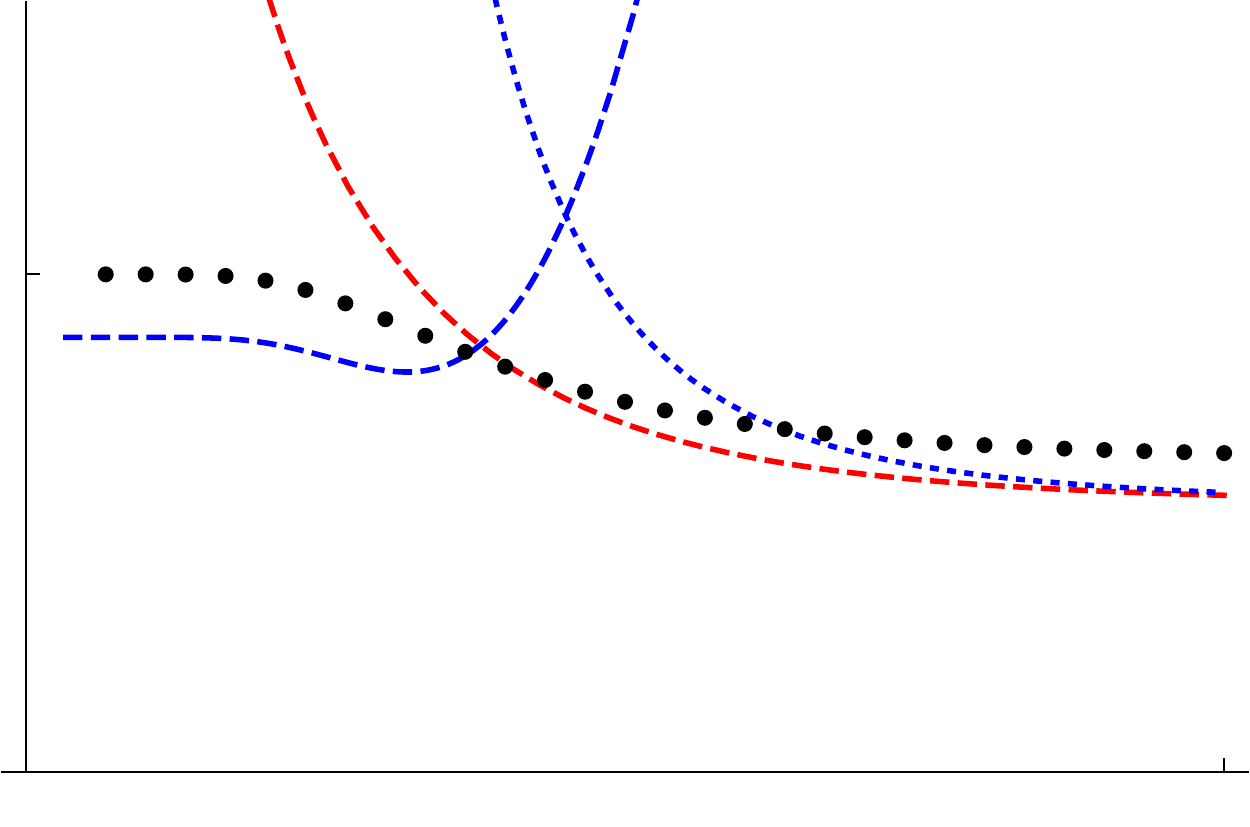}}
\put(53.5,38.5){\includegraphics[width = 0.45\textwidth]{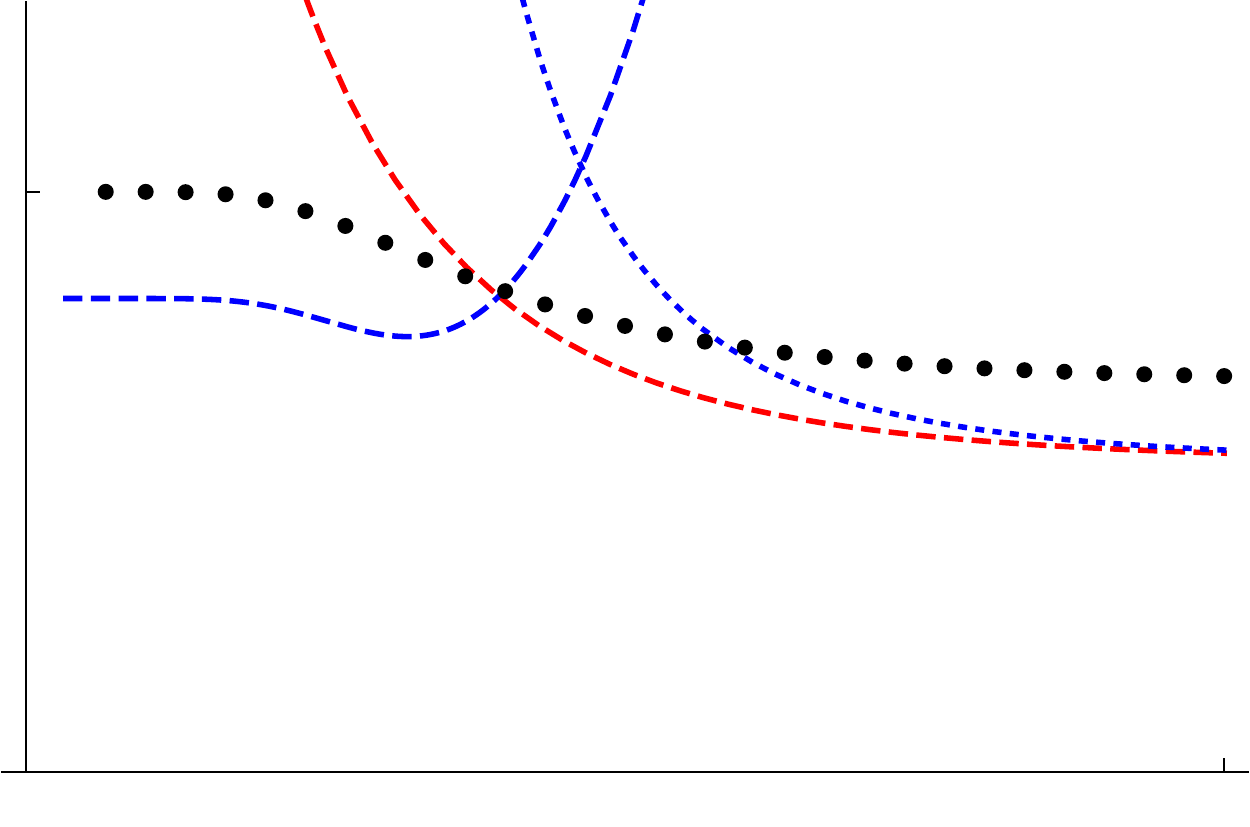}}
\put(3.5,2.5){\includegraphics[width = 0.45\textwidth]{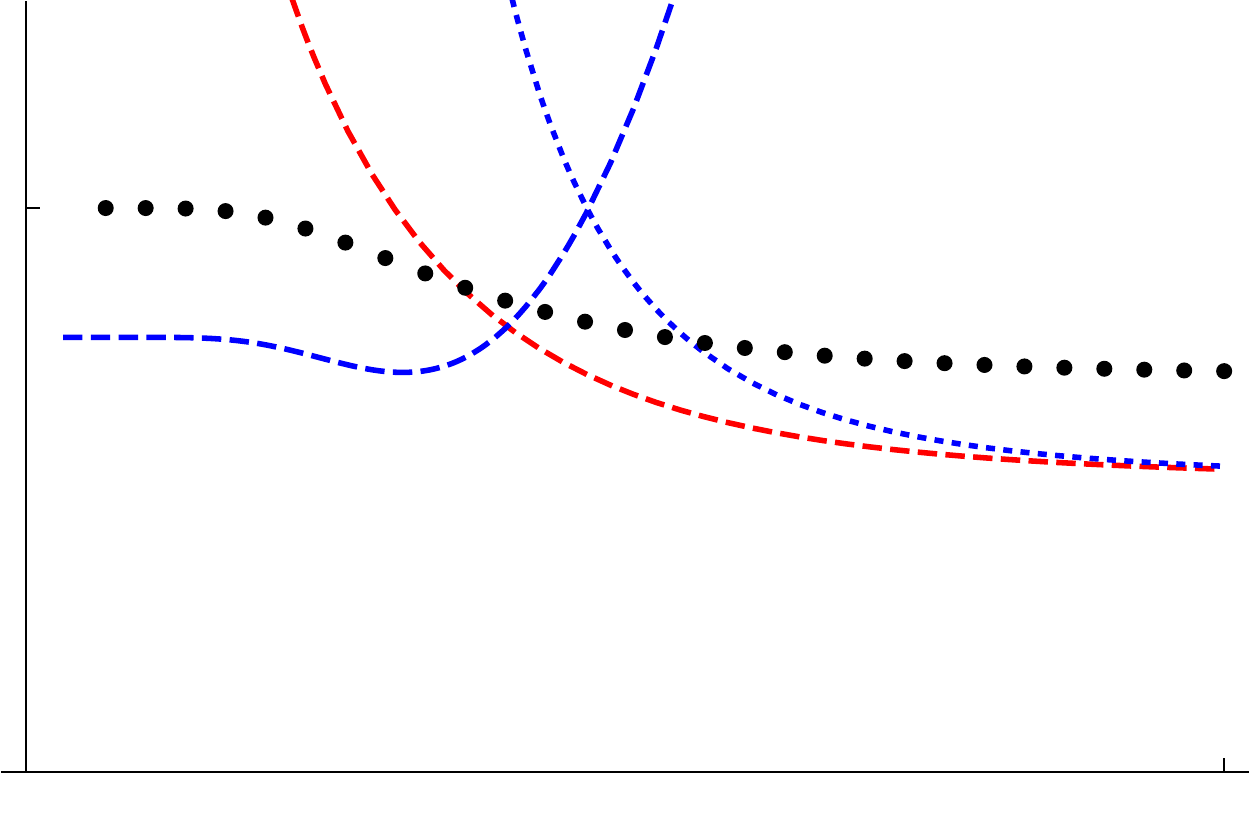}}
\put(53.5,2.5){\includegraphics[width = 0.45\textwidth]{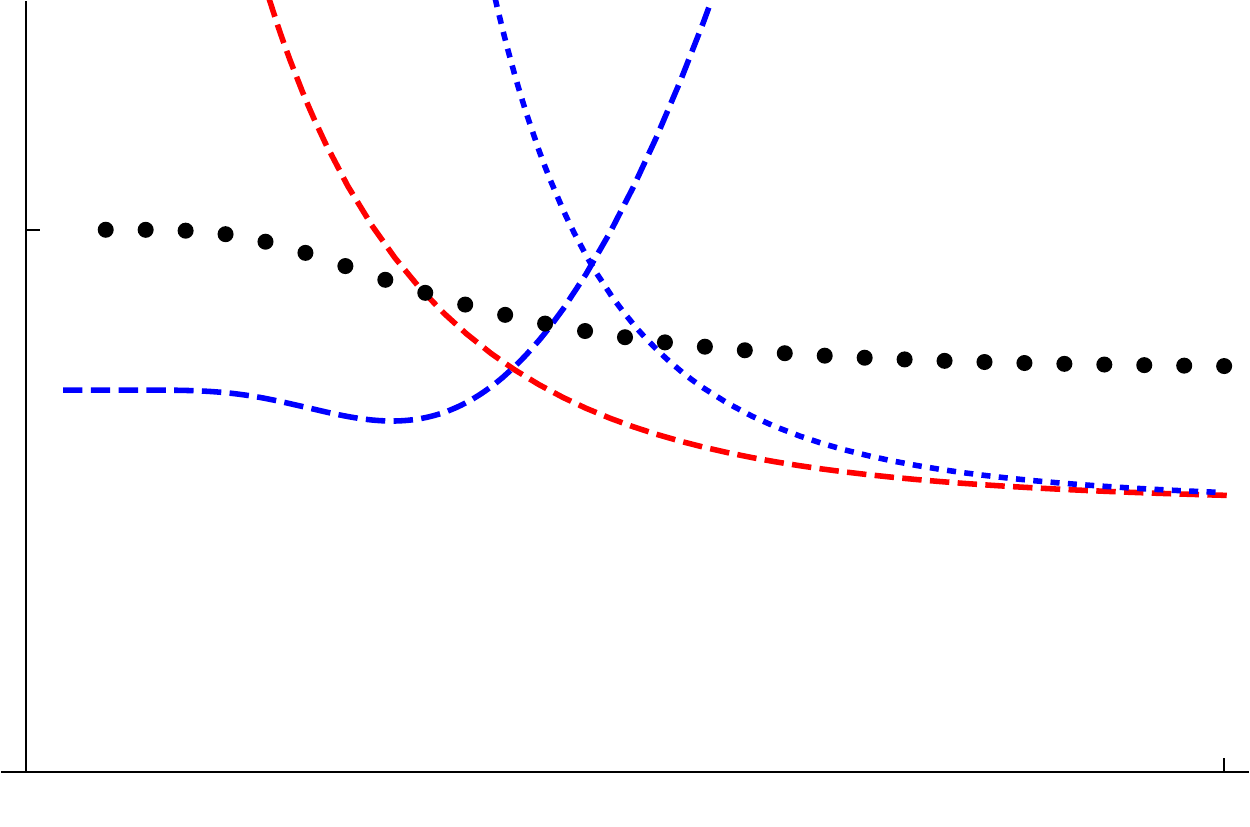}}
\put(71,95){\includegraphics[width = 0.08\textwidth]{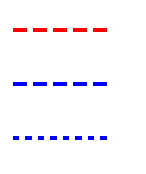}}
\put(73,92.25){\includegraphics[width = 0.04\textwidth]{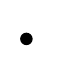}}
\put(78,93.5){numerical}
\put(78,96.5){$\varepsilon$ expansion-high $T$}
\put(78,99.5){$\varepsilon$ expansion-low $T$}
\put(78,102.5){$\varepsilon$ expansion}
\put(23,35){$k = 2.25$}
\put(73,35){$k = 2$}
\put(-0.25,33){$I\left( {A:{A^C}} \right)$}
\put(49.75,32.5){$I\left( {A:{A^C}} \right)$}
\put(48.75,3.75){$T$}
\put(98.75,3.75){$T$}
\put(-0.25,23.75){$2S_A^0$}
\put(49.75,23.25){$2S_A^0$}
\put(42.5,1.5){$\sqrt{k_0}/2$}
\put(92.5,1.5){$\sqrt{k_0}/2$}
\put(23,71){$k = 3$}
\put(73,71){$k = 2.5$}
\put(-0.25,69){$I\left( {A:{A^C}} \right)$}
\put(49.75,68.5){$I\left( {A:{A^C}} \right)$}
\put(48.75,39.75){$T$}
\put(98.75,39.75){$T$}
\put(-0.25,57.75){$2S_A^0$}
\put(49.75,60.25){$2S_A^0$}
\put(42.5,37.5){$\sqrt{k_0}/2$}
\put(92.5,37.5){$\sqrt{k_0}/2$}
\put(23,107){$k = 18$}
\put(73,107){$k = 6$}
\put(-0.25,105){$I\left( {A:{A^C}} \right)$}
\put(49.75,104.5){$I\left( {A:{A^C}} \right)$}
\put(48.75,75.75){$T$}
\put(98.75,75.75){$T$}
\put(-0.25,93.75){$2S_A^0$}
\put(49.75,95.5){$2S_A^0$}
\put(42.5,73.5){$\sqrt{k_0}/2$}
\put(92.5,73.5){$\sqrt{k_0}/2$}
\end{picture}
\caption{The mutual information as function of the temperature for the chain of oscillators \eqref{eq:chains_non_degenerate_def} for various value of the parameter $k$}
\label{fig:non_deg_chain}
\end{figure}
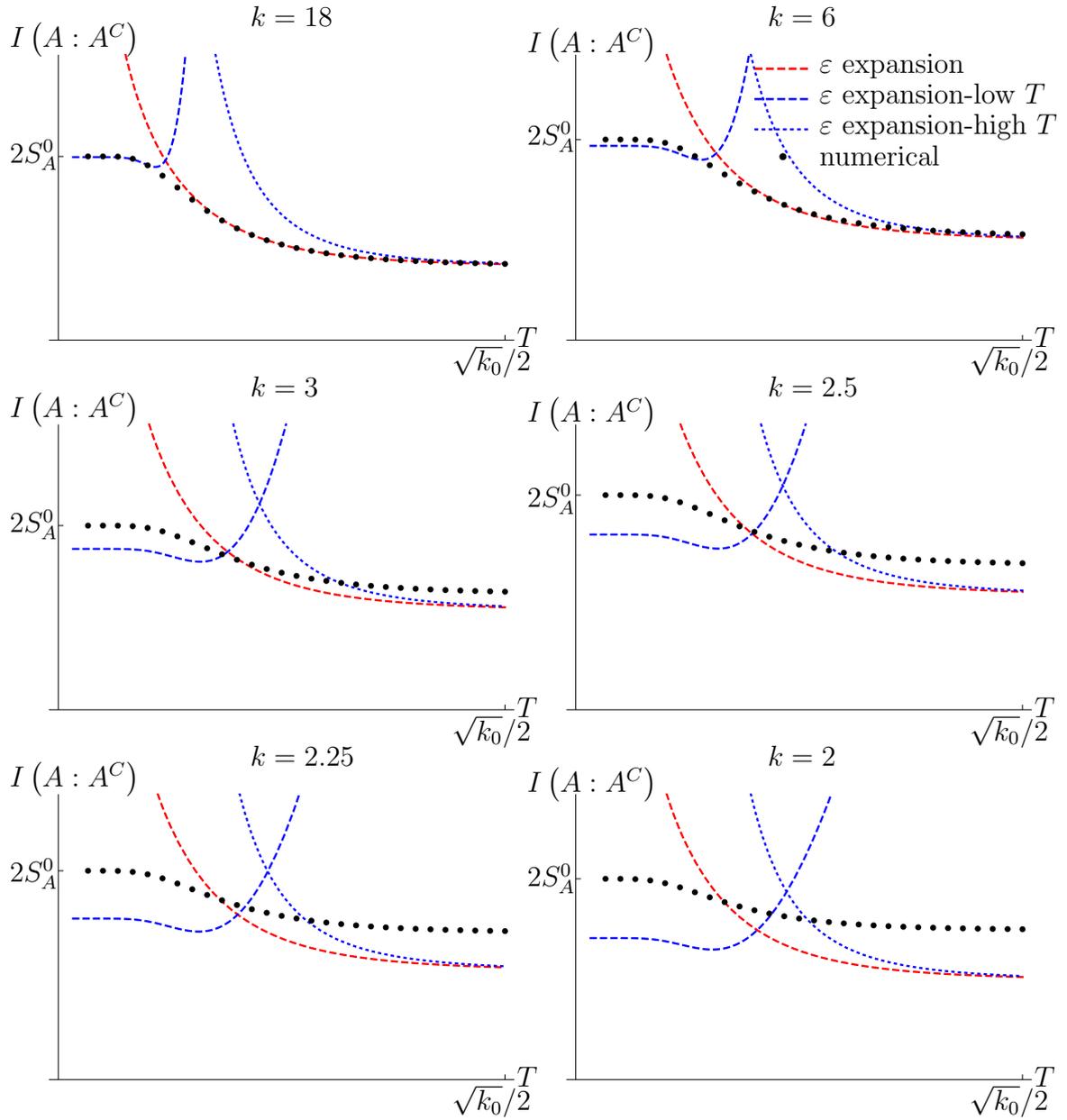

Let us now consider two characteristic example chains of oscillators. The one is a chain, whose couplings matrix is of the form
\begin{equation}
K = \left( {\begin{array}{*{20}{c}}
k&l&0&0& \cdots \\
l&{2k}&l&0& \cdots \\
0&l&k&l& \cdots \\
0&0&l&{2k}&{}\\
 \vdots & \vdots & \vdots &{}& \ddots 
\end{array}} \right) .
\label{eq:chains_non_degenerate_def}
\end{equation}
In an obvious way, this is a chain, where the non-degenerate perturbation theory is appropriate for the determination of the eigenvalues of the matrix $\gamma^{-1}\beta$. We compare the $l/k$ expansion \eqref{eq:chain_I_non_degenerate}, its high temperature expansion \eqref{eq:chain_I_non_degenerate_high_T} and its low temperature expansion \eqref{eq:chain_low_T} to numerical results. The numerical calculation of entanglement entropy and the mutual information is performed with the use of Wolfram's Mathematica. The comparison of the numerical and analytic results for various values of $k$ are shown in figure \ref{fig:non_deg_chain}. In all cases $l$ is considered equal to $-1$. Furthermore, in all cases we assume $N = 60$ and $n = 30$.
It is evident that the perturbative formulae approximate the numerical results successfully, especially for large values of the parameter $k$.

The second chain of oscillators that we consider has a couplings matrix of the form
\begin{equation}
K = \left( {\begin{array}{*{20}{c}}
k&l&0&0& \cdots \\
l&k&l&0& \cdots \\
0&l&k&l& \cdots \\
0&0&l&k&{}\\
 \vdots & \vdots & \vdots &{}& \ddots 
\end{array}} \right) .
\label{eq:chains_degenerate_def}
\end{equation}
Obviously, this is the basic example where the degenerate perturbation theory applies. This case is also very interesting, as it can be obtained from the discritization of the degrees of freedom of $1+1$ dimensional free massive scalar field theory.

In this case one can obtain another analytic formula. Whenever, the couplings matrix is of the form of a chain of oscillators, i.e. only neighbouring oscillators are coupled, the high temperature expansion formula \eqref{eq:N_MI_high_T} assumes a simple form, as the block $K_B$ contains only one non-vanishing element, which is equal to $l_n$, namely
\begin{multline}
I =  - \frac{1}{2}\ln \left( {1 - {{\left( {K_A^{ - 1}} \right)}_{nn}}{{\left( {K_C^{ - 1}} \right)}_{11}}{l_n^2}} \right) \\
- \frac{{{l_n^2}}}{{720{T^4}}}\left[ {{{\left( {l_n{{\left( {K_A^{ - 1}} \right)}_{nn}}} \right)}^2} + {{\left( {l_n{{\left( {K_C^{ - 1}} \right)}_{11}}} \right)}^2} - \frac{1}{2}} \right] + O\left( {\frac{1}{{{T^6}}}} \right) .
\label{eq:chain:degenerate_exact_high_T}
\end{multline}
In the case of the chain \eqref{eq:chains_degenerate_def}, it is possible to calculate exactly the above expression, since the eigenvectors of the block $K_A$ are known (see e.g. appendix \ref{subsec:app_hopping_deg}),
\begin{align}
{\left( {K_A^{ - 1}} \right)_{nn}} &=  - \frac{1}{l}\frac{{\sinh \left( {n \, \arccosh \left( { - \frac{k}{{2l}}} \right)} \right)}}{{\sinh \left( {\left( {n + 1} \right)\arccosh \left( { - \frac{k}{{2l}}} \right)} \right)}}\\
{\left( {K_C^{ - 1}} \right)_{11}} &=  - \frac{1}{l}\frac{{\sinh \left( {\left( {N - n} \right)\arccosh \left( { - \frac{k}{{2l}}} \right)} \right)}}{{\sinh \left( {\left( {N - n + 1} \right)\arccosh \left( { - \frac{k}{{2l}}} \right)} \right)}} .
\end{align}
Therefore, in this case we also have an expression for the high temperature expansion of the mutual information, which is exact in $l/k$.

As in the previous example, the analytic formulae are compared with numerical calculations for various values of $k$. All examples have $l=-1$, $N=60$ and $n=30$. 
\begin{figure}[ph]
\centering
\begin{picture}(100,110)
\put(3.5,74.5){\includegraphics[width = 0.45\textwidth]{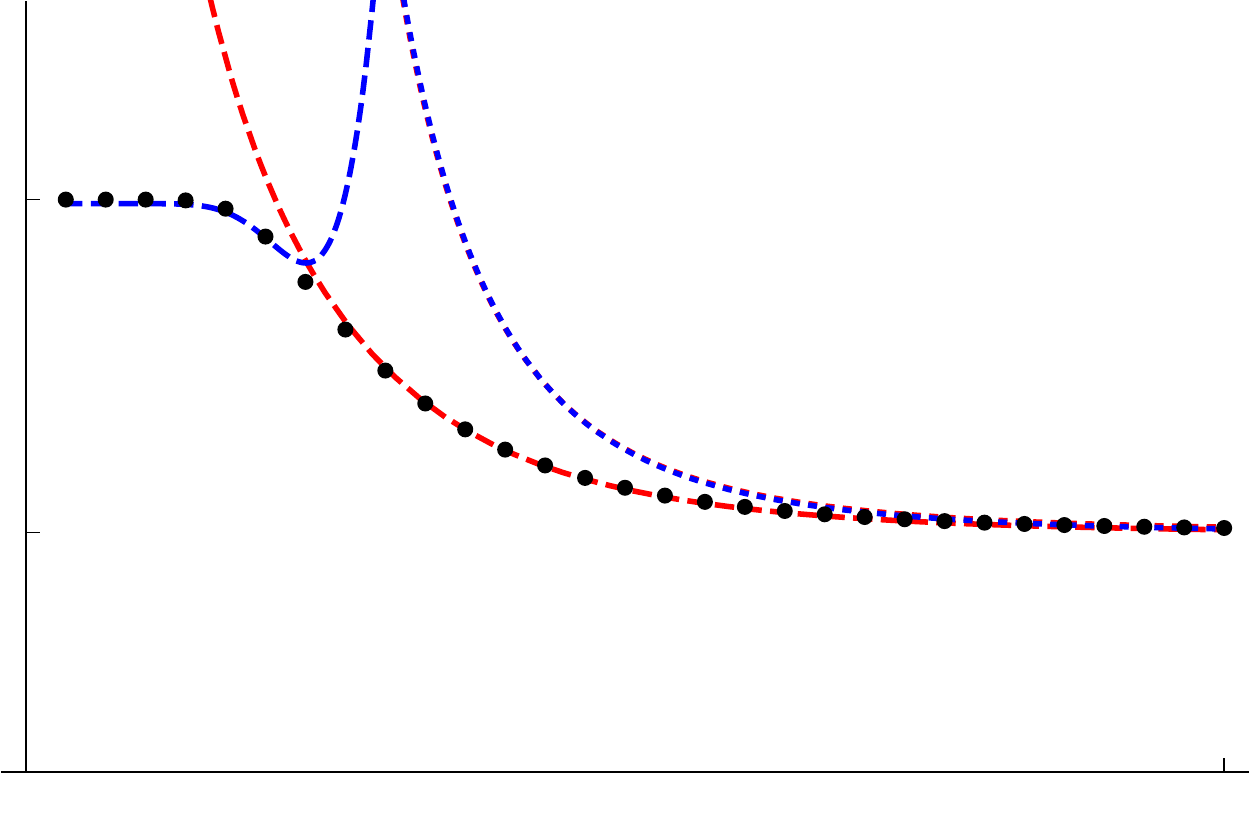}}
\put(53.5,74.5){\includegraphics[width = 0.45\textwidth]{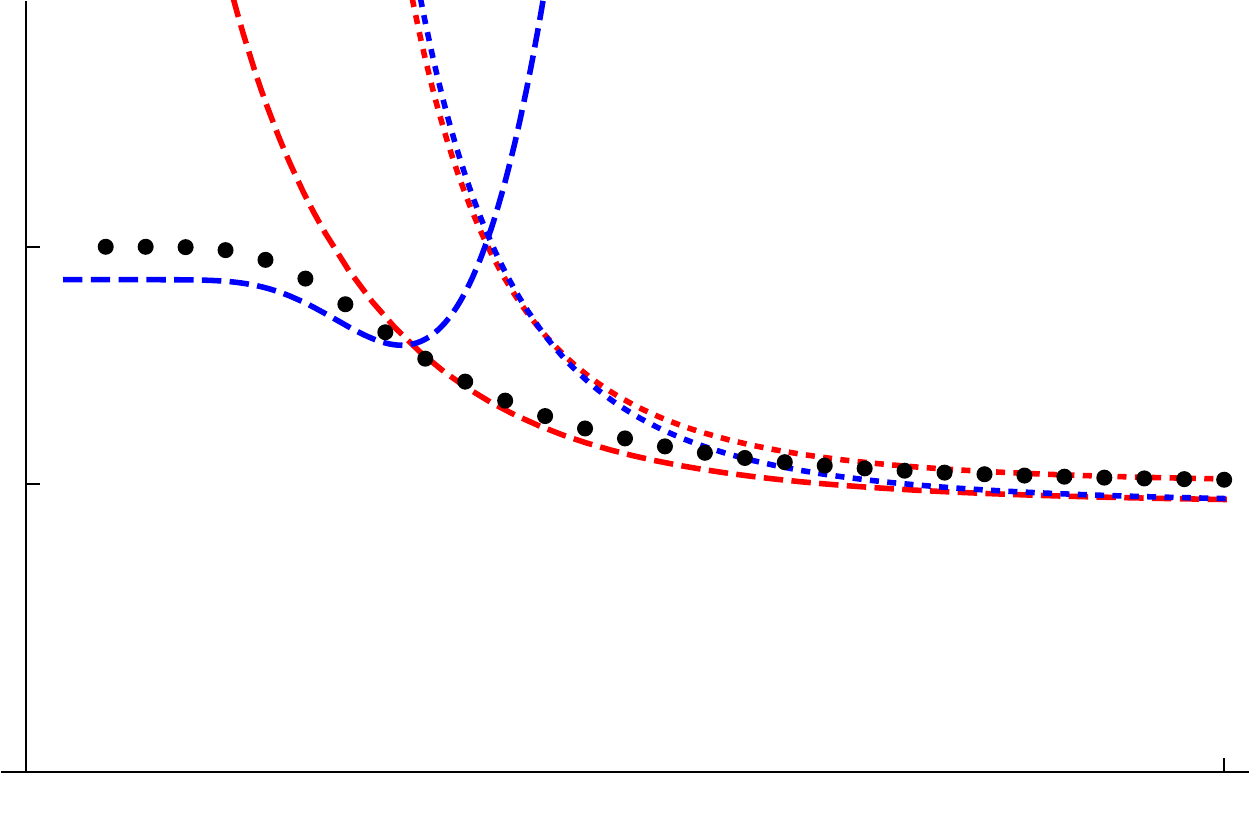}}
\put(3.5,38.5){\includegraphics[width = 0.45\textwidth]{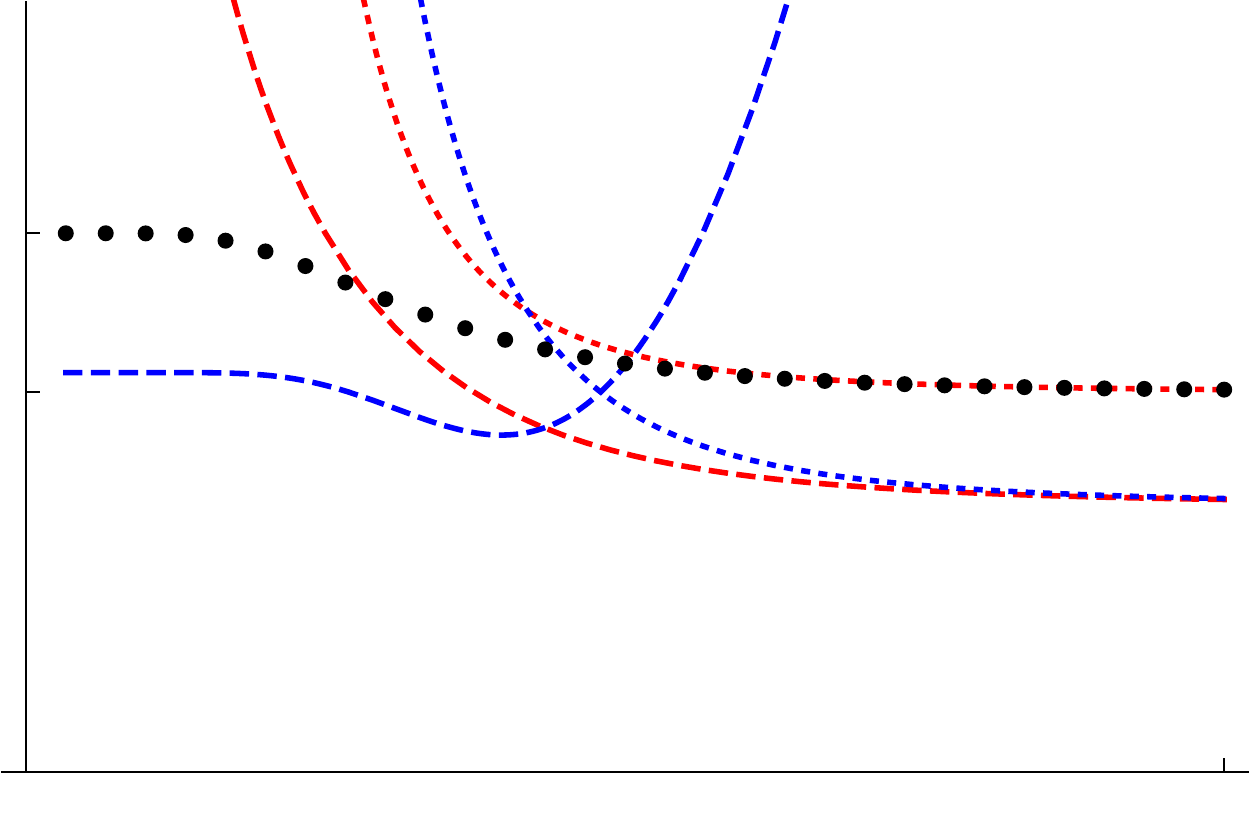}}
\put(53.5,38.5){\includegraphics[width = 0.45\textwidth]{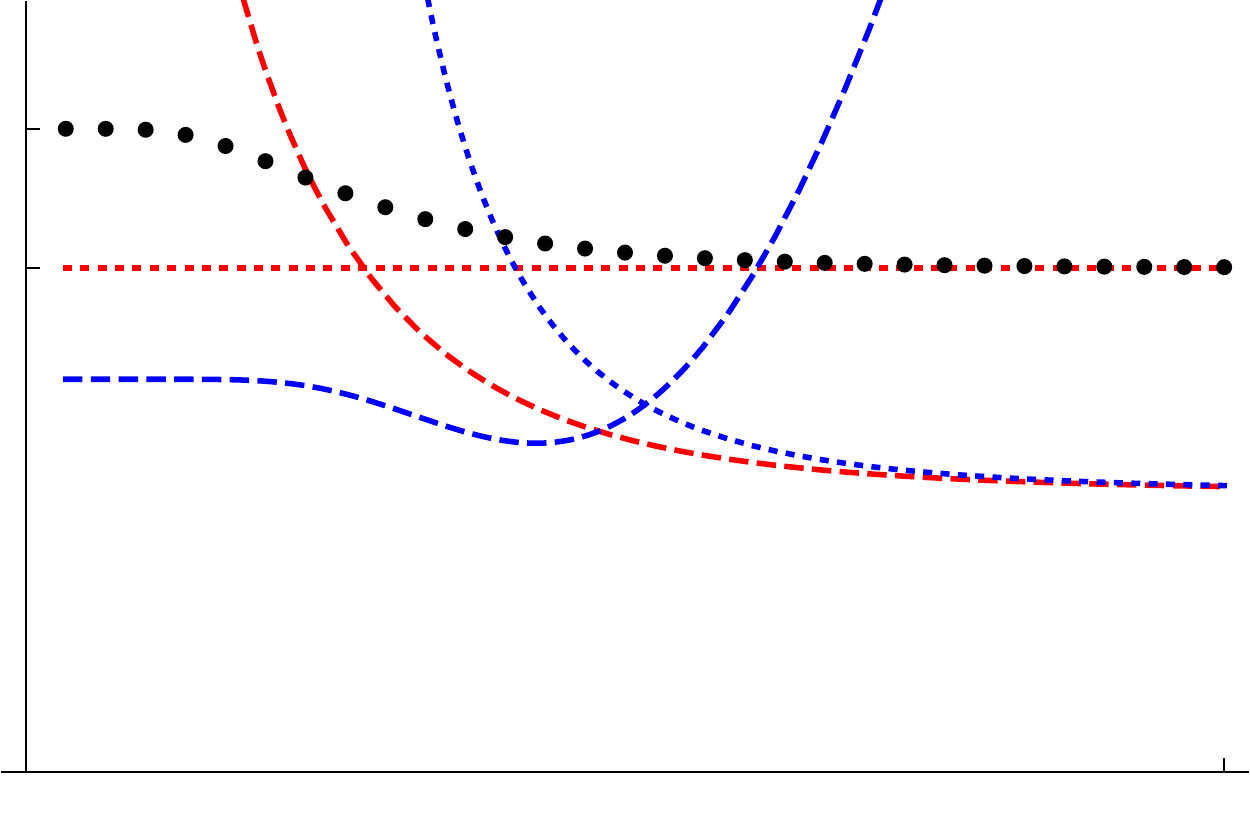}}
\put(3.5,2.5){\includegraphics[width = 0.45\textwidth]{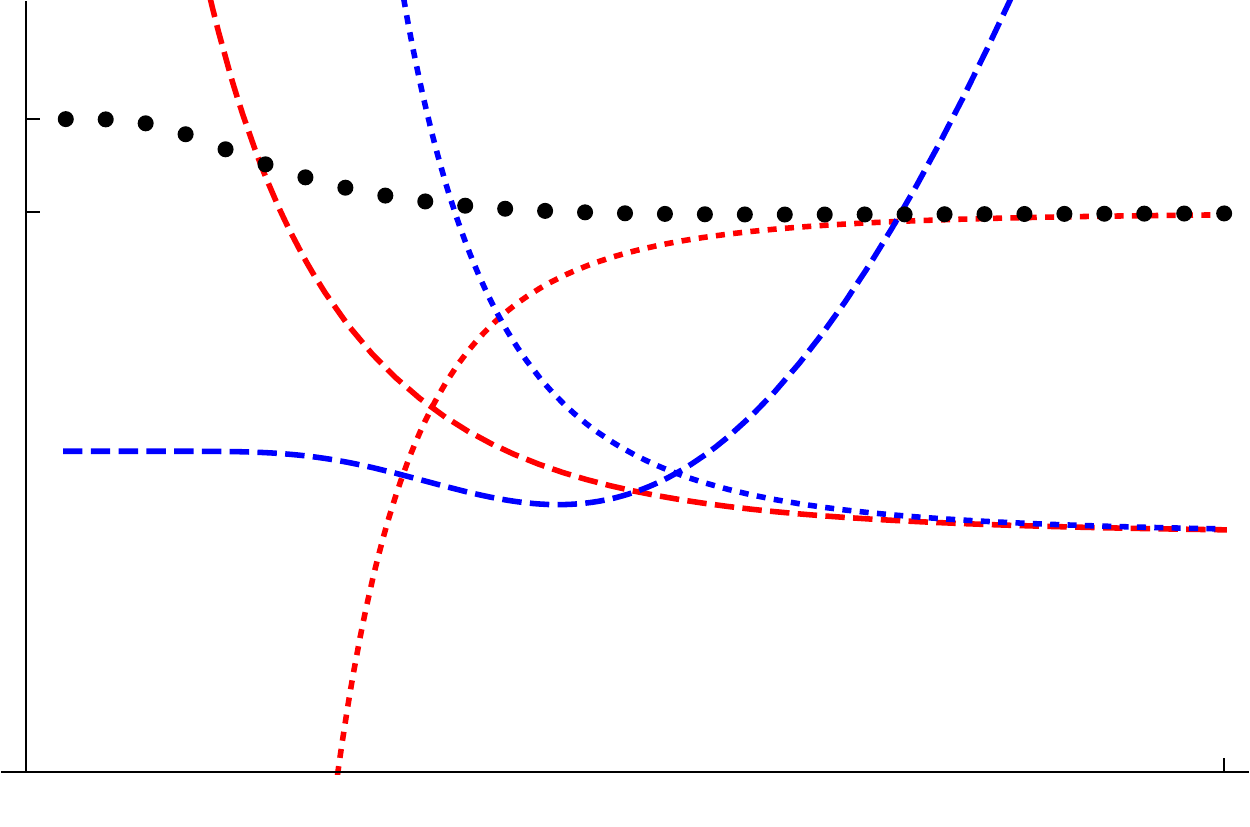}}
\put(53.5,2.5){\includegraphics[width = 0.45\textwidth]{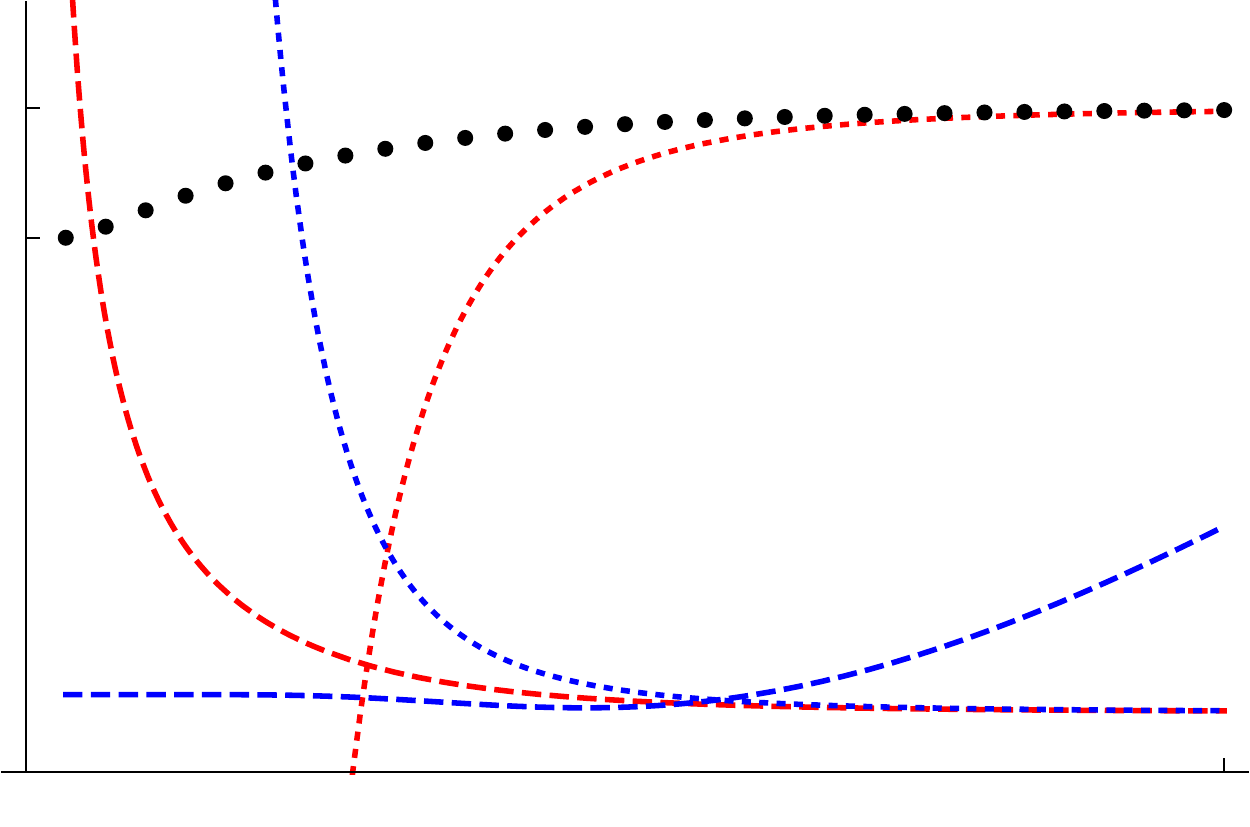}}
\put(73,93){\includegraphics[width = 0.08\textwidth]{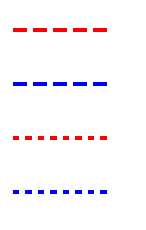}}
\put(75,90.25){\includegraphics[width = 0.04\textwidth]{legend_chain_point.pdf}}
\put(80,91.5){numerical}
\put(80,94.5){$\varepsilon$ expansion-high $T$}
\put(80,97.5){exact-high $T$}
\put(80,100.5){$\varepsilon$ expansion-low $T$}
\put(80,103.5){$\varepsilon$ expansion}
\put(23,35){$k = 2.25$}
\put(73,35){$k = 2$}
\put(-0.25,33){$I\left( {A:{A^C}} \right)$}
\put(49.75,32.5){$I\left( {A:{A^C}} \right)$}
\put(48.75,3.75){$T$}
\put(98.75,3.75){$T$}
\put(-0.25,27.25){$2S_A^0$}
\put(49.75,23.25){$2S_A^0$}
\put(0.75,23.75){$I^\infty$}
\put(50.75,27.5){$I^\infty$}
\put(42.5,1.5){$\sqrt{k_0}/2$}
\put(92.5,1.5){$\sqrt{k_0}/2$}
\put(23,71){$k = 3$}
\put(73,71){$k = 2.5$}
\put(-0.25,69){$I\left( {A:{A^C}} \right)$}
\put(49.75,68.5){$I\left( {A:{A^C}} \right)$}
\put(48.75,39.75){$T$}
\put(98.75,39.75){$T$}
\put(-0.25,59){$2S_A^0$}
\put(49.75,62.75){$2S_A^0$}
\put(0.75,53.75){$I^\infty$}
\put(50.75,58){$I^\infty$}
\put(42.5,37.5){$\sqrt{k_0}/2$}
\put(92.5,37.5){$\sqrt{k_0}/2$}
\put(23,107){$k = 18$}
\put(73,107){$k = 6$}
\put(-0.25,105){$I\left( {A:{A^C}} \right)$}
\put(49.75,104.5){$I\left( {A:{A^C}} \right)$}
\put(48.75,75.75){$T$}
\put(98.75,75.75){$T$}
\put(-0.25,96){$2S_A^0$}
\put(49.75,94.75){$2S_A^0$}
\put(0.75,84.25){$I^\infty$}
\put(50.75,86){$I^\infty$}
\put(42.5,73.5){$\sqrt{k_0}/2$}
\put(92.5,73.5){$\sqrt{k_0}/2$}
\end{picture}
\caption{The mutual information as function of the temperature for the chain of oscillators \eqref{eq:chains_degenerate_def} for viarious values of the parameter $k$}
\label{fig:deg_chain}
\end{figure}
The perturbation theory is in good agreement with the numerical results, whenever the parameter $k$ is large. Notice that there is an interesting change in the behaviour of the mutual information as $k$ gets lower. There is a critical value of $k$, where the dependence of the mutual information on the temperature ceases being monotonous. This is exactly the value where the coefficient of $1/T^4$ in the exact high temperature expansion \eqref{eq:chain:degenerate_exact_high_T} vanishes. This critical $k$, for large values of $n$ and $N$ tends exponentially fast to the value $k = - 5/2 l$. As $k$ further reduces, another more dramatic change occurs. The mutual information at infinite temperature becomes larger than that at zero temperature.

\setcounter{equation}{0}
\section{Free Scalar QFT}
\label{sec:qft}
\subsection{Discritizing the Degrees of Freedom in a Spherical Lattice}
\label{subsec:massless}
In this section, we extend the results of sections \ref{sec:qm_many} and \ref{sec:chain} to quantum field theory. We restrict our attention to the case of a free real scalar field in 3+1 dimensions. The analysis closely follows that of \cite{Srednicki:1993im}. The Hamiltonian equals
\begin{equation}
H = \frac{1}{2}\int {{d^3}x\left[ {{{\pi}\left( {\vec x} \right)}^2 + {{\left| {\vec \nabla \varphi \left( {\vec x} \right)} \right|}^2} + {\mu^2}\varphi {{\left( {\vec x} \right)}^2}} \right]} .
\label{eq:discretize_hamiltonian}
\end{equation}

We define,
\begin{align}
{\varphi _{lm}}\left( x \right) &= x\int {d\Omega {Y_{lm}}\left( {\theta ,\varphi } \right)\varphi \left( {\vec x} \right)} ,\label{eq:discretize_def1}\\
{\pi _{lm}}\left( x \right) &= x\int {d\Omega {Y_{lm}}\left( {\theta ,\varphi } \right)\pi \left( {\vec x} \right)} ,\label{eq:discretize_def2}
\end{align}
where ${Y_{lm}}$ are the real spherical harmonics namely,
\begin{equation}
{Y_{lm}} = \begin{cases}
\sqrt 2 {\left( { - 1} \right)^m}{\mathop{\rm Im}\nolimits} \left[ {Y_l^{ - m}} \right], & m<0 ,\\
Y_l^0, & m=0, \\
\sqrt 2 {\left( { - 1} \right)^m}{\mathop{\rm Re}\nolimits} \left[ {Y_l^m} \right], & m>0 ,
\end{cases}
\end{equation}
which form an orthonormal basis on the sphere $S^2$. The moments ${\varphi _{lm}}\left( x \right)$ and ${\pi _{lm}}\left( x \right)$ obey the canonical commutation relations $
\left[ {{\varphi _{lm}}\left( x \right),{\pi _{lm}}\left( x \right)} \right] = i\delta \left( {x - x'} \right){\delta _{ll'}}{\delta _{mm'}} $.
%
%
The Hamiltonian expressed in terms of ${\varphi _{lm}}\left( x \right)$ and ${\pi _{lm}}\left( x \right)$ assumes the form
\begin{equation}
H = \frac{1}{2}\sum\limits_{l,m} {\int_0^\infty  {dx \left\{ {{\pi _{lm}^2}\left( x \right) + {x^2}{{\left[ {\frac{\partial }{{\partial x}}\left( {\frac{{{\varphi _{lm}}\left( x \right)}}{x}} \right)} \right]}^2} + \left( {\frac{{l\left( {l + 1} \right)}}{{{x^2}}} + {\mu^2}} \right){\varphi _{lm}^2}\left( x \right)} \right\}} } .
\label{eq:free_massive_continuous}
\end{equation}

Had we descritized the radial coordinate appropriately, we would have resulted in an expression of the Hamiltonian containing countably infinite, canonically commuting variables, thus a Hamiltonian that can be dealt with the techniques of section \ref{sec:qm_many}. In order to achieve that, we introduce a lattice of spherical shells with radii $x_i = i a$ with $i \in \mathbb{N}$ and $1 \le i \le N$. The radial distance between consequent spherical shells sets a UV cutoff equal to $1/a$ to our system, while the overall size of the lattice sets an IR cutoff equal to $1/({Na})$. The Hamiltonian of the discretized system can be obtained from equation {\eqref{eq:free_massive_continuous} substituting, 
\begin{equation}
\begin{split}
x \to ja, \quad {\varphi _{lm}}\left( {ja} \right) \to {\varphi _{lm,j}} , &\quad {\pi _{lm}}\left( {ja} \right) \to \frac{{{\pi _{lm,j}}}}{a} ,\\
\left. \frac{{\partial {\varphi _{lm}}\left( x \right)}}{{\partial x}} \right|_{x=ja} \to \frac{{{\varphi _{lm,j + 1}} - {\varphi _{lm,j}}}}{a} , &\quad \int_0^{\left( {N + 1} \right)a} {dx} \to a\sum\limits_{j = 1}^N {} ,
\end{split}
\end{equation}
which results in
\begin{equation}
H = \frac{1}{{2a}}\sum\limits_{l,m} {\sum\limits_{j = 1}^N {\left[ {{\pi _{lm,j}^2} + {{\left( {j + \frac{1}{2}} \right)}^2}{{\left( {\frac{{{\varphi _{lm,j + 1}}}}{{j + 1}} - \frac{{{\varphi _{lm,j}}}}{j}} \right)}^2} + \left( {\frac{{l\left( {l + 1} \right)}}{{{j^2}}} + {\mu^2}{a^2}} \right){\varphi _{lm,j}^2}} \right]} } .
\end{equation}

Different $l$ and $m$ indices do not mix and moreover $m$ does not enter directly in the Hamiltonian, thus, the problem can be split to infinite independent sectors identified by index $l$, each being composed by $2 l + 1$ identical subsectors. Thus, the overall entanglement entropy can be calculated as the series
\begin{equation}
S\left( {N,n} \right) = \sum\limits_l {\left( {2l + 1} \right){S_{l}}\left( {N,n} \right)} ,
\label{eq:qft_entropy_series}
\end{equation}
where ${S_{l}}\left( {N,n} \right)$ is the entanglement entropy corresponding to the Hamiltonian
\begin{equation}
H_l = \frac{1}{{2a}} {\sum\limits_{j = 1}^N {\left[ {{\pi _{l,j}^2} + {{\left( {j + \frac{1}{2}} \right)}^2}{{\left( {\frac{{{\varphi _{l,j + 1}}}}{{j + 1}} - \frac{{{\varphi _{l,j}}}}{j}} \right)}^2} + \left( {\frac{{l\left( {l + 1} \right)}}{{{j^2}}} + {\mu^2}{a^2}} \right){\varphi _{l,j}^2}} \right]} } .
\label{eq:qft_hamiltonian_l}
\end{equation}
The latter contains a finite number of degrees of freedom, thus, ${S_{l}}\left( {N,n} \right)$ can be calculated using the results of section \ref{sec:qm_many}.

For large $l$, the matrix describing the $N$ oscillators becomes almost diagonal and as a result for large $l$ the system is almost disentangled. As a consequence, it can be shown that the series \eqref{eq:qft_entropy_series} is converging \cite{Srednicki:1993im}.

It follows that the mutual information can also be calculated as the series
\begin{equation}
I\left( {N,n} \right) = \sum\limits_l {\left( {2l + 1} \right){I_{l}}\left( {N,n} \right)} ,
\label{eq:qft_mutual_info_series}
\end{equation}
where ${I_{l}}\left( {N,n} \right)$ is the mutual information corresponding to the Hamiltonian \eqref{eq:qft_hamiltonian_l}.

\subsection{The Large $R$ Expansion}

We intend to study the dependence of the entanglement entropy and more importantly the mutual information, as a function of the size of the entangling sphere. For this purpose, we assume that the entangling sphere lies in the middle between the $n$-th and $(n + 1)$-th site of the spherical lattice. It follows that if we define
\begin{equation}
n_R := n + \frac{1}{2} ,
\end{equation}
then the radius of the entangling sphere will be
\begin{equation}
R = n_R a.
\end{equation}
In the following we will study the expansion of the entanglement entropy and the mutual information for large radii $R$ of the entangling sphere, or equivalently for large $n_R$.

The series \eqref{eq:qft_entropy_series} or \eqref{eq:qft_mutual_info_series} cannot be summed directly. Instead we will approximate them using the Euler-MacLaurin formula, closely following \cite{Katsinis:2017qzh}. This reads
\begin{multline}
\sum\limits_{n = a}^b {f\left( n \right)}  = \int_a^b {dxf\left( x \right)}  + \frac{{f\left( a \right) + f\left( b \right)}}{2}\\
 + \sum\limits_{k - 1}^\infty  {\frac{{{B_{2k}}}}{{\left( {2k} \right)!}}\left[ {{{\left. {\frac{{{d^{2k - 1}}f\left( x \right)}}{{d{x^{2k - 1}}}}} \right|}_{x = b}} - {{\left. {\frac{{{d^{2k - 1}}f\left( x \right)}}{{d{x^{2k - 1}}}}} \right|}_{x = a}}} \right]} ,
\label{eq:qft_Euler_MacLaurin}
\end{multline}
where the coefficients $B_k$ are the Bernoulli numbers defined so that $B_1 = 1/2$. Using this formula, we may approximate the series \eqref{eq:qft_mutual_info_series} with the integral
\begin{equation}
I \simeq \int_0^\infty  {d\ell \left( {2\ell  + 1} \right){I_\ell }\left( {N,n,\ell \left( {\ell  + 1} \right)} \right)} .
\label{eq:qft_mutual_info_integral_1}
\end{equation}
We are interested in the behaviour of this integral for large $R$. This behaviour cannot be isolated trivially, since $n_R$ appears in the integrand in the form of the fraction $ \ell (\ell + 1)/n_R^2$ and $\ell$ takes arbitrarily large values within the integration range. This can be bypassed performing the change of variables $ \ell (\ell + 1)/n_R^2 = y$. Then the integral formula \eqref{eq:qft_mutual_info_integral_1} assumes the form
\begin{equation}
I \simeq n_R^2\int_0^\infty  {dy{I_\ell }\left( {N,{n_R} - \frac{1}{2},yn_R^2} \right)} ,
\label{eq:qft_mutual_info_integral_2}
\end{equation}
which can be expanded for large $n_R$.

The term that is proportional to the highest power of $n_R$ that appears in this expansion is the one which is proportional to $n^2_R$, i.e. the ``area law'' term. When the size of the entangling sphere is sufficiently large, the mutual information is dominated by the area law term, in agreement with \cite{thermal_short}. This term receives contributions only from the integral term of the Euler-MacLaurin formula \eqref{eq:qft_Euler_MacLaurin}. Therefore, the large $R$ behaviour of the mutual information is determined by equation \eqref{eq:qft_mutual_info_integral_2}.

\subsection{The Hopping Expansion for the Area Law Term}

The Hamiltonian \eqref{eq:qft_hamiltonian_l} describes a system of coupled oscillators with couplings matrix, which can be approximated as
\begin{equation}
{K_{ij}} = \frac{1}{a}\left[ {\left( {2 + \frac{{l\left( {l + 1} \right)}}{{{i^2}}} + {\mu ^2}{a^2}} \right){\delta _{ij}} - {\delta _{i + 1,j}} - {\delta _{i,j + 1}}} \right] ,
\label{eq:qft_K}
\end{equation}
for the purpose of the determination of the leading ``area law'' term in the large $R$ expansion. Trivially, the Hamiltonian \eqref{eq:qft_hamiltonian_l} describes a chain of oscillators and we may use the results of section \ref{sec:chain}. Substituting the hopping expansion of the mutual information for a chain of oscillators \eqref{eq:chain_I_non_degenerate} with the couplings \eqref{eq:qft_K} to the integral formula \eqref{eq:qft_mutual_info_integral_2} and expanding for large $n_R$ yields
\begin{equation}
\begin{split}
I &= n_R^2\int\limits_0^\infty  {\frac{{\sqrt {2 + {a^2}{\mu ^2} + y}  + aT\sinh \left[ {\frac{1}{T}\sqrt {2 + {a^2}{\mu ^2} + y} } \right]}}{{8{a^2T^2}{{\left( {2 + {a^2}{\mu ^2} + y} \right)}^{\frac{3}{2}}} \left( \cosh \left[ {\frac{1}{T}\sqrt {2 + {a^2}{\mu ^2} + y} } \right] - 1 \right) }} + \mathcal{O}\left( {{n_R}} \right)} \\
 &= n_R^2\frac{{\coth \left[ {\frac{1}{{2aT}}\sqrt {2 + {a^2}{\mu ^2}} } \right]}}{{4aT\sqrt {2 + {a^2}{\mu ^2}} }} + \mathcal{O}\left( {{n_R}} \right) .
\end{split}
\label{eq:field_MI_expansion}
\end{equation}

This formula has the high temperature expansion
\begin{equation}
I = n_R^2\left( {\frac{1}{{2\left( {2 + {a^2}{\mu ^2}} \right)}} + \frac{1}{{24{a^2T^2}}} - \frac{{2 + {a^2}{\mu ^2}}}{{1440{a^4T^4}}} + \mathcal{O}\left( {\frac{1}{{{T^6}}}} \right)} \right) + \mathcal{O}\left( {{n_R}} \right) ,
\label{eq:field_MI_highT}
\end{equation}
which unlike the general formula for coupled oscillators contains an $1/T^2$ term. This seeming contradiction is due to the fact that we have integrated contributions from arbitrary high angular momenta $\ell$. The high temperature expansion \eqref{eq:N_MI_high_T} holds for temperatures higher than the eigenvalues of the matrix $K$. However, when one considers arbitrarily high angular momenta, these eigenvalues become arbitrarily large. This would be resolved had one introduced a physical cutoff to the angular momenta. We will return to this at the next subsection.

As we have seen in section \ref{sec:chain}, the $1/\mu$ expansion fails at low temperatures. In the same section, we obtained the appropriate low temperature expansion for the mutual information \eqref{eq:chain_low_T}. Substituting this low temperature expansion into the Euler MacLaurin formula yields
\begin{multline}
I = I_{T=0} + n_R^2\int_0^\infty  {dy\Bigg[ {2\log \left( {4\left( {2 + {\mu ^2}{a^2} + y} \right)} \right)\left( {1 + \frac{1}{{8{{\left( {2 + {\mu ^2}{a^2} + y} \right)}^2}}}} \right)} } \\
{ - \left( {1 + \frac{{\sqrt {2 + {\mu ^2}{a^2} + y} }}{T}\left( {1 + \frac{3}{{4y\left( {2 + {\mu ^2}{a^2} + y} \right)}}} \right)} \right)} \Bigg] \\
\times \exp \left[ - {\frac{{\sqrt {2 + {\mu ^2}{a^2} + y} }}{T}\left( {1 + \frac{3}{{4y\left( {2 + {\mu ^2}{a^2} + y} \right)}}} \right)} \right] .
\end{multline}
The first term, $I_{T=0}$, is the zero temperature mutual information, which is simply twice the zero temperature entanglement entropy. Perturbative expressions for this term in the $l/k$ expansion may be obtained from \cite{Katsinis:2017qzh}. Unlike the general case, the integral in the above formula cannot be performed analytically. However, its behaviour is dominated by the exponential part. The exponent, i.e. the function
\begin{equation}
f\left( y \right) = \frac{{\sqrt {2 + {\mu ^2}{a^2} + y} }}{T}\left( {1 + \frac{3}{{4y\left( {2 + {\mu ^2}{a^2} + y} \right)}}} \right)
\end{equation}
has only one minimum in $\left( 0 , \infty \right)$, which lies at
\begin{equation}
{y_{\min }} = \sqrt {\frac{3}{2}} ,
\end{equation}
at this order in $l/k$. Therefore, a saddle point approximation may be performed. The value of the function $f$ and its second derivative at the minimum, at this order equal
\begin{equation}
f\left( {{y_{\min }}} \right) = \frac{{\sqrt {2 + {\mu ^2}{a^2}} }}{T},\quad {\left. {\frac{{{d^2}f\left( y \right)}}{{d{y^2}}}} \right|_{y = {y_{\min }}}} = \sqrt {\frac{2}{{3\left( {2 + {\mu ^2}{a^2}} \right)}}} \frac{1}{T} .
\end{equation}
It is then a matter of algebra to show that
\begin{multline}
I \simeq {I_{T = 0}} + 2n_R^2\sqrt {2\pi T} \sqrt[4]{{\frac{{3\left( {2 + {\mu ^2}{a^2}} \right)}}{2}}} \\
\times\left[ {2\log \left( {4\left( {2 + {\mu ^2}{a^2}} \right)} \right) - 1 - \frac{{\sqrt {2 + {\mu ^2}{a^2}} }}{T}} \right]\exp \left[ { - \frac{{\sqrt {2 + {\mu ^2}{a^2}} }}{T}} \right] .
\label{eq:field_MI_lowT}
\end{multline}

Figure \ref{fig:MI3d} shows the dependence of the coefficient of the ``area law'' term of the mutual information on the temperature, for various values of the mass parameter. For each mass, the first order result in the $l/k$ \eqref{eq:field_MI_expansion}, as well as the high temperature \eqref{eq:field_MI_highT} and low temperature \eqref{eq:field_MI_lowT} expansions are displayed. The analytic formulas are compared with a numerical calculation of the mutual information, which is based on the direct numerical specification of the eigenvalues of the matrix $\gamma^{-1} \beta$ and is performed with the use of Wolfram's Mathematica. We used the third order result for the entanglement entropy at zero temperature, derived in \cite{Katsinis:2017qzh}, in order to approximate the $I_{T=0}$ term in the low temperature formula \eqref{eq:field_MI_lowT}. It is evident that the analytic formulae that we obtained in this section are in good agreement to the numerical results, especially for large masses.
\begin{figure}[p]
\centering
\begin{picture}(100,110)
\put(3.5,74.5){\includegraphics[width = 0.45\textwidth]{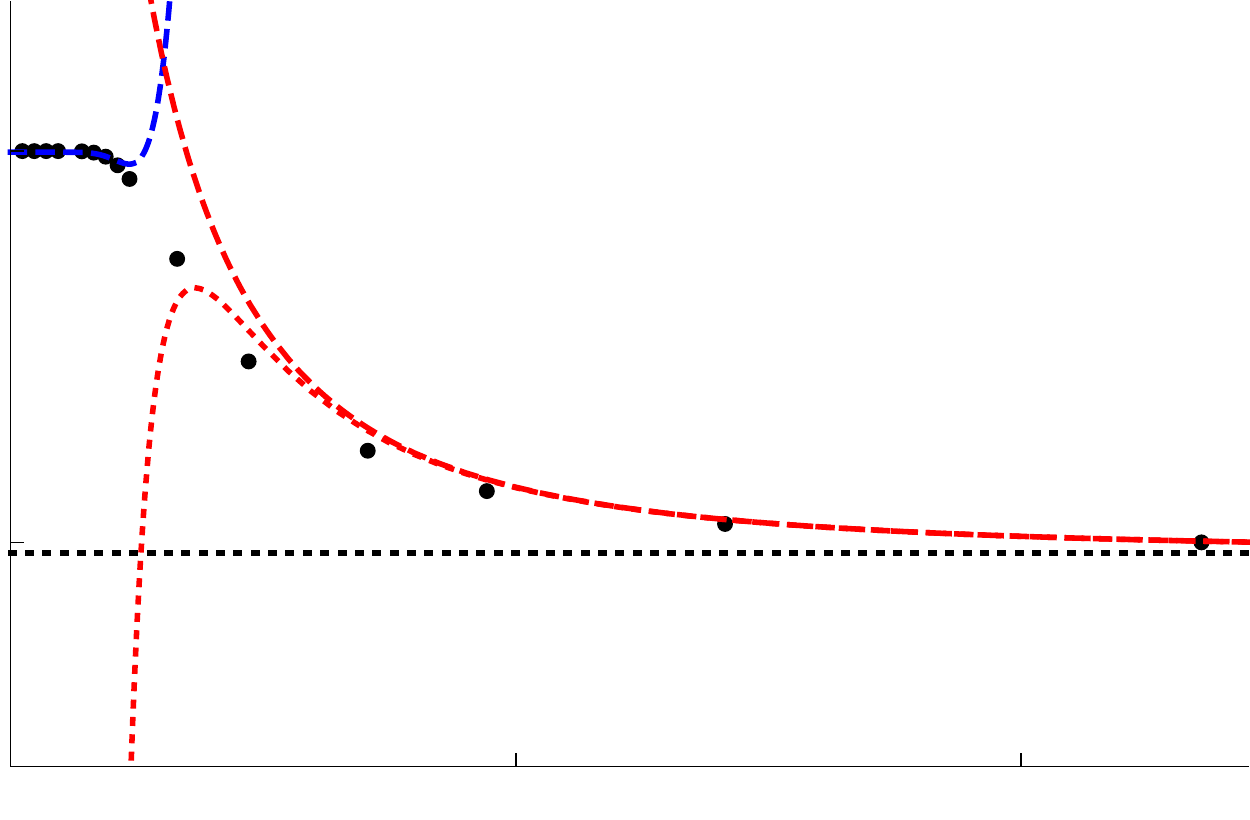}}
\put(53.5,74.5){\includegraphics[width = 0.45\textwidth]{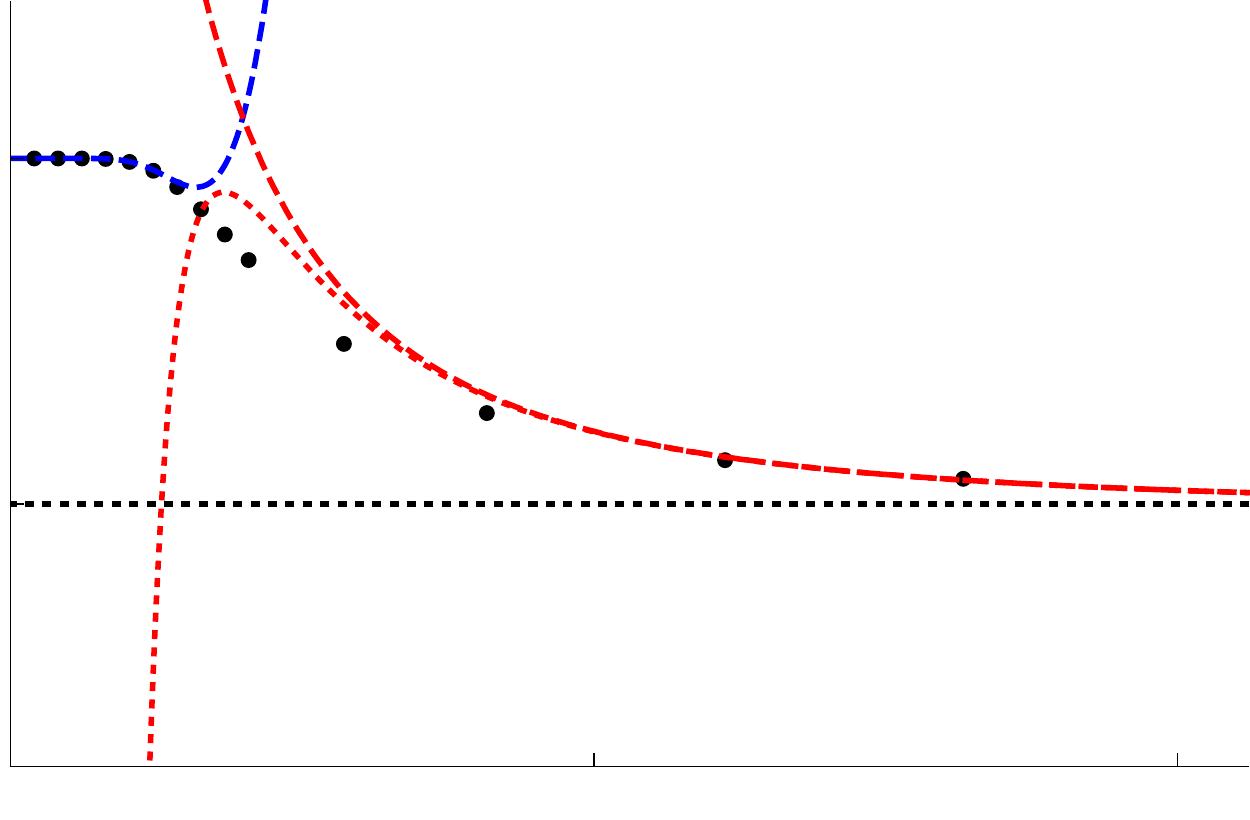}}
\put(3.5,38.5){\includegraphics[width = 0.45\textwidth]{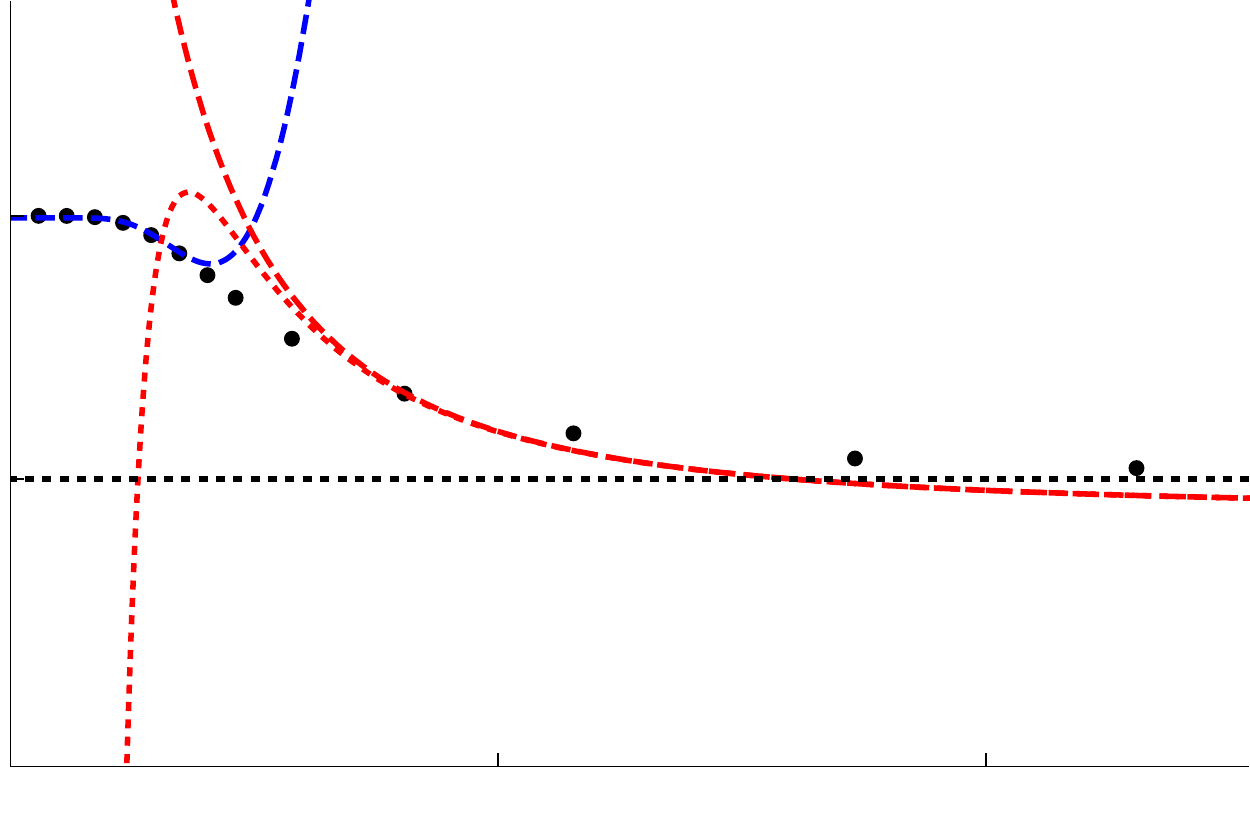}}
\put(53.5,38.5){\includegraphics[width = 0.45\textwidth]{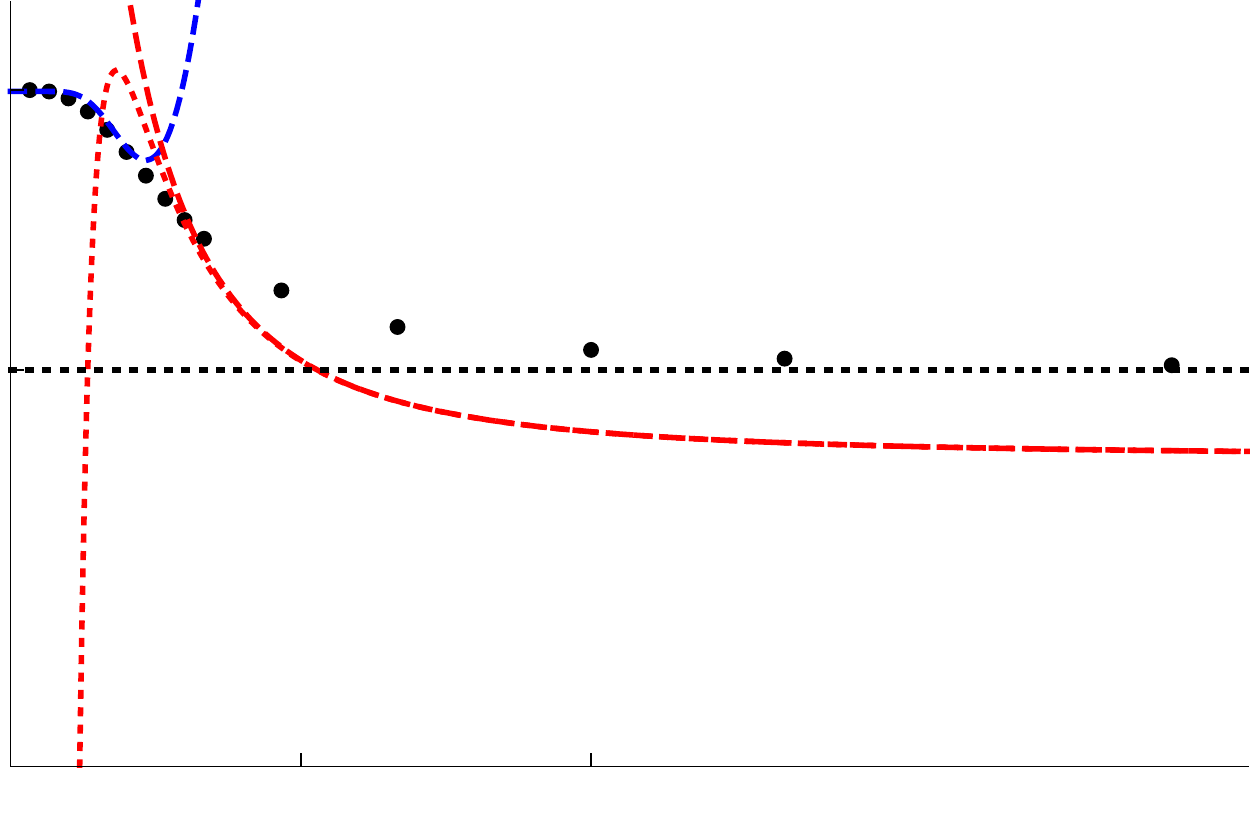}}
\put(3.5,2.5){\includegraphics[width = 0.45\textwidth]{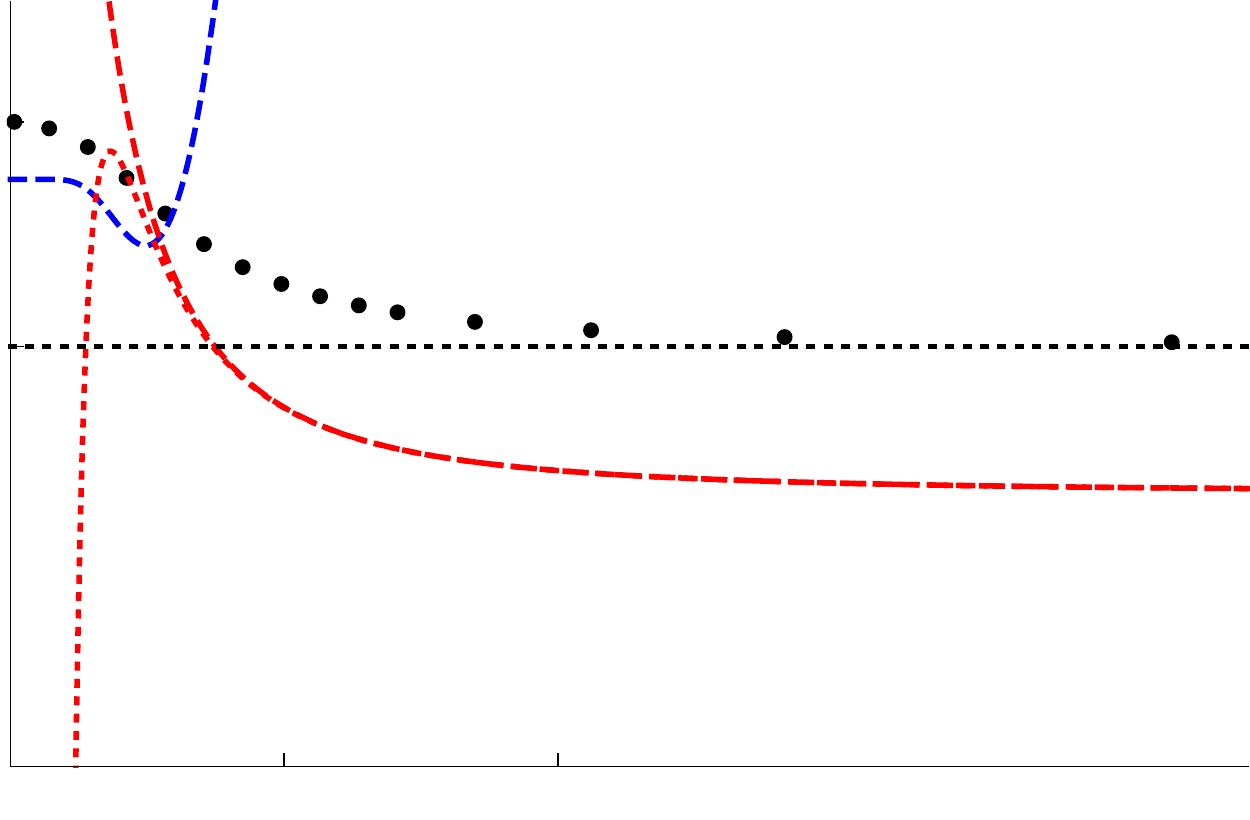}}
%
\put(61,15){\includegraphics[width = 0.08\textwidth]{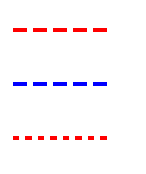}}
\put(63,12.25){\includegraphics[width = 0.04\textwidth]{legend_chain_point.pdf}}
\put(68,13.5){numerical}
\put(68,16.5){high $T$ of $\varepsilon$ expansion}
\put(68,19.5){$\varepsilon$ expansion at low $T$}
\put(68,22.5){$\varepsilon$ expansion}
\put(23,33){$\mu = 0$}
\put(0.75,33){$I/n_R^2$}
\put(48.75,3.5){$T$}
\put(-0.25,27){$\frac{2S_A^0}{n_R^2}$}
\put(0.5,19){$\frac{I^\infty}{n_R^2}$}
\put(19.5,1.5){$\sqrt{2}/a$}
%
\put(23,69){$\mu a = 1$}
\put(72.5,69){$\mu a = 1/2$}
\put(0.5,69){$I/n_R^2$}
\put(50.5,68.5){$I/n_R^2$}
\put(48.75,39.5){$T$}
\put(98.75,39.5){$T$}
\put(-0.25,59.5){$\frac{2S_A^0}{n_R^2}$}
\put(49.75,64.25){$\frac{2S_A^0}{n_R^2}$}
\put(0.75,50){$\frac{I^\infty}{n_R^2}$}
\put(50.75,54){$\frac{I^\infty}{n_R^2}$}
\put(31.25,37.5){$\sqrt{\mu^2+2/a^2}$}
\put(67,37.5){$\sqrt{\mu^2+2/a^2}$}
\put(23,105){$\mu a = 4$}
\put(73,105){$\mu a = 2$}
\put(0.75,105){$I/n_R^2$}
\put(50.75,104.5){$I/n_R^2$}
\put(48.75,75.5){$T$}
\put(98.75,75.5){$T$}
\put(-0.25,97.75){$\frac{2S_A^0}{n_R^2}$}
\put(49.75,97.5){$\frac{2S_A^0}{n_R^2}$}
\put(0.5,83.25){$\frac{I^\infty}{n_R^2}$}
\put(50.5,85){$\frac{I^\infty}{n_R^2}$}
\put(32.5,73.5){$\sqrt{\mu^2+2/a^2}$}
\put(84.25,73.5){$\sqrt{\mu^2+2/a^2}$}
\end{picture}
\caption{The area law term coefficient of the mutual information as function of the temperature. The dashed lines are the low and high temperature expansions of the mutual information, whereas the dotted lines are the asymptotic values for $T \to \infty$.}
\label{fig:MI3d}
\end{figure}

\subsection{Dependence on the Regularization}
\label{subsec:area_reg}
As explained in \cite{Katsinis:2017qzh}, the regularization scheme that we use in this section is quite peculiar. The radial and angular excitations of the field are treated differently; while there is a UV cutoff equal to $1/a$ for the radial ones, the angular ones are integrated up to infinite scale. One can enforce a more uniform regularization introducing a cutoff at the angular momenta of the form ${l_{\max }} = cR/a$. The appropriate selection for $c$ in $3+1$ dimensions, so that the density of the degrees of freedom at the region of the entangling surface is homogeneous, is $c = 2 \sqrt{\pi}$. Then, the results of the previous subsection serve as an upper bound for the area law term. It has to be noted that had one desired to generalize these results to an arbitrary number of dimensions, they would have found that the integral without the angular momentum cutoff diverges at $4+1$ and higher dimensions; this upper bound exists only in $2+1$ and $3+1$ dimensions. Obviously, the introduction of this cutoff yields the coefficient of the area law term of the mutual information finite at all dimensions. Returning to $3+1$ dimensions, such a regularization yields
\begin{equation}
I = n_R^2\left( {\frac{{\coth \left[ {\frac{1}{{2aT}}\sqrt {2 + {a^2}{\mu ^2}} } \right]}}{{4aT\sqrt {2 + {a^2}{\mu ^2}} }} - \frac{{\coth \left[ {\frac{1}{{2aT}}\sqrt {2 + {a^2}{\mu ^2} + {c^2}} } \right]}}{{4aT\sqrt {2 + {a^2}{\mu ^2} + {c^2}} }}} \right) + \mathcal{O}\left( {{n_R}} \right) .
\end{equation}
This formula has the high temperature expansion
\begin{equation}
I = n_R^2\left( {\frac{1}{{2\left( {2 + {a^2}{\mu ^2}} \right)}} - \frac{1}{{2\left( {2 + {a^2}{\mu ^2} + {c^2}} \right)}} + \frac{{{c^2}}}{{1440{a^4T^4}}} + \mathcal{O} \left( {\frac{1}{{{T^6}}}} \right)} \right) + \mathcal{O}\left( {{n_R}} \right) .
\end{equation}
This is exactly what should be expected from the general high temperature formula \eqref{eq:N_MI_high_T}. The $1/T^2$ term is vanishing, whereas the $1/T^4$ contains only the leading term in the $1/\mu$ expansion (the last term of equation \eqref{eq:N_MI_high_T}), which is equal to $1/(1440a^4T^4)$ from each angular momentum sector. As we have cutoff the angular momenta at ${l_{\max }} = cR/a \simeq c (n_R + 1/2)$, at leading order in $n_R$ there are $c^2 n_R^2$ such sectors, which is consistent with our result.

The low temperature behaviour is determined by the low angular momenta. Naturally, the introduction of the angular momenta cutoff does not alter the procedure of deriving the low temperature expansion of the mutual information, as long as $c> \sqrt{3/2}$. For these values of $c$ the formula \eqref{eq:field_MI_lowT} provides a good approximation of the mutual information at low temperature.

Figure \ref{fig:MI3d_reg} shows the dependence of the coefficient of the ``area law'' term of the mutual information on the temperature, with the use of an angular momentum cutoff ${l_{\max }} = 2 \sqrt{\pi} R/a$, for various values of the mass parameter. The first order expansion, as well as the low and high temperature expansions are compared to numerical calculations performed with the use of Wolfram's Mathematica.
\begin{figure}[p]
\centering
\begin{picture}(100,110)
\put(3.5,74.5){\includegraphics[width = 0.45\textwidth]{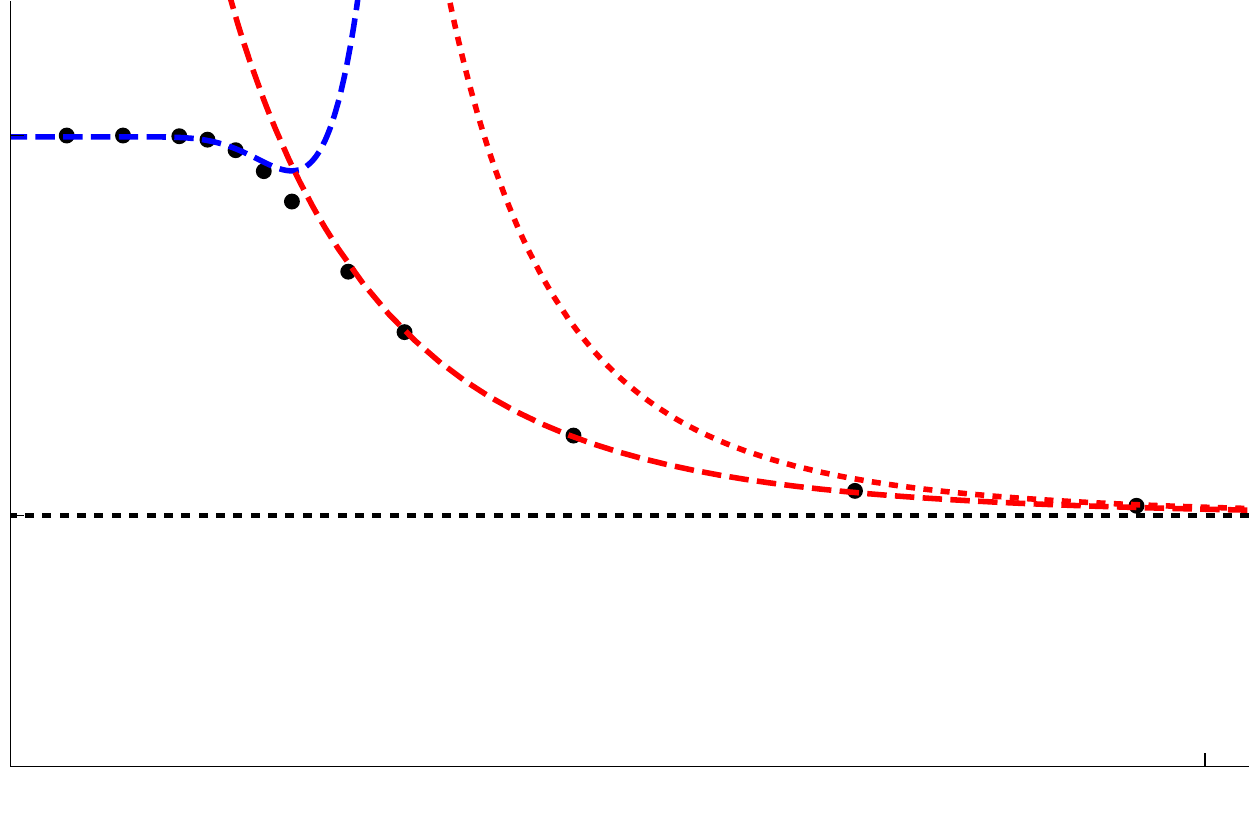}}
\put(53.5,74.5){\includegraphics[width = 0.45\textwidth]{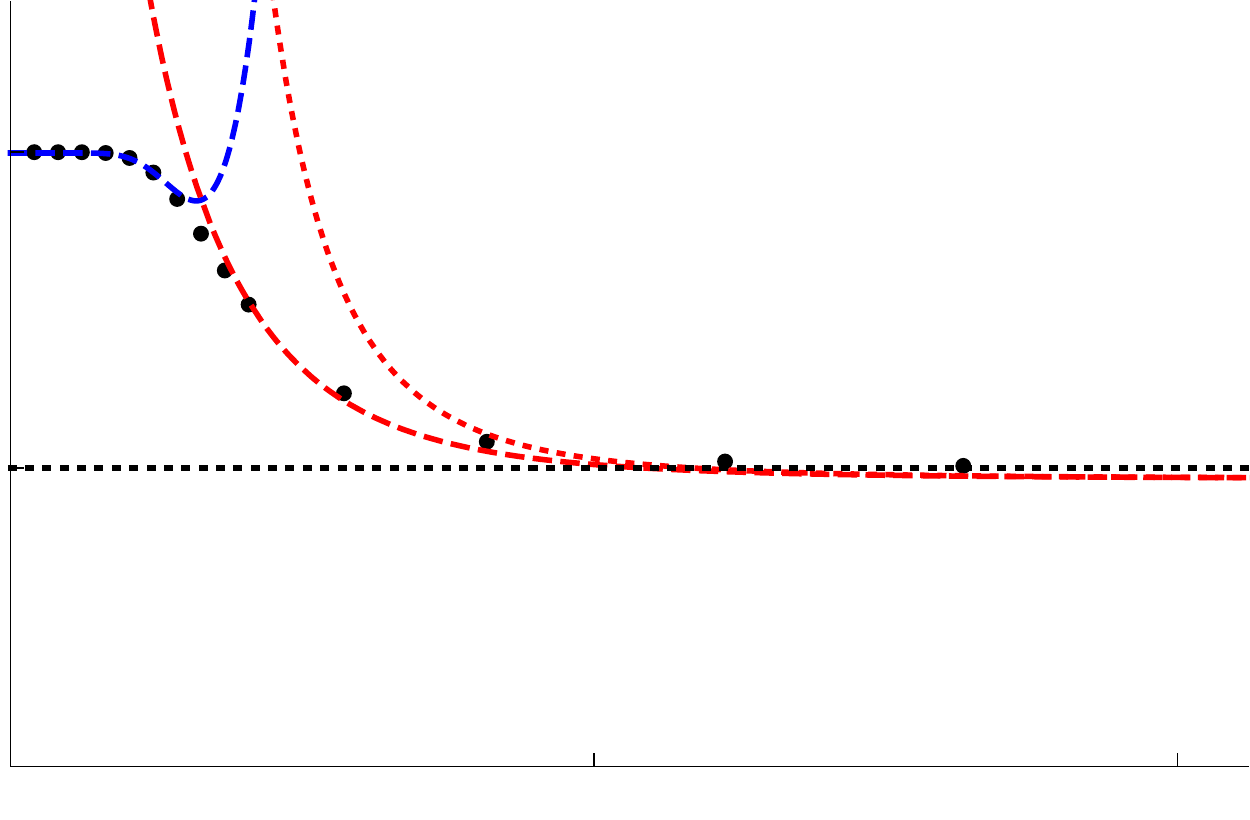}}
\put(3.5,38.5){\includegraphics[width = 0.45\textwidth]{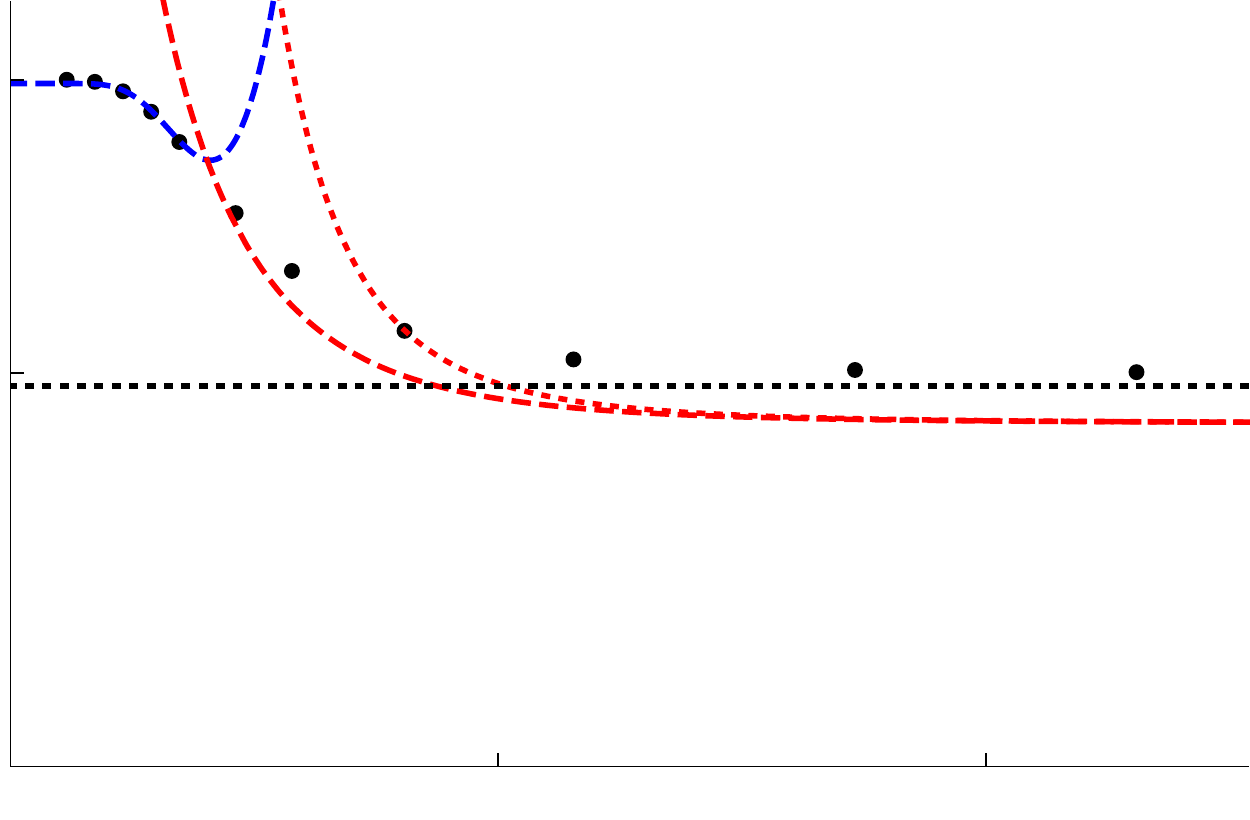}}
\put(53.5,38.5){\includegraphics[width = 0.45\textwidth]{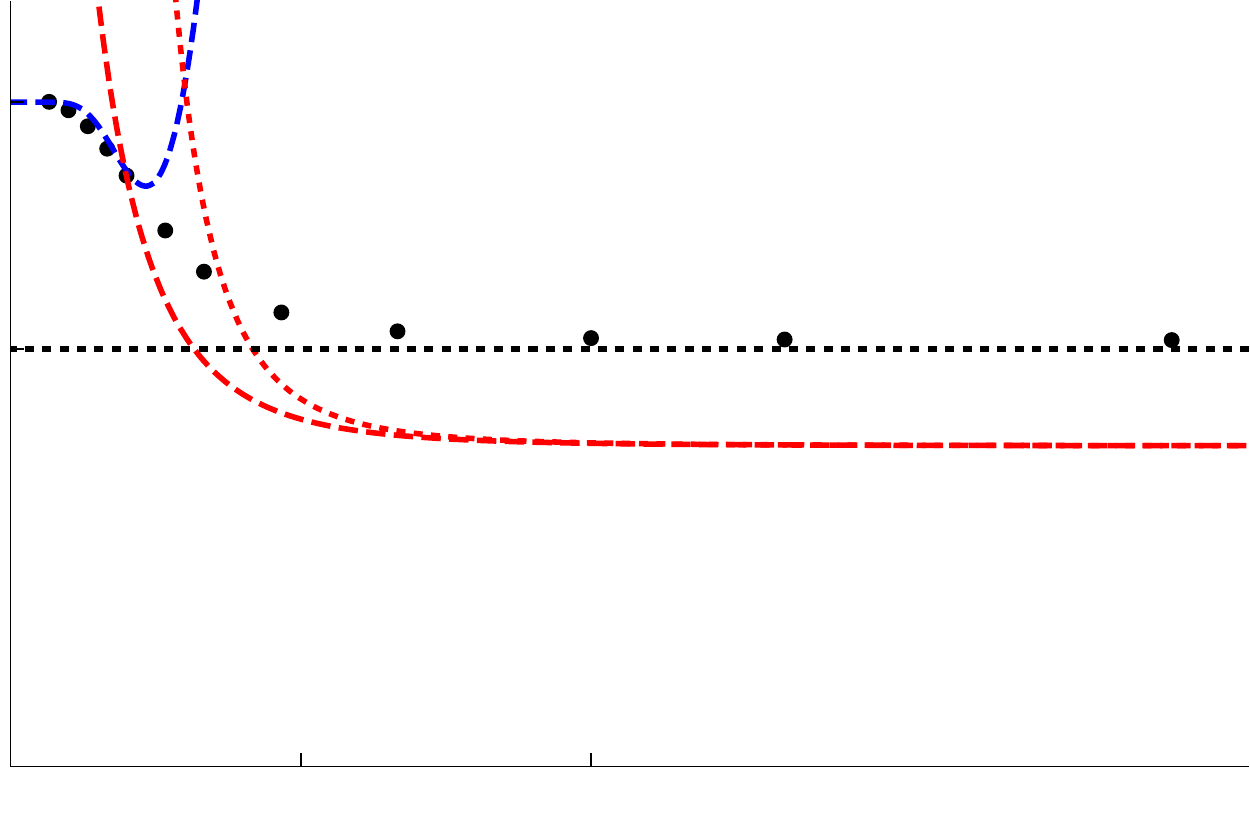}}
\put(3.5,2.5){\includegraphics[width = 0.45\textwidth]{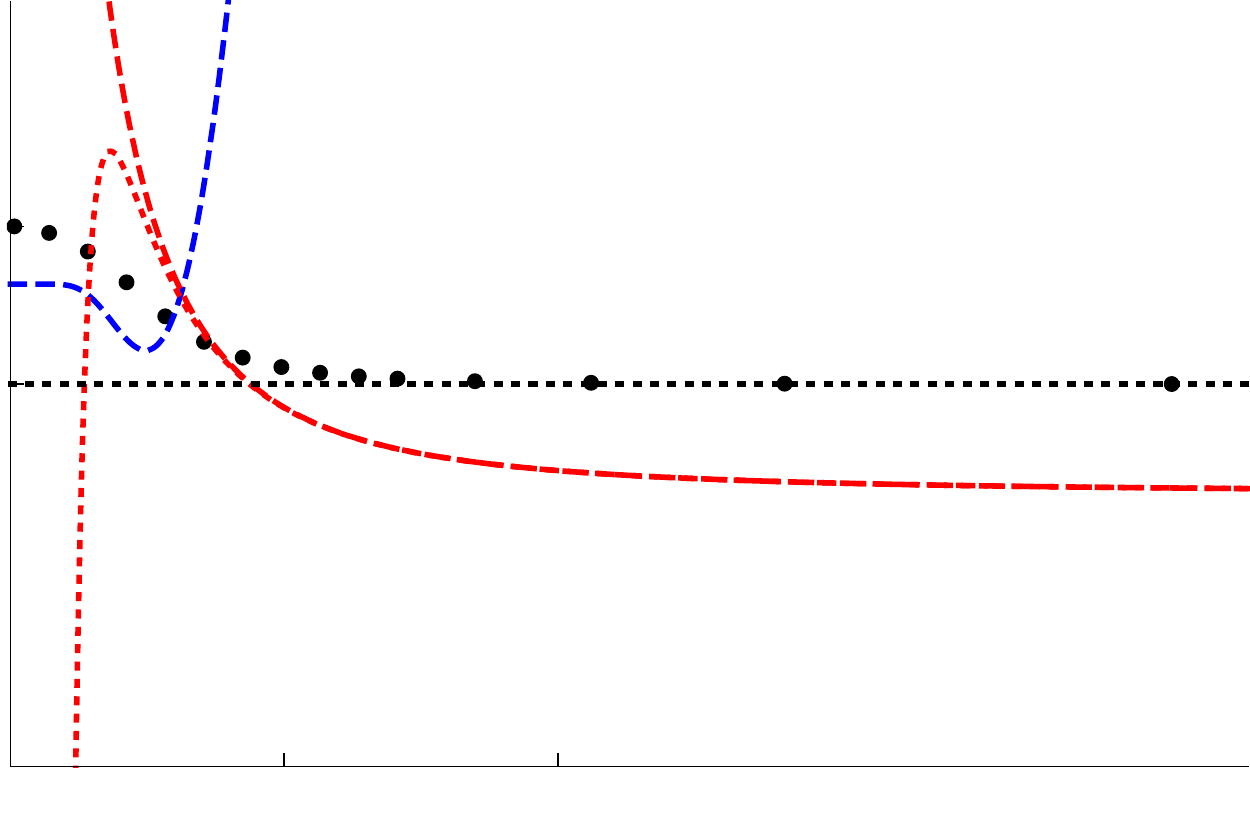}}
%
\put(61,15){\includegraphics[width = 0.08\textwidth]{LegendMI3d.pdf}}
\put(63,12.25){\includegraphics[width = 0.04\textwidth]{legend_chain_point.pdf}}
\put(68,13.5){numerical}
\put(68,16.5){high $T$ of $\varepsilon$ expansion}
\put(68,19.5){$\varepsilon$ expansion at low $T$}
\put(68,22.5){$\varepsilon$ expansion}
\put(23,33){$\mu = 0$}
\put(0.75,33){$I/n_R^2$}
\put(48.75,3.5){$T$}
\put(-0.21,23.25){$\frac{2S_A^0}{n_R^2}$}
\put(0.5,17.5){$\frac{I^\infty}{n_R^2}$}
\put(19.5,1.5){$\sqrt{2}/a$}
%
\put(23,69){$\mu a = 1$}
\put(72.5,69){$\mu a = 1/2$}
\put(0.5,69){$I/n_R^2$}
\put(50.5,68.5){$I/n_R^2$}
\put(48.75,39.5){$T$}
\put(98.75,39.5){$T$}
\put(-0.25,64.5){$\frac{2S_A^0}{n_R^2}$}
\put(49.75,63.75){$\frac{2S_A^0}{n_R^2}$}
\put(0.75,53.5){$\frac{I^\infty}{n_R^2}$}
\put(50.75,54.75){$\frac{I^\infty}{n_R^2}$}
\put(31.25,37.5){$\sqrt{\mu^2+2/a^2}$}
\put(67,37.5){$\sqrt{\mu^2+2/a^2}$}
\put(23,105){$\mu a = 4$}
\put(73,105){$\mu a = 2$}
\put(0.75,105){$I/n_R^2$}
\put(50.75,104.5){$I/n_R^2$}
\put(48.75,75.5){$T$}
\put(98.75,75.5){$T$}
\put(-0.25,98.5){$\frac{2S_A^0}{n_R^2}$}
\put(49.75,97.75){$\frac{2S_A^0}{n_R^2}$}
\put(0.5,84.75){$\frac{I^\infty}{n_R^2}$}
\put(50.5,86.5){$\frac{I^\infty}{n_R^2}$}
\put(31.5,73.5){$\sqrt{\mu^2+2/a^2}/2$}
\put(84.25,73.5){$\sqrt{\mu^2+2/a^2}$}
\end{picture}
\caption{The area law term coefficient of the mutual information as function of the temperature with an angular momentum cutoff ${l_{\max }} = 2 \sqrt{\pi} R/a$. The dashed lines are the low and high temperature expansions of the mutual information, whereas the dotted lines are the asymptotic values for $T \to \infty$.}
\label{fig:MI3d_reg}
\end{figure}

\setcounter{equation}{0}
\section{Discussion}
\label{sec:discussion}

In a seminal paper \cite{Srednicki:1993im}, Srednicki calculated the entanglement entropy in massless scalar field theory at its ground state when the entangling surface is a sphere. It turns out that the entanglement entropy is proportional to the area of the sphere and not its volume, resembling the well-known property of the black hole entropy. This behaviour continues to hold at massive scalar field theory, where perturbative methods have been applied to calculate the entanglement entropy for a spherical entangling surface \cite{Katsinis:2017qzh}.

When the mass of the field is very large, the area law can be understood as a result of the locality. In such cases only correlations between nearest neighbours are important, therefore the entanglement entropy should be expected to be proportional to the number of neighbouring degrees of freedom that have been separated by the entangling surface. These are obviously proportional to the area of the entangling surface. However, the area law holds in the massless case, too. The underlying cause of this behaviour is the symmetric property of the entanglement entropy. Whenever the composite system lies in a pure state it holds that $S_A = S_{A^C}$. Therefore, a volume term cannot appear as it should be proportional to the volume of the interior \emph{and simultaneously} to the volume of the exterior of the sphere. Naturally, the entanglement entropy has to depend on the geometric characteristics of the only common feature that the interior and exterior of the sphere share, i.e. the sphere itself.

In this work we study free scalar field theory at a thermal state, generalizing the perturbative methods of \cite{Katsinis:2017qzh}. It turns out that the entanglement entropy contains volume terms, which are inherited from the thermal entropy of the overall system. The presence of such terms should not be considered surprising, since the symmetry property of the entanglement entropy does not hold whenever the composite system lies in a mixed state. The entanglement entropy is not a good measure of quantum entanglement in such cases; a better measure of the correlations between a subsystem and its complement is the mutual information. This, by definition obeys the symmetry property, and, thus, it should be expected that in field theory it obeys an area law, even at finite temperature. Indeed, our perturbative calculations, as well as the numerical calculations that we performed, verify this intuitive prediction.

The coefficient of the area law term of the mutual information exposes an interesting behaviour as a function of the temperature. This coefficient reduces as the temperature increases; this is expected as the thermal effects tend to wash out the quantum correlations between the considered subsystems. However, as the temperature tends to infinity, the coefficient does not vanish, but it rather tends to a given finite value. This is a property of any harmonic oscillatory system. It turns out that the asymptotic value of the mutual information at infinite temperature is identical to the mutual information of the equivalent classical system of coupled oscillators at finite temperature.

Following the approach of \cite{Katsinis:2017qzh}, we found a perturbative expression for the area law coefficient, expanding in the inverse mass of the scalar field. The calculation is performed in the lowest order. It is in good agreement with the numerical calculations, especially for large values for the field mass. The calculation, although significantly more complicated than the zero temperature one, can be directly performed at higher orders, improving the accuracy of the analytic results.

Similarly to the zero temperature case, due to the particular discretization of the field degrees of freedom in radial shells, the expansion continues to work even at the massless field limit in $3+1$ dimensions. This is due to the fact that the angular momentum effectively acts as a mass term for the corresponding moments of the field. However, it fails in $1+1$ dimensions at the massless limit.

The original calculation of Srednicki implements a peculiar regularization. Although a lattice of spherical shells is used, introducing a UV cutoff at the radial field excitations, the angular momenta are integrated up to infinity. This scheme provides a finite result only at $2+1$ and $3+1$ dimensions. One may apply a more uniform scheme, introducing an angular momentum cutoff so that a similar UV cutoff applies at the angular degrees of freedom on the entangling surface. Such a regularization scheme exposes the fact that the area law term is regularization scheme dependent. Furthermore, similarly to the zero temperature case, the Srednicki regularization in $2+1$ and $3+1$ provides an upper bound for the coefficient of the area law term. In higher dimensions there is no such bound, however, the introduction of this more uniform regularization leads to a finite result for the area law coefficient.

Finally, another interesting property concerns the high temperature expansion of the mutual information in any harmonic oscillatory system. This expansion naturally contains even powers of $1/T$. However, it turns out that the first term, namely the $1/T^2$ term, always vanishes.


\subsection*{Acknowledgements}
The research of G.P. has received funding from the Hellenic Foundation for Research and Innovation (HFRI) and the General Secretariat for Research and Technology (GSRT), in the framework of the ``First Post-doctoral researchers support'', under grant agreement No 2595. The research of D.K. is co-financed by Greece and the European Union (European Social Fund- ESF) through the Operational Programme ``Human Resources Development, Education and Lifelong Learning'' in the context of the project ``Strengthening Human Resources Research Potential via Doctorate Research'' (MIS-5000432), implemented by the State Scholarships Foundation (IKY). The authors would like to thank M. Axenides and E. Floratos for useful discussions.

\appendix

\renewcommand{\theequation}{\Alph{section}.\arabic{equation}}

\setcounter{equation}{0}
\section{The Classical Mutual Information for a Pair of Coupled Oscillators}
\label{sec:classical_mutual}

In order to understand the nature of the remnant of the mutual information at infinite temperature, we present the classical analysis \cite{Cramer:2005mx}. First we consider a single harmonic oscillator with eigenfrequency $\omega$. Without loss of generality we assume that the mass of the oscillator is equal to one. In the classical limit, the probability of finding the particle at position $x$ is inverse proportional to the magnitude of the velocity.
\begin{equation}
{p}\left( x \right) \sim \frac{1}{\left| v \right|}.
\end{equation}
It follows from energy conservation, $\frac{1}{2}{v^2} + \frac{1}{2}{\omega ^2}{x^2} = E$, that when the system has energy $E$, the above probability distribution assumes the form
\begin{equation}
{p_E}\left( x \right) = \frac{\omega }{{\pi \sqrt {2E - {\omega ^2}{x^2}} }} .
\end{equation}

Now we turn on the temperature, introducing a canonical ensemble of harmonic oscillators. As a consequence of the  fact that the period of the motion is independent of the energy, the phase space volume per energy is constant. It follows that the appropriately normalized probability distribution for the energies is
\begin{equation}
p\left( E \right) = \frac{1}{T}{e^{ - \frac{E}{T}}} .
\end{equation}
This implies that the spatial probability distribution at finite temperature $T$ is
\begin{equation}
{p_{{\rm{can}}}}\left( {x;\omega ,T} \right) = \int_{\frac{1}{2}{\omega ^2}{x^2}}^\infty  {p\left( E \right){p_E}\left( x \right)dE}  = \frac{\omega }{{\sqrt {2\pi T} }}{e^{ - \frac{{{\omega ^2}{x^2}}}{{2T}}}} ,
\end{equation}
where the lower bound of the integration was taken equal to $\frac{1}{2}{\omega ^2}{x^2}$, since at least that much energy is required is order to reach the position $x$.

Let us now consider the system of two coupled oscillators of section \ref{sec:qm_2}, which is described by the Hamiltonian \eqref{eq:hamiltonian_original_coordinates}. As usual, one may introduce the canonical coordinates \eqref{eq:normal_coordinates}, which allow the re-expression of the Hamiltonian in the form \eqref{eq:hamiltonian_normal_coordinates}, which describes two decoupled oscillators, one for each mode. Therefore,
\begin{equation}
\begin{split}
p\left( {{x_1},{x_2};T} \right) &= {p_{{\rm{can}}}}\left( {\frac{{{x_1} + {x_2}}}{{\sqrt 2 }};{\omega _ + },T} \right){p_{{\rm{can}}}}\left( {\frac{{{x_1} + {x_2}}}{{\sqrt 2 }};{\omega _ - },T} \right)\\
 &= \frac{{{\omega _ + }{\omega _ - }}}{{2\pi T}}{e^{ - \frac{{\omega _ + ^2{{\left( {{x_1} + {x_2}} \right)}^2} + \omega _ - ^2{{\left( {{x_1} - {x_2}} \right)}^2}}}{{4T}}}} .
\end{split}
\end{equation}
The probability distribution of the position of the first of the two coupled oscillators can be calculated integrating out the position of the second one. Simple algebra yields
\begin{equation}
p\left( {{x_1};T} \right) = \int {p\left( {{x_1},{x_2};T} \right)d{x_2}}  = \frac{{\omega _{{\rm{eff}}}^\infty }}{{\sqrt {2\pi T} }}{e^{ - \frac{{{{\left( {\omega _{{\rm{eff}}}^\infty } \right)}^2}x_1^2}}{{2T}}}} ,
\end{equation}
where $\omega _{{\rm{eff}}}^\infty  = \sqrt {\frac{{2\omega _ + ^2\omega _ - ^2}}{{\omega _ + ^2 + \omega _ - ^2}}} $. We remind the reader that this is not the first time we meet this frequency. It is identical to the limiting value at infinite temperature \eqref{eq:effective_omega_infinity} of the eigenfrequency of the effective single oscillator \eqref{eq:effective_omega} that reproduces the reduced density matrix at the appropriate effective temperature \eqref{eq:effective_T}.

It is now straightforward to find the classical version of the ``entanglement'' entropy, i.e. the Shannon entropy of the classical probability distribution $p\left( {{x_1};T} \right)$,
\begin{equation}
S_A^{{\rm{cl}}} = S_{{A^C}}^{{\rm{cl}}} = - \int {p\left( {{x_1};T} \right)\ln p\left( {{x_1};T} \right)d{x_1}}  = \frac{1}{2}\left( {1 - \ln \frac{{{{\left( {\omega _{{\rm{eff}}}^\infty } \right)}^2}}}{{2\pi T}}} \right)
\end{equation}
and the thermal entropy
\begin{equation}
S_{{\rm A} \cup {A^C}}^{{\rm{cl}}} = - \int {p\left( {{x_1},{x_2};T} \right)\ln p\left( {{x_1},{x_2};T} \right)d{x_1}d{x_2}}  = 1 - \ln \frac{{{\omega _ + }{\omega _ - }}}{{2\pi T}} .
\end{equation}
It follows that the classical mutual information is equal to
\begin{equation}
{I^{{\rm{cl}}}}\left( {A:{A^C}} \right) = \ln \frac{{{\omega _ + }{\omega _ - }}}{{{{\left( {\omega _{{\rm{eff}}}^\infty } \right)}^2}}} = \ln \frac{{\omega _ + ^2 + \omega _ - ^2}}{{2{\omega _ + }{\omega _ - }}} = {I^\infty } .
\end{equation}
This does not depend on the temperature and is equal to the asymptotic value of the quantum mutual information at infinite temperature \eqref{eq:2_mutual_infinity}. It follows that the quantum mutual information at infinite temperature should be attributed to classical correlations. One can trivially show that in a similar manner the classical mutual information coincides with the infinite temperature limit of the quantum mutual information in the case of an arbitrary number of coupled harmonic oscillators \cite{Cramer:2005mx}.

\setcounter{equation}{0}
\section{Entanglement Negativity in Systems of Coupled Oscillators}
\label{sec:negativity}

In section \ref{sec:qm_2}, we showed that there is a finite remnant of mutual information at infinite temperature, unlike the usual behaviour in qubit systems. This remnant can be attributed to classical correlations, as we showed in appendix \ref{sec:classical_mutual}. A verification check is the specification of entanglement negativity. This is defined as the opposite of the sum of the negative eigenvalues of the partially transposed density matrix, ${{\rho ^{{T_A}}}}$, i.e. if $\lambda_i$ are the eigenvalues of ${{\rho ^{{T_A}}}}$, then the negativity $\mathcal{N}$ will be equal to
\begin{equation}
\mathcal{N} = \sum\limits_i {\frac{1}{2}\left( {\left| {{\lambda _i}} \right| - {\lambda _i}} \right)} .
\end{equation}
The entanglement negativity is a measure of quantum entanglement\footnote{Strictly speaking, a measure of quantum entanglement should reduce to the entanglement entropy in the case of pure states of the composite system, which is not the case for entanglement negativity.}. Although a non-vanishing negativity implies the presence of quantum entanglement, the opposite does not hold, when the subsystems have sufficiently high-dimensional Hilbert spaces \cite{Horodecki:1996nc}. Obviously, this is the case for harmonic oscillators, since the corresponding Hilbert spaces are infinite dimensional. Thus, finding vanishing negativity at infinite temperature is not a proof of the classical origin of the mutual information, but it is consistent with such an interpretation.

In qubit systems, typically negativity vanishes at a given finite temperature and it remains vanishing at temperatures higher than that. We will show that this also holds in harmonic oscillatory systems. The techniques of section \ref{sec:qm_many} can be easily generalized for the calculation of entanglement negativity.

The density matrix of a system of $N$ oscillators in a thermal state reads (see equation \eqref{eq:many_thermal_rho}),
\begin{equation}
\rho=\left(\frac{\det\left(a+b\right)}{\pi^N}\right)^{1/2}\exp\left\{-\frac{1}{2}{\bf{x}}^T a {\bf{x}}-\frac{1}{2}{\bf{x}}^{\prime T}a {\bf{x}}^\prime-{\bf{x}}^T b {\bf{x}}^\prime\right\},
\end{equation}
where 
\begin{equation}
a=\begin{pmatrix}
a_A & a_B\\
a_B^T & a_C
\end{pmatrix} \qquad b=\begin{pmatrix}
b_A & b_B\\
b_B^T & b_C
\end{pmatrix}
\end{equation}
We calculate the entanglement negativity between the first $n$ (system $A$) and the last $N-n$ (system $A^C$) oscillators. As usually, we decompose ${\bf{x}}$ as
\begin{equation}
{\bf{x}}=\begin{pmatrix}
x \\ x^C
\end{pmatrix}
\end{equation}
Taking the partial transpose ${{\rho ^{{T_A}}}}$ is equivalent to the interchange of $x^C$ and $x^{C \prime}$, which is also equivalent to the interchange of $x$ and $x^ \prime$. It is easy to show that after this action the density matrix assumes the form
\begin{equation}
{{\rho ^{{T_A}}}}=\left(\frac{\det\left(\gamma-\beta\right)}{\pi^N}\right)^{1/2}\exp\left\{-\frac{1}{2}{\bf{x}}^T\gamma {\bf{x}}-\frac{1}{2}{\bf{x}}^{\prime T}\gamma {\bf{x}}^\prime+{\bf{x}}^T\beta  {\bf{x}}^\prime\right\},
\end{equation}
where
\begin{equation}
\gamma=\begin{pmatrix}
a_A & b_B\\
b_B^T & a_C
\end{pmatrix} , \quad 
\beta=-\begin{pmatrix}
b_A & a_B\\
a_B^T & b_C
\end{pmatrix} 
\label{eq:negativity_gamma_beta}
\end{equation}
The spectrum of the density matrix is given by 
\begin{equation}
{p_{{n_1}, \ldots ,{n_N}}} = \prod\limits_{i = 1}^N {\left( {1 - {\xi _i}} \right)\xi _i^{{n_i}}} ,\quad {n_i} \in \mathbb{Z} ,
\end{equation}
where the quantities $\xi_i$ are related to the eigenvalues $\lambda_i$ of the matrix $\gamma^{-1}\beta$ as
\begin{equation}
\xi_i = \frac{\lambda_i}{1+\sqrt{1 - \lambda_i^2}}.
\end{equation}

Let us consider first the case of two coupled harmonic oscillators. In this case the elements of the matrices $\gamma$ and $\beta$ in the expressions \eqref{eq:negativity_gamma_beta} are not blocks but single elements. These matrices equal
\begin{equation}
\gamma  = \frac{1}{2}\left( {\begin{array}{*{20}{c}}
{{a_ + } + {a_ - }}&{{b_ + } - {b_ - }}\\
{{b_ + } - {b_ - }}&{{a_ + } + {a_ - }}
\end{array}} \right),\quad \beta  =  - \frac{1}{2}\left( {\begin{array}{*{20}{c}}
{{b_ + } + {b_ - }}&{{a_ + } - {a_ - }}\\
{{a_ + } - {a_ - }}&{{b_ + } + {b_ - }}
\end{array}} \right) .
\end{equation}
The eigenvalues of the matrix $\gamma^{-1} \beta$ are
\begin{equation}
{\lambda _1} = \frac{{{\omega _ - } - {\omega _ + }\tanh \frac{{{\omega _ + }}}{{2T}}\tanh \frac{{{\omega _ - }}}{{2T}}}}{{{\omega _ - } + {\omega _ + }\tanh \frac{{{\omega _ + }}}{{2T}}\tanh \frac{{{\omega _ - }}}{{2T}}}},\quad {\lambda _2} = \frac{{{\omega _ + } - {\omega _ - }\tanh \frac{{{\omega _ + }}}{{2T}}\tanh \frac{{{\omega _ - }}}{{2T}}}}{{{\omega _ + } + {\omega _ - }\tanh \frac{{{\omega _ + }}}{{2T}}\tanh \frac{{{\omega _ - }}}{{2T}}}} .
\end{equation}
Clearly, one of those, namely $\lambda_2$, is negative at zero temperature, since
\begin{equation}
\mathop {\lim }\limits_{T \to 0} {\lambda _1} =  - \mathop {\lim }\limits_{T \to 0} {\lambda _2} = \frac{{{\omega _ - } - {\omega _ + }}}{{{\omega _ - } + {\omega _ + }}},
\end{equation}
whereas they are both positive at infinite temperature since
\begin{equation}
\mathop {\lim }\limits_{T \to \infty } {\lambda _1} = \mathop {\lim }\limits_{T \to \infty } {\lambda _2} = 1 .
\end{equation}
Both eigenvalues are monotonous functions of the temperature, therefore there is a specific finite critical temperature ${T_{{\rm{neg}}}}$, defined as the single solution of the equation
\begin{equation}
{\omega _ + } - {\omega _ - }\tanh \frac{{{\omega _ + }}}{{2{T_{{\rm{neg}}}}}}\tanh \frac{{{\omega _ - }}}{{2{T_{{\rm{neg}}}}}} = 0 ,
\end{equation}
where $\lambda_2$ vanishes. At temperatures higher than this critical temperature, the negativity vanishes. Figure \ref{fig:neg_Tcr} shows the dependence of ${T_{{\rm{neg}}}}$ on the ratio ${\omega_-}/{\omega_+}$.
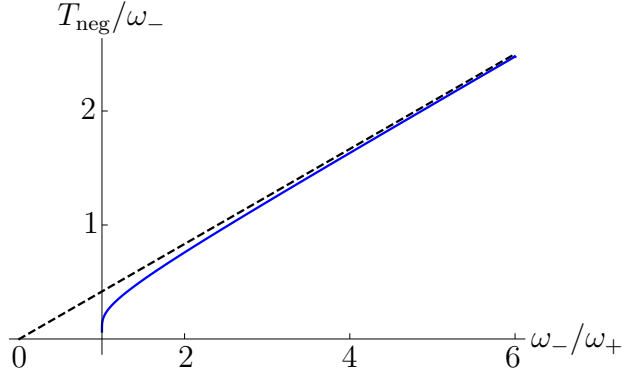
\begin{figure}[ht]
\centering
\begin{picture}(60,35)
\put(3.5,2.5){\includegraphics[width = 0.45\textwidth]{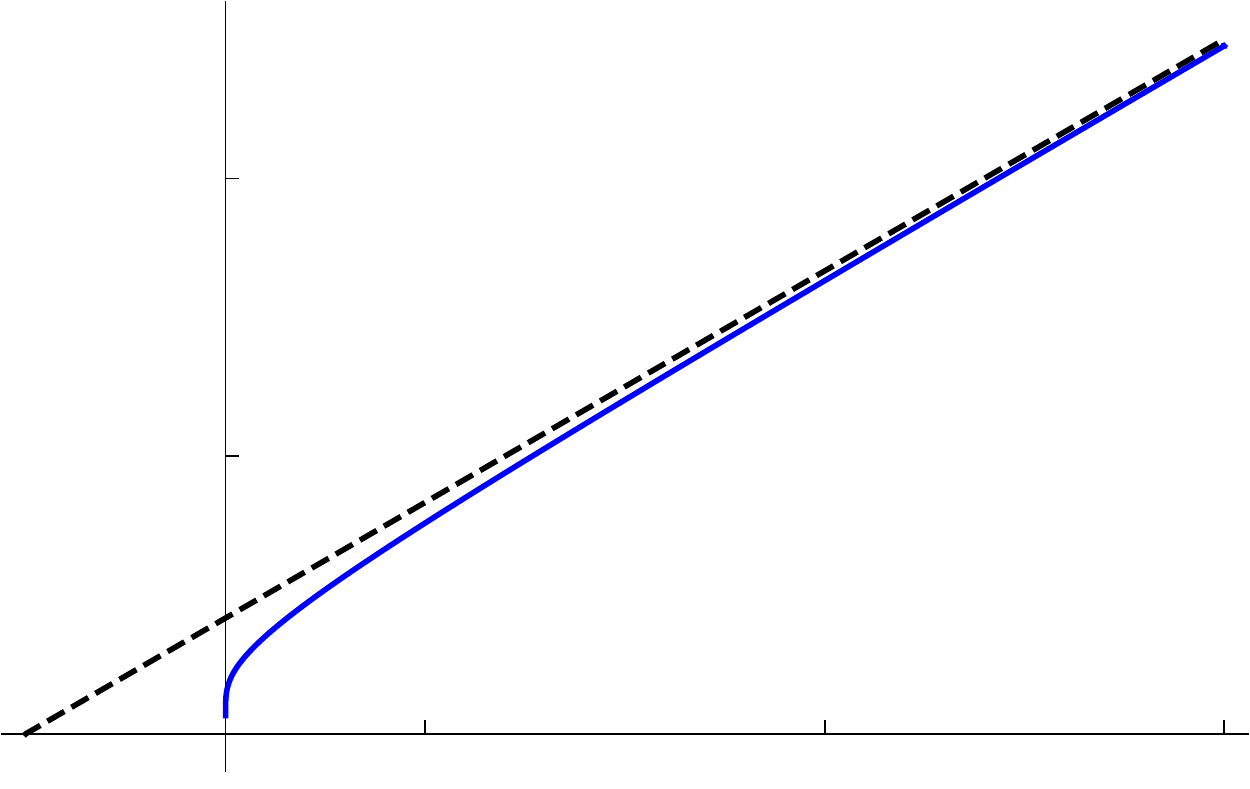}}
\put(7.75,32){${T_{{\rm{neg}}}} / \omega_-$}
\put(49,3.75){$\omega_- / \omega_+$}
\put(9.75,23.75){$2$}
\put(10,13.75){$1$}
\put(3.75,2){$0$}
\put(18.25,2){$2$}
\put(32.5,2){$4$}
\put(46.75,2){$6$}
\end{picture}
\caption{The critical temperature ${T_{{\rm{neg}}}}$, as function of the ratio $\omega_- / \omega_+$}
\label{fig:neg_Tcr}
\end{figure}
Appropriate expansions can be used to show that the critical temperature for large values of the ratio $\omega_- / \omega_+$ is approximately equal to
\begin{equation}
{T_{{\rm{neg}}}} \simeq c \frac{\omega_-}{\omega_+} ,
\end{equation}
where $c$ is the solution of the equation $\tanh \frac{1}{2c} = 2c$, namely $c \simeq 0.41678$. 

It is a matter of simple algebra to show that below the critical temperature ${T_{{\rm{neg}}}}$, the entanglement negativity equals
\begin{equation}
\mathcal{N} =  - \frac{{{\lambda _2}}}{{\sqrt {1 + {\lambda _2}} }}\frac{1}{{\sqrt {1 + {\lambda _2}}  + \sqrt {1 - {\lambda _2}} }}.
\end{equation}
Figure \ref{fig:neg_negativity} shows the dependence of the entanglement negativity on the temperature.
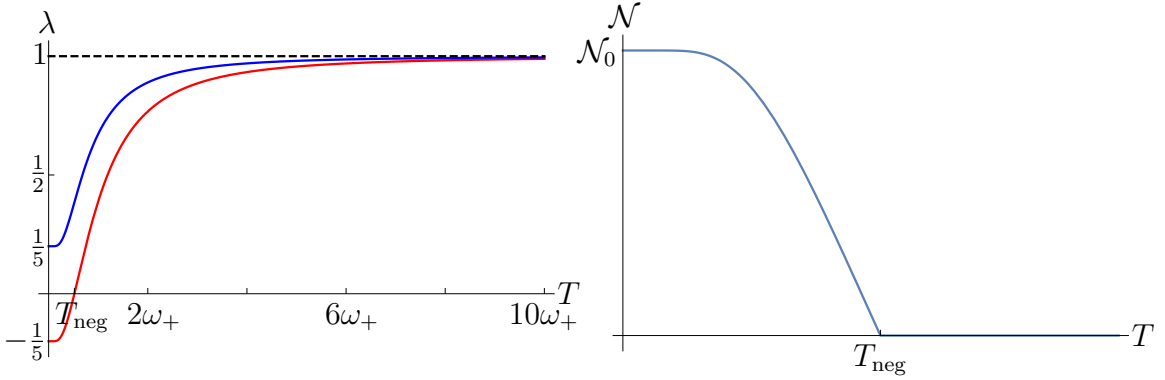
\begin{figure}[ht]
\centering
\begin{picture}(100,34)
\put(3.5,2.5){\includegraphics[width = 0.45\textwidth]{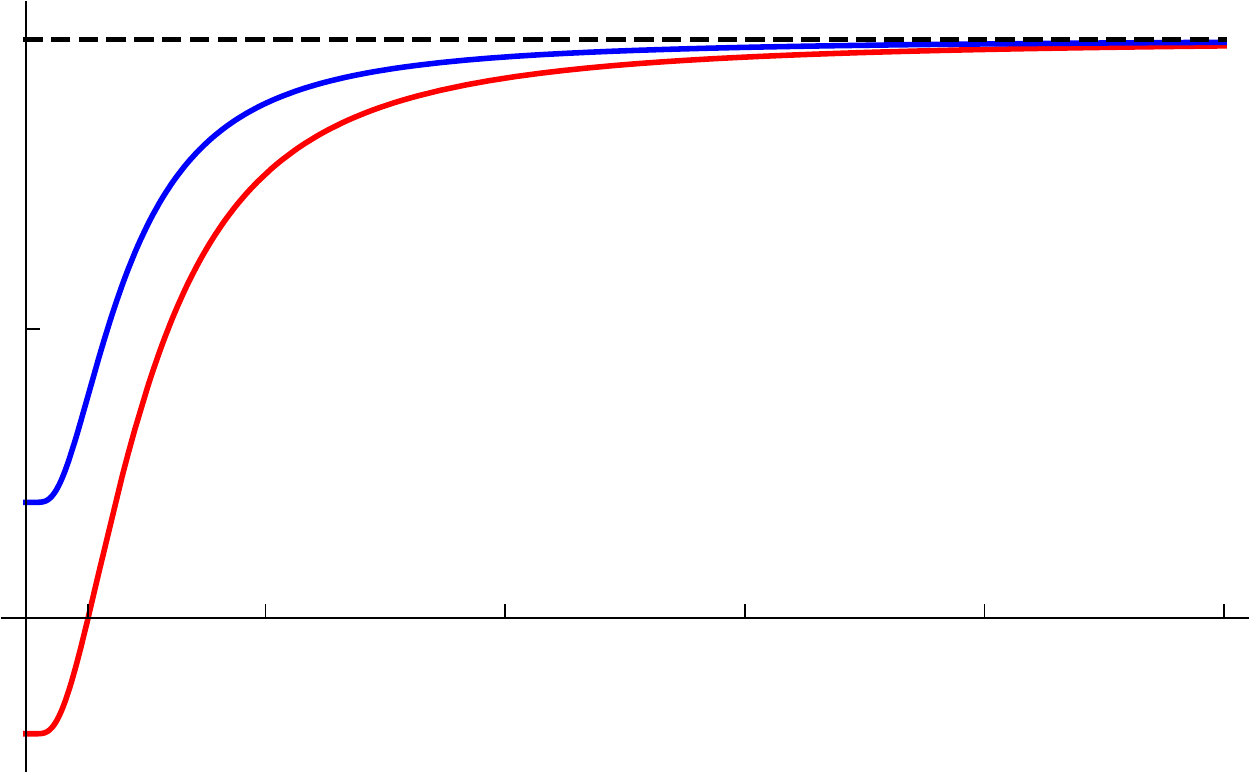}}
\put(53.5,2.5){\includegraphics[width = 0.45\textwidth]{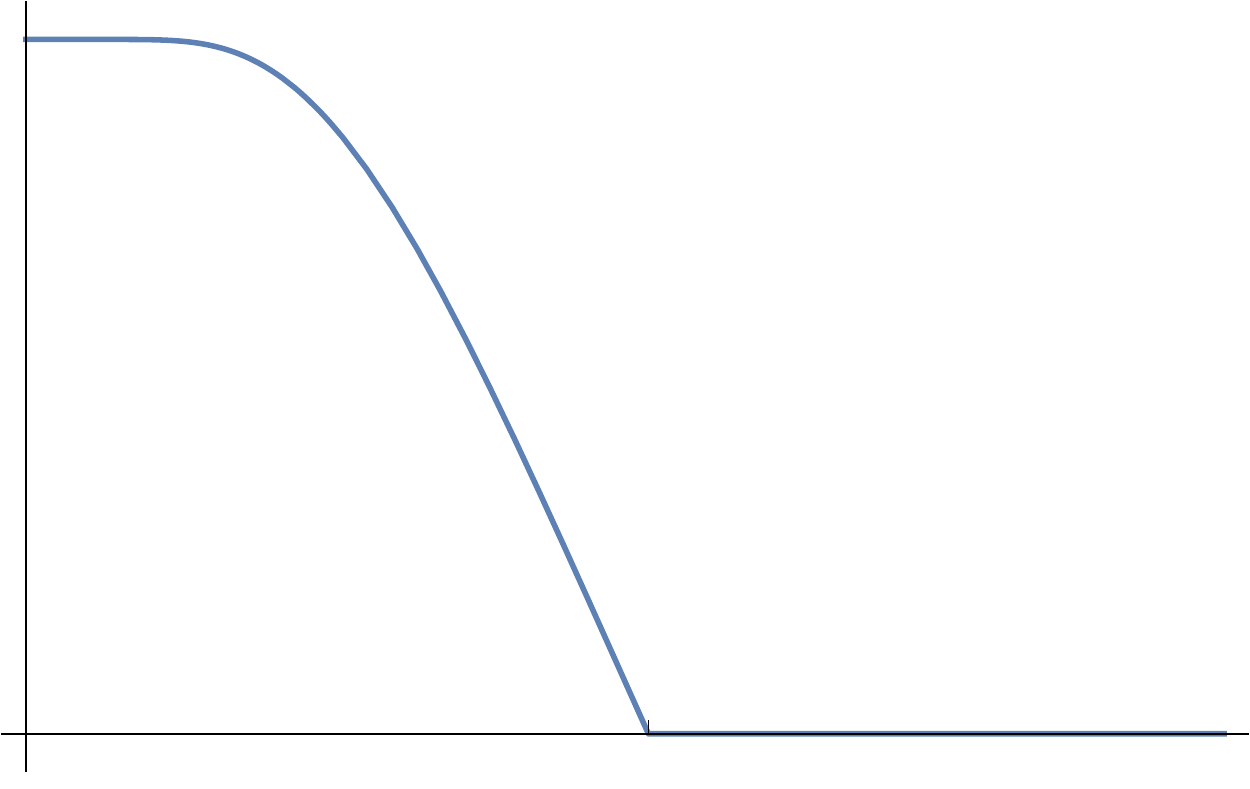}}
\put(3.5,31){$\lambda$}
\put(53.25,31.5){$\mathcal{N}$}
\put(48.75,7.25){$T$}
\put(98.75,3.5){$T$}
\put(3,28.25){$1$}
\put(50.5,28.5){${\mathcal{N}}_0$}
\put(2.75,18){$\frac{1}{2}$}
\put(2.75,11.5){$\frac{1}{5}$}
\put(0.5,3.25){$-\frac{1}{5}$}
\put(5,5.5){${T_{{\rm{neg}}}}$}
\put(11.25,5.5){$2\omega_+$}
\put(28.5,5.5){$6\omega_+$}
\put(44.5,5.5){$10\omega_+$}
\put(74.5,1.75){${T_{{\rm{neg}}}}$}
\end{picture}
\caption{The eigenvalues of the partially transposed density matrix (left) and the entanglement negativity (right), as functions of the temperature. For these plots it is assumed that $\omega_- / \omega_+ = 3/2$, which implies that $\mathop {\lim }\limits_{T \to 0} {\lambda _{1/2}} = \pm \frac{1}{5}$.}
\label{fig:neg_negativity}
\end{figure}

In the case of a system of $N$ coupled oscillators, the eigenvalues $\lambda_i$ are determined by the equation
\begin{equation}
\det\begin{pmatrix}
b_A+\lambda a_A & a_B+\lambda b_B\\
a_B^T+\lambda b_B^T & b_C+\lambda a_C
\end{pmatrix}=0
\end{equation} 
or equivalently by
\begin{equation}
\det\begin{pmatrix}
(1+\lambda)(a+b)_A-(1-\lambda)(a-b)_A & (1+\lambda)(a+b)_B+(1-\lambda)(a-b)_B\\
(1+\lambda)(a+b)_B^T+(1-\lambda)(a-b)_B^T & (1+\lambda)(a+b)_C-(1-\lambda)(a-b)_C
\end{pmatrix}=0.
\end{equation}
The eigenvalues $\lambda_i$ can be re-expressed as
\begin{equation}
\frac{1 + \lambda_i}{1 - \lambda_i}=\Lambda_i,
\end{equation}
where $\Lambda_i$ are the eigenvalues of the matrix
\begin{equation}
\begin{pmatrix}
(a+b)_A & (a+b)_B\\
(a+b)_B^T & (a+b)_C
\end{pmatrix}^{-1}\begin{pmatrix}
(a-b)_A & -(a-b)_B\\
-(a-b)_B^T & (a-b)_C
\end{pmatrix} .
\end{equation}
Since the matrix $a+b$ tends to the zero matrix at infinite temperature, it follows that all eigenvalues $\Lambda_i$ tend to infinity, or equivalently all eigenvalues $\lambda_i$ tend to one. This implies that the negativity vanishes at infinite temperature. Actually, since all $\lambda_i$'s tend to one, it follows that they all become positive at a finite critical temperature, similarly to the two oscillators case. 

One the contrary at zero temperature, the $b$ matrix vanishes and the $a$ matrix tends to the matrix $\Omega = \sqrt{K}$. Therefore, the eigenvalues $\lambda_i$ are determined by the equation
\begin{equation}
\det\begin{pmatrix}
\lambda I_n & \Omega_A^{-1}\Omega_B\\
\Omega_C^{-1}\Omega_B^T & \lambda I_{N-n}
\end{pmatrix}=0
\end{equation} 
or equivalently by
\begin{equation}
\det\left(\lambda^2I_{N-n}- \Omega_C^{-1} \Omega_B^T \Omega_A^{-1} \Omega_B \right)=0 .
\end{equation}
These eigenvalues come in $\min(n,N-n)$ pairs in view of Sylvester's determinant identity. There are always negative eigenvalues, therefore the system exhibits quantum entanglement. This is obviously expected, since the system lies in a pure state and has non-vanishing entanglement entropy.

\setcounter{equation}{0}
\section{The High Temperature Expansion for Coupled Oscillators}
\label{sec:highT}

In this appendix, we obtain the high temperature expansion for the entanglement entropy and the mutual information for systems of coupled harmonic oscillators. For this purpose, we first need to expand the matrices $a$, $b$ and $a+b$, which are defined in equation \eqref{eq:N_ab_defs}, for high temperatures. It is simple to show that
\begin{align}
a &= T\left( {I + \frac{1}{{3{T^2}}}K - \frac{1}{{45{T^4}}}{K^2} + \frac{2}{{945{T^6}}}{K^3}} + \mathcal{O} \left( \frac{1}{T^8} \right) \right) , \label{eq:highT_a_expansion} \\
b &=  - T\left( {I - \frac{1}{{6{T^2}}}K + \frac{7}{{360{T^4}}}{K^2} - \frac{{31}}{{15120{T^6}}}{K^3}} + \mathcal{O} \left( \frac{1}{T^8} \right) \right) , \label{eq:highT_b_expansion} \\
a + b &= \frac{1}{{2T}}K\left( {I - \frac{1}{{12{T^2}}}K + \frac{1}{{120{T^4}}}{K^2}} + \mathcal{O} \left( \frac{1}{T^6} \right) \right) . \label{eq:highT_ab_expansion}
\end{align}
In the following, we will need the $A$, $B$ and $C$ blocks of the matrices $K^2$ and $K^3$, in order to substitute them into formulae \eqref{eq:highT_a_expansion}, \eqref{eq:highT_b_expansion} and \eqref{eq:highT_ab_expansion}. These are given in terms of the corresponding blocks of the matrix $K$ by
\begin{align}
{\left( {{K^2}} \right)_A} &= K_A^2 + {K_B}K_B^T , \\
{\left( {{K^2}} \right)_B} &= {K_A}{K_B} + {K_B}{K_C} , \\
\left( {{K^2}} \right)_B^T &= K_B^T{K_A} + {K_C}K_B^T , \\
{\left( {{K^2}} \right)_C} &= K_B^T{K_B} + K_C^2 \label{eq:highT_KC2}
\end{align}
and
\begin{align}
{\left( {{K^3}} \right)_A} &= K_A^3 + {K_B}K_B^T{K_A} + {K_A}{K_B}K_B^T + {K_B}{K_C}K_B^T , \\
{\left( {{K^3}} \right)_B} &= K_A^2{K_B} + {K_B}K_B^T{K_B} + {K_A}{K_B}{K_C} + {K_B}K_C^2 , \\
\left( {{K^3}} \right)_B^T &= K_B^TK_A^2 + {K_C}K_B^T{K_A} + K_B^T{K_B}K_B^T + K_C^2K_B^T , \\
{\left( {{K^3}} \right)_C} &= K_B^T{K_A}{K_B} + {K_C}K_B^T{K_B} + K_B^T{K_B}{K_C} + K_C^3 .
\end{align}

We need to specify the high temperature expansion of the eigenvalues of the matrix $\gamma^{-1} \beta$. We recall that the matrices $\gamma$ and $\beta$ are defined as $\gamma = a_C - d / 2$ and $\beta = - b_C + d/2$, where $d = \left( {{a_B^T} + b_B^T} \right){\left( {a_A + b_A} \right)^{ - 1}}\left( {a_B + b_B} \right)$. As a direct consequence of the equation \eqref{eq:highT_ab_expansion}, we have
\begin{equation}
{\left( {a + b} \right)_A} = \frac{1}{{2T}}\left( {{K_A} - \frac{1}{{12{T^2}}}{{\left( {{K^2}} \right)}_A} + \frac{1}{{120{T^4}}}{{\left( {{K^3}} \right)}_A}} + \mathcal{O} \left( \frac{1}{T^6} \right) \right)
\end{equation}
and
\begin{multline}
{\left( {{{\left( {a + b} \right)}_A}} \right)^{ - 1}} = 2T\left[ {{{\left( {{K_A}} \right)}^{ - 1}} + \frac{1}{{12{T^2}}}{{\left( {{K_A}} \right)}^{ - 1}}{{\left( {{K^2}} \right)}_A}{{\left( {{K_A}} \right)}^{ - 1}}} \right. \\
 + \frac{1}{{24{T^4}}}\left( {\frac{1}{6}{{\left( {{K_A}} \right)}^{ - 1}}{{\left( {{K^2}} \right)}_A}{{\left( {{K_A}} \right)}^{ - 1}}{{\left( {{K^2}} \right)}_A}{{\left( {{K_A}} \right)}^{ - 1}}} \right.  \\
\left. {\left. { - \frac{1}{5}{{\left( {{K_A}} \right)}^{ - 1}}{{\left( {{K^3}} \right)}_A}{{\left( {{K_A}} \right)}^{ - 1}}} \right)} + \mathcal{O} \left( \frac{1}{T^6} \right) \right] .
\end{multline}
Then, defining ${{\tilde K}_C} \equiv {K_C} - K_B^T{\left( {{K_A}} \right)^{ - 1}}{K_B}$ and using the notation
\begin{equation}
d  = T\left( {0 + \frac{1}{{{T^2}}}{d ^{\left( 1 \right)}} + \frac{1}{{{T^4}}}{d ^{\left( 2 \right)}} + \frac{1}{{{T^6}}}{d ^{\left( 3 \right)}} + \mathcal{O}\left( {\frac{1}{{{T^8}}}} \right)} \right) ,
\end{equation}
we find
\begin{align}
{d ^{\left( 1 \right)}} &= \frac{1}{2}\left( {{K_C} - {{\tilde K}_C}} \right) , \\
{d ^{\left( 2 \right)}} &= \frac{1}{{24}}\left[ {{{\left( {{{\tilde K}_C}} \right)}^2} - {{\left( {{K^2}} \right)}_C}} \right] , \\
{d ^{\left( 3 \right)}} &= \frac{1}{{240}}\left[ {{{\left( {{K^3}} \right)}_C} - \frac{5}{6}{{\tilde K}_C}\left( {\frac{1}{5}{K_C} + {{\tilde K}_C}} \right){{\tilde K}_C}} \right] .
\end{align}
Adopting a similar notation for the high temperature expansions of the matrices $\beta$ and $\gamma$,
their definitions \eqref{eq:N_gamma_def} and \eqref{eq:N_beta_def} yield

\begin{align}
{\beta ^{\left( 1 \right)}} &= \frac{1}{{12}}{K_C} - \frac{1}{4}{{\tilde K}_C} ,\\
{\beta ^{\left( 2 \right)}} &= - \frac{1}{{720}}{\left( {{K^2}} \right)_C} + \frac{1}{{48}}{\left( {{{\tilde K}_C}} \right)^2} ,\\
{\beta ^{\left( 3 \right)}} &= \frac{1}{{30240}}{\left( {{K^3}} \right)_C} - \frac{1}{{576}}{{\tilde K}_C}\left( {\frac{1}{5}{K_C} + {{\tilde K}_C}} \right){{\tilde K}_C}
\end{align}
and
\begin{align}
{\gamma ^{\left( 1 \right)}} &= \frac{1}{{12}}{K_C} + \frac{1}{4}{{\tilde K}_C} , \\
{\gamma ^{\left( 2 \right)}} &=  - \frac{1}{{720}}{\left( {{K^2}} \right)_C} - \frac{1}{{48}}{\left( {{{\tilde K}_C}} \right)^2} , \\
{\gamma ^{\left( 3 \right)}} &= \frac{1}{{30240}}{\left( {{K^3}} \right)_C} + \frac{1}{{576}}{{\tilde K}_C}\left( {\frac{1}{5}{K_C} + {{\tilde K}_C}} \right){{\tilde K}_C} .
\end{align}
The calculation of the high temperature expansion of the matrix $\gamma^{-1} \beta$,
\begin{equation}
{\gamma ^{ - 1}}\beta  = I + \frac{1}{{{T^2}}}{\left( {{\gamma ^{ - 1}}\beta } \right)^{\left( 1 \right)}} + \frac{1}{{{T^4}}}{\left( {{\gamma ^{ - 1}}\beta } \right)^{\left( 2 \right)}} + \frac{1}{{{T^6}}}{\left( {{\gamma ^{ - 1}}\beta } \right)^{\left( 3 \right)}} + \mathcal{O}\left( {\frac{1}{{{T^8}}}} \right),
\end{equation}
is facilitated by the use of the iterative formulae
\begin{align}
{\left( {{\gamma ^{ - 1}}\beta } \right)^{\left( 1 \right)}} &= {\beta ^{\left( 1 \right)}} - {\gamma ^{\left( 1 \right)}} , \\
{\left( {{\gamma ^{ - 1}}\beta } \right)^{\left( 2 \right)}} &= {\beta ^{\left( 2 \right)}} - {\gamma ^{\left( 2 \right)}} - {\gamma ^{\left( 1 \right)}}{\left( {{\gamma ^{ - 1}}\beta } \right)^{\left( 1 \right)}} , \\
{\left( {{\gamma ^{ - 1}}\beta } \right)^{\left( 3 \right)}} &= {\beta ^{\left( 3 \right)}} - {\gamma ^{\left( 3 \right)}} - {\gamma ^{\left( 1 \right)}}{\left( {{\gamma ^{ - 1}}\beta } \right)^{\left( 2 \right)}} - {\gamma ^{\left( 2 \right)}}{\left( {{\gamma ^{ - 1}}\beta } \right)^{\left( 1 \right)}} ,
\end{align}
which yield
\begin{align}
{\left( {{\gamma ^{ - 1}}\beta } \right)^{\left( 1 \right)}} &=  - \frac{1}{2}{{\tilde K}_C} , \label{eq:highT_gb1}\\
{\left( {{\gamma ^{ - 1}}\beta } \right)^{\left( 2 \right)}} &= \frac{1}{6}\left( {\frac{1}{4}{K_C} + {{\tilde K}_C}} \right){{\tilde K}_C} , \\
{\left( {{\gamma ^{ - 1}}\beta } \right)^{\left( 3 \right)}} &=  - \frac{1}{{18}}\left[ {\left( {\frac{1}{4}{K_C} + {{\tilde K}_C}} \right)\left( {\frac{1}{5}{K_C} + {{\tilde K}_C}} \right) + \frac{1}{{80}}\left( {{{\left( {{K^2}} \right)}_C} + K_C^2} \right)} \right]{{\tilde K}_C} .
\end{align}

The specification of the high temperature expansion of the eigenvalues of the matrix $\gamma^{-1} \beta$ is now a straightforward perturbation theory problem. The zeroth order result is obviously $1$ and the eigenvectors are arbitrary. Let $\left| {{v_i}} \right\rangle $ be the eigenvectors of the matrix ${{\tilde K}_C}$, i.e.
\begin{equation}
{{\tilde K}_C}\left| {{v_i}} \right\rangle  = {\lambda _i}\left| {{v_i}} \right\rangle .
\end{equation}
We expand the eigenvalues of the matrix $\gamma^{-1} \beta$ as
\begin{equation}
{\beta _{Di}} = 1 - \frac{{\beta _{Di}^{\left( 1 \right)}}}{{{T^2}}} - \frac{{\beta _{Di}^{\left( 2 \right)}}}{{{T^4}}} - \frac{{\beta _{Di}^{\left( 3 \right)}}}{{{T^6}}} + \mathcal{O} \left( {\frac{1}{{{T^8}}}} \right) .
\label{eq:highT_eigenvalue_expansion}
\end{equation}
As a direct consequence of the equation \eqref{eq:highT_gb1}, we have
\begin{equation}
\beta _D^{\left( 1 \right)} = \frac{{{\lambda _i}}}{2} .
\end{equation}
The specification of the next corrections to the eigenvalues is a problem identical to the usual perturbation theory in quantum mechanics. The role of the unperturbed Hamiltonian is played by $ - {\left( {{\gamma ^{ - 1}}\beta } \right)^{\left( 1 \right)}}$ and there are two perturbations, one which is of first order in the expansive parameter $1 / T^2$, namely $ - {\left( {{\gamma ^{ - 1}}\beta } \right)^{\left( 2 \right)}}$, and a second order one, namely $ - {\left( {{\gamma ^{ - 1}}\beta } \right)^{\left( 3 \right)}}$. Therefore,
\begin{equation}
\beta _D^{\left( 2 \right)} =  - \frac{1}{6}\left\langle {{v_i}} \right|\left( {\frac{1}{4}{K_C} + {{\tilde K}_C}} \right){{\tilde K}_C}\left| {{v_i}} \right\rangle  =  - \frac{{\lambda _i^2}}{6} - \frac{{{\lambda _i}\left\langle {{v_i}} \right|{K_C}\left| {{v_i}} \right\rangle }}{{24}} ,
\end{equation}
while $\beta _D^{\left( 3 \right)}$ gets contributions from both perturbations
\begin{multline}
\beta _D^{\left( 3 \right)} = \frac{1}{{18}}\left\langle {{v_i}} \right|\left[ {\left( {\frac{1}{4}{K_C} + {{\tilde K}_C}} \right)\left( {\frac{1}{5}{K_C} + {{\tilde K}_C}} \right) + \frac{1}{{80}}\left( {{{\left( {{K^2}} \right)}_C} + K_C^2} \right)} \right]{{\tilde K}_C}\left| {{v_i}} \right\rangle \\
 + \frac{1}{{18}}\sum\limits_{j \ne i} {\frac{{\left\langle {{v_i}} \right|\left( {\frac{1}{4}{K_C} + {{\tilde K}_C}} \right){{\tilde K}_C}\left| {{v_j}} \right\rangle \left\langle {{v_j}} \right|\left( {\frac{1}{4}{K_C} + {{\tilde K}_C}} \right){{\tilde K}_C}\left| {{v_i}} \right\rangle }}{{{\lambda _i} - {\lambda _j}}}} \\
 = \frac{1}{{18}}\left( {\lambda _i^3 + \frac{9}{{20}}\lambda _i^2\left\langle {{v_i}} \right|{K_C}\left| {{v_i}} \right\rangle  + \frac{1}{{16}}{\lambda _i}\left\langle {{v_i}} \right|K_C^2\left| {{v_i}} \right\rangle  + \frac{1}{{80}}{\lambda _i}\left\langle {{v_i}} \right|{{\left( {{K^2}} \right)}_C}\left| {{v_i}} \right\rangle } \right)\\
 + \frac{1}{{288}}\sum\limits_{j \ne i} {\frac{{{\lambda _i}{\lambda _j}\left\langle {{v_i}} \right|{K_C}\left| {{v_j}} \right\rangle \left\langle {{v_j}} \right|{K_C}\left| {{v_i}} \right\rangle }}{{{\lambda _i} - {\lambda _j}}}} .
\end{multline}

Given the expansion \eqref{eq:highT_eigenvalue_expansion}, the corresponding quantities $\xi_i$ and the contribution of each eigenvalue to the entanglement entropy are
\begin{multline}
\xi_i  = 1 - \sqrt {2\beta _{Di}^{\left( 1 \right)}} \frac{1}{T} + \beta _{Di}^{\left( 1 \right)}\frac{1}{{{T^2}}} - \frac{{3{{\left( {\beta _{Di}^{\left( 1 \right)}} \right)}^2} + 2\beta _{Di}^{\left( 2 \right)}}}{{2\sqrt {2\beta _{Di}^{\left( 1 \right)}} }}\frac{1}{{{T^3}}} + \left( {{{\left( {\beta _{Di}^{\left( 1 \right)}} \right)}^2} + \beta _{Di}^{\left( 2 \right)}} \right)\frac{1}{{{T^4}}}\\
 - \frac{{23{{\left( {\beta _{Di}^{\left( 1 \right)}} \right)}^4} + 36{{\left( {\beta _{Di}^{\left( 1 \right)}} \right)}^2}\beta _{Di}^{\left( 2 \right)} - 4{{\left( {\beta _{Di}^{\left( 2 \right)}} \right)}^2} + 16\beta _{Di}^{\left( 1 \right)}\beta _{Di}^{\left( 3 \right)}}}{{8{{\left( {\sqrt {2\beta _{Di}^{\left( 1 \right)}} } \right)}^3}}}\frac{1}{{{T^5}}}\\
 + \left( {{{\left( {\beta _{Di}^{\left( 1 \right)}} \right)}^3} + 2\beta _{Di}^{\left( 1 \right)}\beta _{Di}^{\left( 2 \right)} + \beta _{Di}^{\left( 3 \right)}} \right)\frac{1}{{{T^6}}} + \mathcal{O}\left( {\frac{1}{{{T^7}}}} \right)
\end{multline}
and
\begin{multline}
S_i = \frac{1}{2}\ln \frac{{{T^2}}}{{2\beta _{Di}^{\left( 1 \right)}}} + 1 - \left( {\frac{{\beta _{Di}^{\left( 1 \right)}}}{3} + \frac{{\beta _{Di}^{\left( 2 \right)}}}{{2\beta _{Di}^{\left( 1 \right)}}}} \right)\frac{1}{{{T^2}}}\\
 - \left( {\frac{{7{{\left( {\beta _{Di}^{\left( 1 \right)}} \right)}^2}}}{{60}} + \frac{{\beta _{Di}^{\left( 2 \right)}}}{3} - \frac{{{{\left( {\beta _{Di}^{\left( 2 \right)}} \right)}^2}}}{{4{{\left( {\beta _{Di}^{\left( 1 \right)}} \right)}^2}}} + \frac{{\beta _{Di}^{\left( 3 \right)}}}{{2\beta _{Di}^{\left( 1 \right)}}}} \right)\frac{1}{{{T^4}}} + \mathcal{O}\left( {\frac{1}{{{T^6}}}} \right) ,
\end{multline}
respectively. Notice that although odd powers of $T$ are absent in the expansion of $\beta_{Di}$, they appear in $\xi_i$ due to the presence of $\sqrt{1 - \beta^2_{Di}}$ in the definition of $\xi_i$.

We expand the entanglement entropy as
\begin{equation}
{S_A} = \left( {N - n} \right)\ln T + S_A^{\left( 0 \right)} + \frac{{S_A^{\left( 1 \right)}}}{{{T^2}}} + \frac{{S_A^{\left( 2 \right)}}}{{{T^4}}} + \mathcal{O} \left( {\frac{1}{{{T^6}}}} \right) .
\end{equation}
We recall the definition of the mutual information $I\left( {A:{A^C}} \right) = {S_A} + {S_{A^C}} - S_{\textrm{th}}$. The formula \eqref{eq:N_thermal_highT} implies that in the case of $N$ coupled oscillators the thermal entropy has a high temperature expansion of the form
\begin{equation}
{S_{\textrm{th}}} = \frac{1}{2}\ln \frac{{{T^2}}}{{\det K}} + N + \frac{{\tr K}}{{24}}\frac{1}{{{T^2}}} - \frac{{\tr{K^2}}}{{960}}\frac{1}{{{T^4}}} + \mathcal{O}\left( {\frac{1}{{{T^6}}}} \right) .
\end{equation}
It follows that the logarithmic terms cancel and the mutual information has a high temperature expansion of the form
\begin{equation}
I\left( {A:{A^C}} \right) = {I^{\left( 0 \right)}} + \frac{{{I^{\left( 1 \right)}}}}{{{T^2}}} + \frac{{{I^{\left( 2 \right)}}}}{{{T^4}}} + \mathcal{O}\left( {\frac{1}{{{T^6}}}} \right) .
\end{equation}

At zeroth order we find
\begin{equation}
S_A^{\left( 0 \right)} = \sum\limits_i {\frac{1}{2}\left( {\ln \frac{1}{{2\beta _D^{\left( 1 \right)}}} + 1} \right)}  =  - \frac{1}{2}\ln \prod\limits_i {{\lambda _i}}  + N - n =  - \frac{1}{2}\ln \det {{\tilde K}_C} + N - n .
\end{equation}
In an obvious manner, $S_{{A^C}}^{\left( 0 \right)} =  - \frac{1}{2}\ln \det {{\tilde K}_A} + n$, where ${{\tilde K}_A} = {K_A} - {K_B}{\left( {{K_C}} \right)^{ - 1}}K_B^T$. Then the zeroth order contribution to the mutual information is
\begin{equation}
\begin{split}
{I^{\left( 0 \right)}} =  - \frac{1}{2}\ln \frac{{\det {{\tilde K}_A}\det {{\tilde K}_C}}}{{\det K}} &=  - \frac{1}{2}\ln \det \left( {I - {{\left( {{K_C}} \right)}^{ - 1}}K_B^T{{\left( {{K_A}} \right)}^{ - 1}}{K_B}} \right)\\
 &=  - \frac{1}{2}\ln \det \left( {I - {{\left( {{K_A}} \right)}^{ - 1}}{K_B}{{\left( {{K_C}} \right)}^{ - 1}}K_B^T} \right) ,
\end{split}
\label{eq:highT_I0}
\end{equation}
since $\det K = \det {K_A}\det {{\tilde K}_C} = \det {{\tilde K}_A}\det {K_C}$. The two last forms for ${I^{\left( 0 \right)}}$, although they are expressed as determinants of matrices of different dimensions, they are equal and they are connected through the Sylvester's determinant formula.

Similarly,
\begin{equation}
S_A^{\left( 1 \right)} =  - \sum\limits_i {\left( {\frac{{\beta _D^{\left( 1 \right)}}}{3} + \frac{{\beta _D^{\left( 2 \right)}}}{{2\beta _D^{\left( 1 \right)}}}} \right)}  = \frac{1}{{24}}\sum\limits_i {\left\langle {{v_i}} \right|{K_C}\left| {{v_i}} \right\rangle }  = \frac{1}{{24}} \tr {K_C} .
\end{equation}
Obviously, $S_{{A^C}}^{\left( 1 \right)} = \frac{1}{{24}} \tr {K_A}$ and thus,
\begin{equation}
{I^{\left( 1 \right)}} = \frac{1}{{24}}\left( {\tr {K_A} + \tr {K_C} - \tr K} \right) = 0 .
\end{equation}

Finally,
\begin{multline}
S_A^{\left( 2 \right)} =  - \sum\limits_i {\left( {\frac{{7{{\left( {\beta _D^{\left( 1 \right)}} \right)}^2}}}{{60}} + \frac{{\beta _D^{\left( 2 \right)}}}{3} - \frac{{{{\left( {\beta _D^{\left( 2 \right)}} \right)}^2}}}{{4{{\left( {\beta _D^{\left( 1 \right)}} \right)}^2}}} + \frac{{\beta _D^{\left( 3 \right)}}}{{2\beta _D^{\left( 1 \right)}}}} \right)} \\
 = \sum\limits_i {\Bigg( { - \frac{1}{{720}}\lambda _i^2 + \frac{1}{{360}}{\lambda _i}\left\langle {{v_i}} \right|{K_C}\left| {{v_i}} \right\rangle  + \frac{1}{{576}}{{\left( {\left\langle {{v_i}} \right|{K_C}\left| {{v_i}} \right\rangle } \right)}^2}} } - \frac{1}{{288}}\left\langle {{v_i}} \right|K_C^2\left| {{v_i}} \right\rangle \\
  - \frac{1}{{1440}}\left\langle {{v_i}} \right|{\left( {{K^2}} \right)_C}\left| {{v_i}} \right\rangle 
 { - \frac{1}{{288}}\sum\limits_{j \ne i} {\frac{{{\lambda _j}\left\langle {{v_i}} \right|{K_C}\left| {{v_j}} \right\rangle \left\langle {{v_j}} \right|{K_C}\left| {{v_i}} \right\rangle }}{{{\lambda _i} - {\lambda _j}}}} } \Bigg)
\end{multline}
or
\begin{multline}
S_A^{\left( 2 \right)} =  - \frac{1}{{720}}\tr \tilde K_C^2 + \frac{1}{{360}}\tr \left( {{{\tilde K}_C}{K_C}} \right) + \frac{1}{{576}}\sum\limits_i {{{\left( {\left\langle {{v_i}} \right|{K_C}\left| {{v_i}} \right\rangle } \right)}^2}} \\
 - \frac{1}{{288}}\tr K_C^2 - \frac{1}{{1440}}\tr {\left( {{K^2}} \right)_C} - \frac{1}{{288}}\sum\limits_{i,j,j \ne i} {\frac{{{\lambda _j}\left\langle {{v_i}} \right|{K_C}\left| {{v_j}} \right\rangle \left\langle {{v_j}} \right|{K_C}\left| {{v_i}} \right\rangle }}{{{\lambda _i} - {\lambda _j}}}}  .
\end{multline}
The two terms that are written as a sum, simplify if we write the double sum term as the symmetrized sum,
\begin{equation}
\begin{split}
\sum\limits_i &{{{\left( {\left\langle {{v_i}} \right|{K_C}\left| {{v_i}} \right\rangle } \right)}^2}} - 2 \sum\limits_{i,j,j \ne i} {\frac{{{\lambda _j}\left\langle {{v_i}} \right|{K_C}\left| {{v_j}} \right\rangle \left\langle {{v_j}} \right|{K_C}\left| {{v_i}} \right\rangle }}{{{\lambda _i} - {\lambda _j}}}} \\
 &\quad\quad = \sum\limits_i {{{\left( {\left\langle {{v_i}} \right|{K_C}\left| {{v_i}} \right\rangle } \right)}^2}}  - \sum\limits_{i,j,j \ne i} {\frac{{\left( {{\lambda _j} - {\lambda _i}} \right)\left\langle {{v_i}} \right|{K_C}\left| {{v_j}} \right\rangle \left\langle {{v_j}} \right|{K_C}\left| {{v_i}} \right\rangle }}{{{\lambda _i} - {\lambda _j}}}} \\
 &\quad\quad = \sum\limits_i {{{\left( {\left\langle {{v_i}} \right|{K_C}\left| {{v_i}} \right\rangle } \right)}^2}}  + \sum\limits_{i,j,j \ne i} {\left\langle {{v_i}} \right|{K_C}\left| {{v_j}} \right\rangle \left\langle {{v_j}} \right|{K_C}\left| {{v_i}} \right\rangle } \\
 &\quad\quad = \sum\limits_{i,j} {\left\langle {{v_i}} \right|{K_C}\left| {{v_j}} \right\rangle \left\langle {{v_j}} \right|{K_C}\left| {{v_i}} \right\rangle } = \sum\limits_i {\left\langle {{v_i}} \right|K_C^2\left| {{v_i}} \right\rangle }  = \tr K_C^2 .
\end{split}
\end{equation}
The latter implies
\begin{equation}
{S^{\left( 2 \right)}} =  - \frac{1}{{720}}\tr \tilde K_C^2 + \frac{1}{{360}}\tr \left( {{{\tilde K}_C}{K_C}} \right) - \frac{1}{{576}}\tr K_C^2 - \frac{1}{{1440}}\tr {\left( {{K^2}} \right)_C} .
\end{equation}
Using the definition of ${{{\tilde K}_C}}$ and expressing $K_C^2$ in terms of ${\left( {{K^2}} \right)_C}$, using formula \eqref{eq:highT_KC2}, yields
\begin{equation}
{S^{\left( 2 \right)}} =  - \frac{1}{{960}}\tr {\left( {{K^2}} \right)_C} - \frac{1}{{720}}\tr \left[ {{{\left( {K_B^T{{\left( {{K_A}} \right)}^{ - 1}}{K_B}} \right)}^2}} \right] + \frac{1}{{2880}}\tr \left( {K_B^T{K_B}} \right) .
\end{equation}
Finally, the above equation implies that
\begin{multline}
{I^{\left( 2 \right)}} =  - \frac{1}{{960}}\left[ {\tr {{\left( {{K^2}} \right)}_C} + \tr {{\left( {{K^2}} \right)}_A} - \tr \left( {{K^2}} \right)} \right]  + \frac{1}{{2880}}\left[ {\tr \left( {K_B^T{K_B}} \right) + \tr \left( {{K_B}K_B^T} \right)} \right] \\
 - \frac{1}{{720}}\left( {\tr \left[ {{{\left( {K_B^T{{\left( {{K_A}} \right)}^{ - 1}}{K_B}} \right)}^2}} \right] + \tr \left[ {{{\left( {{K_B}{{\left( {{K_C}} \right)}^{ - 1}}K_B^T} \right)}^2}} \right]} \right)\\
 =  - \frac{1}{{720}}\left( {\tr \left[ {{{\left( {K_B^T{{\left( {{K_A}} \right)}^{ - 1}}{K_B}} \right)}^2}} \right] + \tr \left[ {{{\left( {{K_B}{{\left( {{K_C}} \right)}^{ - 1}}K_B^T} \right)}^2}} \right] + \frac{1}{2}\tr \left( {K_B^T{K_B}} \right)} \right) .
\end{multline}

Putting everything together, the high temperature expansions of the entanglement entropy and the mutual information are given by the equations \eqref{eq:N_EE_high_T} and \eqref{eq:N_MI_high_T}, respectively.

\setcounter{equation}{0}
\section{The Low Temperature Expansion for Coupled Oscillators}
\label{sec:lowT}

At zero temperature, the matrices $a$ and $b$ \eqref{eq:N_ab_defs} are not analytic functions of the temperature. Acquiring a low temperature expansion of the entanglement entropy or the mutual information is not as straightforward as the respective high temperature expansion presented in appendix \ref{sec:highT}. In an obvious manner, at exactly $T=0$, $a=\sqrt{K}$ and $b=0$, resulting in the usual results for the ground state of the system, presented in \cite{Srednicki:1993im}. Beyond that, we may obtain an asymptotic expansion, approximating the hyperbolic functions as a series of exponentials. More specifically,
\begin{align}
a&=\Omega\left(I+2\sum_{n=1}^\infty \tilde{\Omega}^{2n}\right),\\
b&=-2\Omega\tilde{\Omega}\left(I+\sum_{n=1}^\infty \tilde{\Omega}^{2n}\right),\\
a+b&=\Omega \left(I+2\sum_{n=1}^\infty(-1)^n\tilde{\Omega}^{n}\right),
\end{align}
where
\begin{equation}
\tilde{\Omega}=\exp \left(-\Omega/T\right) .
\end{equation}

Only even powers of $\tilde{\Omega}$ appear in $a$, whereas only odd powers of $\tilde{\Omega}$ appear in $b$,
\begin{align}
a_C&=a_C^{(0)}+a_C^{(2)}+\dots, \label{eq:lowT_aexp}\\
b_C&=b_C^{(1)}+b_C^{(3)}+\dots, \label{eq:lowT_bexp}
\end{align}
where the superscript in parentheses indicates the power of $\tilde{\Omega}$ that appears in each term. Using the same notation for the matrices $\gamma$, $\beta$, $\gamma^{-1}$ and $\gamma^{-1}\beta$,
it is easy to show that
\begin{equation}
\gamma^{-1}=\left(\gamma^{-1}\right)^{(0)}\left[\gamma^{(0)}-\gamma^{(1)}+\left(\gamma^{(1)}\right)^2-\gamma^{(2)} +\dots \right]\left(\gamma^{-1}\right)^{(0)} ,
\end{equation}
thus, at leading order one recovers the zero temperature result
\begin{equation}
\left(\gamma^{-1}\beta\right)^{(0)}=\left(\gamma^{-1}\right)^{(0)}\beta^{(0)}.
\end{equation}

At next to leading order it holds
\begin{equation}
\left(\gamma^{-1}\beta\right)^{(1)}=\left(\gamma^{-1}\right)^{(0)}\beta^{(1)}-\left(\gamma^{-1}\right)^{(0)}\gamma^{(1)}\left(\gamma^{-1}\right)^{(0)}\beta^{(0)} .
\end{equation}
We recall that the matrices $\gamma$ and $\beta$ are defined as $\gamma = a_C - d / 2$ and $\beta = - b_C + d/2$, where $d = \left( {{a_B^T} + b_B^T} \right){\left( {a_A + b_A} \right)^{ - 1}}\left( {a_B + b_B} \right)$.
As a direct consequence of the form of the expansions \eqref{eq:lowT_aexp} and \eqref{eq:lowT_bexp}, it holds
\begin{equation}
\beta^{(0)}=\frac{1}{2}d^{(0)} \quad \textrm{and} \quad \gamma^{(1)}=-\frac{1}{2}d^{(1)}.
\end{equation}
As a result, we obtain
\begin{equation}
\left(\gamma^{-1}\beta\right)^{(1)}=-\left(\gamma^{-1}\right)^{(0)}b_C^{(1)}+\frac{1}{2}\left(\gamma^{-1}\right)^{(0)}d^{(1)}+\frac{1}{2}\left(\gamma^{-1}\right)^{(0)}d^{(1)}\left(\gamma^{-1}\right)^{(0)}\beta^{(0)} .
\label{eq:low_gammabeta1}
\end{equation}
At leading order it holds
\begin{equation}
d^{(0)}=\Omega_B^T\Omega_A^{-1}\Omega_B .
\end{equation}
At next to leading order it holds
\begin{equation}
\begin{split}
\left( (a_A + b_A)^{-1} \right)^{(1)} = 2\Omega_A^{-1}\left(\Omega \tilde{\Omega}\right)_A\Omega_A^{-1}, &\quad
(a_B + b_B)^{(1)} = - 2\left(\Omega \tilde{\Omega}\right)_B,\\
(a_B^T + b_B^T)^{(1)} = -2\left(\Omega \tilde{\Omega}\right)_{B^T}, &\quad
b_C^{(1)} = - 2\left(\Omega \tilde{\Omega}\right)_C ,
\end{split}
\end{equation}
where we used the following shorthand notation
\begin{equation}
\begin{split}
\left(\Omega \tilde{\Omega}\right)_A=\Omega_A \tilde{\Omega}_A+\Omega_B \tilde{\Omega}_B^T, &\quad
\left(\Omega \tilde{\Omega}\right)_B=\Omega_A \tilde{\Omega}_B+\Omega_B \tilde{\Omega}_C,\\
\left(\Omega \tilde{\Omega}\right)_{B^T}=\Omega_B^T \tilde{\Omega}_A+\Omega_C \tilde{\Omega}_B^T, &\quad
\left(\Omega \tilde{\Omega}\right)_C=\Omega_C \tilde{\Omega}_C+\Omega_B^T \tilde{\Omega}_B.
\end{split}
\end{equation}
After some algebra, we obtain
\begin{equation}
d^{(1)}=2\left[ \left(\gamma^{(0)}-\beta^{(0)}\right)\left(\tilde{\Omega}_C-\tilde{\Omega}_B^T\Omega_A^{-1}\Omega_B\right)-\left(\Omega \tilde{\Omega}\right)_C\right]
\end{equation}
and
\begin{equation}
\beta^{(1)}=\left(\gamma^{(0)}-\beta^{(0)}\right)\left(\tilde{\Omega}_C-\tilde{\Omega}_B^T\Omega_A^{-1}\Omega_B\right)+\left(\Omega \tilde{\Omega}\right)_C.
\end{equation}
It is straightforward to substitute the above into \eqref{eq:low_gammabeta1} and show that
\begin{multline}
\left(\gamma^{-1}\beta\right)^{(1)}=\left(1-\left(\gamma^{-1}\beta\right)^{(0)}\right)\left(\tilde{\Omega}_C-\tilde{\Omega}_B^T\Omega_A^{-1}\Omega_B\right)\left(1+\left(\gamma^{-1}\beta\right)^{(0)}\right)\\+\left(\gamma^{-1}\right)^{(0)}\left(\Omega \tilde{\Omega}\right)_C\left(1-\left(\gamma^{-1}\beta\right)^{(0)}\right) .
\end{multline}

It is not possible to obtain analytic expressions for the eigenvalues of $\left(\gamma^{-1}\beta\right)$ in the low temperature expansion. However, the above formula implies that the corrections to the zero temperature result are exponentially suppressed.

\setcounter{equation}{0}
\section{The Hopping Expansion in a Chain of Oscillators}
\label{sec:app_hopping}

In this appendix, we provide some details on the perturbative calculation of the mutual information in chains of oscillators in the inverse mass expansion. First we perturbatively calculate the matrix $\gamma^{-1} \beta$ and then we proceed to the specification of its eigenvalues.

\subsection{The Matrix $\gamma^{-1} \beta$ in the Hopping Expansion}
\label{subsec:app_hopping_matrix}

In order to find a perturbative expansion for the matrix $\gamma^{-1} \beta$, first we need to expand the matrices
\begin{equation}
a = T{f_1}\left( {\frac{K}{{{T^2}}}} \right),\quad b = T{f_2}\left( {\frac{K}{{{T^2}}}} \right),
\end{equation}
where
\begin{align}
{f_1}\left( x \right) &= \sqrt x \coth \sqrt x  = \sum\limits_{n = 0}^\infty  {{a_n}{x^n}} , \label{eq:app:f1_Taylor} \\
{f_2}\left( x \right) &=  - \sqrt x {\mathop{\rm \csch}\nolimits} \sqrt x  = \sum\limits_{n = 0}^\infty  {{b_n}{x^n}} , \label{eq:app:f2_Taylor}
\end{align}
since both $\coth x$ and $\csch x$ are odd functions of $x$, and, thus, the Taylor expansions of $f_1 \left( x \right)$ and $f_2 \left( x \right)$ contain only even powers of $\sqrt{x}$. It obviously holds that
\begin{align}
\sum\limits_{n = 0}^\infty  {n{a_n}{x^{n - 1}}}  &= {f_1}'\left( x \right) = \frac{1}{2}\left( {\frac{1}{{\sqrt x }}\coth \sqrt x  - \csch^2 \sqrt x } \right) , \label{eq:app:f1p_Taylor} \\
\sum\limits_{n = 0}^\infty  {n{b_n}{x^{n - 1}}}  &= {f_2}'\left( x \right) =  - \frac{1}{2}\left( {\frac{1}{{\sqrt x }}{\mathop{\rm csch}\nolimits} \sqrt x  - \coth \sqrt x \csch \sqrt x } \right) . \label{eq:app:f2p_Taylor}
\end{align}
Moreover the following identities are obeyed
\begin{align}
{f_1^2}\left( x \right) - {f_2^2}\left( x \right) &= x , \\
{f_1}'\left( x \right){f_2}\left( x \right) - {f_1}\left( x \right){f_2}'\left( x \right) &= \frac{1}{2}{f_2}\left( x \right) ,
\end{align}
which will become handy later.

In order to obtain the expansions of the matrices $a$ and $b$ in $\varepsilon$, we first need to find the corresponding expansion of the powers of the matrix $K$. The latter equals,
\begin{equation}
{K_{ij}} = \frac{1}{\varepsilon }\left[ {{k_i}{\delta _{ij}} + \varepsilon \left( {{l_i}{\delta _{i,j + 1}} + {l_j}{\delta _{i + 1,j}}} \right)} \right] \equiv \frac{1}{\varepsilon }{K^{\left( 0 \right)}} + {K^{\left( 1 \right)}} .
\end{equation}
Therefore, writing
\begin{equation}
{K^N} = \frac{1}{{{\varepsilon ^N}}}\left[ {{{\left( {{K^N}} \right)}^{\left( 0 \right)}} + \varepsilon {{\left( {{K^N}} \right)}^{\left( 1 \right)}} + {\varepsilon ^2}{{\left( {{K^N}} \right)}^{\left( 2 \right)}} + O\left( {{\varepsilon ^3}} \right)} \right] ,
\end{equation}
it follows that
\begin{align}
{\left( {{K^N}} \right)^{\left( 0 \right)}} &= {\left( {{K^{\left( 0 \right)}}} \right)^N},\\
{\left( {{K^N}} \right)^{\left( 1 \right)}} &= \sum\limits_{n = 0}^{N - 1} {{{\left( {{K^{\left( 0 \right)}}} \right)}^n}{K^{\left( 1 \right)}}{{\left( {{K^{\left( 0 \right)}}} \right)}^{N - 1 - n}}} ,\\
{\left( {{K^N}} \right)^{\left( 2 \right)}} &= \sum\limits_{n = 0}^{N - 2} {\sum\limits_{m = 0}^{N - 2 - n} {{{\left( {{K^{\left( 0 \right)}}} \right)}^n}{K^{\left( 1 \right)}}{{\left( {{K^{\left( 0 \right)}}} \right)}^m}{K^{\left( 1 \right)}}{{\left( {{K^{\left( 0 \right)}}} \right)}^{N - 2 - n - m}}} } .
\end{align}
Since $K^{\left( 0 \right)}$ is diagonal, it is trivial to find its powers. Therefore it is a matter of simple algebra to show that at zeroth order
\begin{equation}
\left( {{K^N}} \right)_{ij}^{\left( 0 \right)} = \left( {{K^N}} \right)_i^{0\left( 0 \right)}{\delta _{ij}} ,
\end{equation}
where
\begin{equation}
\left( {{K^N}} \right)_i^{0\left( 0 \right)} = k_i^N .
\end{equation}
At first order
\begin{equation}
\left( {{K^N}} \right)_{ij}^{\left( 1 \right)} = \left( {{K^N}} \right)_i^{1\left( 1 \right)}{\delta _{i + 1,j}} + \left( {{K^N}} \right)_j^{1\left( 1 \right)}{\delta _{i,j + 1}} ,
\end{equation}
where
\begin{equation}
\left( {{K^N}} \right)_i^{1\left( 1 \right)} = \frac{{k_i^N - k_{i + 1}^N}}{{{k_i} - {k_{i + 1}}}}{l_i} .
\end{equation}
Finally, at second order
\begin{equation}
\left( {{K^N}} \right)_{ij}^{\left( 2 \right)} = \left( {{K^N}} \right)_i^{0\left( 2 \right)}{\delta _{ij}} + \left( {{K^N}} \right)_i^{2\left( 2 \right)}{\delta _{i + 2,j}} + \left( {{K^N}} \right)_j^{2\left( 2 \right)}{\delta _{i,j + 2}} ,
\end{equation}
where
\begin{align}
\left( {{K^N}} \right)_i^{0\left( 2 \right)} &= Nk_i^{N - 1}\left( {\frac{{l_i^2}}{{{k_i} - {k_{i + 1}}}} - \frac{{l_{i - 1}^2}}{{{k_{i - 1}} - {k_i}}}} \right) \nonumber\\
&\quad\quad\quad\quad\quad\quad\quad\quad + \left( {\frac{{l_{i - 1}^2\left( {k_{i - 1}^N - k_i^N} \right)}}{{{{\left( {{k_{i - 1}} - {k_i}} \right)}^2}}} - \frac{{l_i^2\left( {k_i^N - k_{i + 1}^N} \right)}}{{{{\left( {{k_i} - {k_{i + 1}}} \right)}^2}}}} \right) ,\\
\left( {{K^N}} \right)_i^{2\left( 2 \right)} &= {l_i}{l_{i + 1}}\left( {\frac{{k_i^N}}{{\left( {{k_i} - {k_{i + 1}}} \right)\left( {{k_i} - {k_{i + 2}}} \right)}} } \right. \nonumber \\
&\quad\quad\left. { - \frac{{k_{i + 1}^N}}{{\left( {{k_i} - {k_{i + 1}}} \right)\left( {{k_{i + 1}} - {k_{i + 2}}} \right)}} + \frac{{k_{i + 2}^N}}{{\left( {{k_i} - {k_{i + 2}}} \right)\left( {{k_{i + 1}} - {k_{i + 2}}} \right)}}} \right) .
\end{align}

Throughout this appendix, we will use the shorthand notation
\begin{equation}
f_n \left( \frac{k_i}{T^2} \right) \equiv f_{n,i} .
\end{equation}

Writing the matrix $a$ as
\begin{equation}
a = T\left( {{a^{\left( 0 \right)}} + \varepsilon {a^{\left( 1 \right)}} + {\varepsilon ^2}{a^{\left( 2 \right)}} + O\left( {{\varepsilon ^3}} \right)} \right) ,
\end{equation}
one can make use of the Taylor series of the functions $f_1$ and $f_2$ \eqref{eq:app:f1_Taylor} and \eqref{eq:app:f2_Taylor}, to show that
\begin{equation}
a_{ij}^{\left( 0 \right)} = a_i^{0\left( 0 \right)}{\delta _{ij}} ,
\end{equation}
where
\begin{equation}
a_i^{0\left( 0 \right)} = {f_{1,i}} .
\end{equation}
Similarly, at first order
\begin{equation}
a_{ij}^{\left( 1 \right)} = a_i^{1\left( 1 \right)}{\delta _{i + 1,j}} + a_j^{1\left( 1 \right)}{\delta _{i,j + 1}} ,
\end{equation}
where
\begin{equation}
a_i^{1\left( 1 \right)} = \frac{{{f_{1,i}} - {f_{1,i+1}}}}{{{k_i} - {k_{i + 1}}}}{l_i} .
\end{equation}
Finally, at second order
\begin{equation}
a_{ij}^{\left( 2 \right)} = a_i^{0\left( 2 \right)}{\delta _{ij}} + a_i^{2\left( 2 \right)}{\delta _{i + 2,j}} + a_j^{2\left( 2 \right)}{\delta _{i,j + 2}} ,
\end{equation}
where
\begin{align}
a_i^{0\left( 2 \right)} &= \frac{{{f_{1,i}}'}}{{{T^2}}}\left( {\frac{{l_i^2}}{{{k_i} - {k_{i + 1}}}} - \frac{{l_{i - 1}^2}}{{{k_{i - 1}} - {k_i}}}} \right) + \frac{{l_{i - 1}^2\left( {{f_{1,i-1}} - {f_{1,i}}} \right)}}{{{{\left( {{k_{i - 1}} - {k_i}} \right)}^2}}} - \frac{{l_i^2\left( {{f_{1,i}} - {f_{1,i+1}}} \right)}}{{{{\left( {{k_i} - {k_{i + 1}}} \right)}^2}}} ,\\
a_i^{2\left( 2 \right)} &= {l_i}{l_{i + 1}}\left( {\frac{{{f_{1,i}}}}{{\left( {{k_i} - {k_{i + 1}}} \right)\left( {{k_i} - {k_{i + 2}}} \right)}} } \right. \nonumber \\
&\quad\quad\quad\quad \left. { - \frac{{{f_{1,i+1}}}}{{\left( {{k_i} - {k_{i + 1}}} \right)\left( {{k_{i + 1}} - {k_{i + 2}}} \right)}} + \frac{{{f_{1,i+2}}}}{{\left( {{k_i} - {k_{i + 2}}} \right)\left( {{k_{i + 1}} - {k_{i + 2}}} \right)}}} \right) .
\end{align}
In a similar manner, one can obtain the expansion for the matrix $b$. The formulae are identical upon the substitution of the function $f_1$ with the function $f_2$.

We proceed to calculate the matrix $\gamma^{-1} \beta$. We define
\begin{align}
{f_3}\left( x \right) &:= {f_1}\left( x \right) + {f_2}\left( x \right) = \sqrt x \tanh \frac{{\sqrt x }}{2} ,\\
{f_4}\left( x \right) &:=  - \frac{{{f_2}\left( x \right)}}{{{f_1}\left( x \right)}} = \frac{1}{{\cosh \sqrt x }} .
\end{align}
Similarly to the previous steps of this calculation, we expand $\gamma^{-1} \beta$ as
\begin{equation}
{\gamma ^{ - 1}}\beta  = {\left( {{\gamma ^{ - 1}}\beta } \right)^{\left( 0 \right)}} + \varepsilon {\left( {{\gamma ^{ - 1}}\beta } \right)^{\left( 1 \right)}} + {\varepsilon ^2}{\left( {{\gamma ^{ - 1}}\beta } \right)^{\left( 2 \right)}} + O\left( {{\varepsilon ^3}} \right).
\end{equation}
Although $\gamma^{-1}$ and $\beta$ are symmetric, this is not the case for $\gamma^{-1} \beta$. At zeroth order we get
\begin{equation}
\left( {{\gamma ^{ - 1}}\beta } \right)_i^{0\left( 0 \right)} = {f_{4,n+i}} .
\end{equation}
At first order we get
\begin{align}
\left( {{\gamma ^{ - 1}}\beta } \right)_i^{1\left( 1 \right)} &= \frac{{{l_{n + i}}}}{{{k_{n + i}} - {k_{n + i + 1}}}}\left( {{f_{4,n+i}} - {f_{4,n+i+1}}} \right) , \\
\left( {{\gamma ^{ - 1}}\beta } \right)_i^{ - 1\left( 1 \right)} &= \frac{{{l_{n + i}}}}{{{k_{n + i}} - {k_{n + i + 1}}}}\left( {{f_{4,n+i}} - {f_{4,n+i+1}}} \right) .
\end{align}
Finally, at second order we get
\begin{multline}
\left( {{\gamma ^{ - 1}}\beta } \right)_i^{0\left( 2 \right)} = \frac{{l_{n + i - 1}^2}}{{{k_{n + i - 1}} - {k_{n + i}}}}\left( {\frac{{{f_{4,n+i-1}} - {f_{4,n+i}}}}{{{k_{n + i - 1}} - {k_{n + i}}}} + \frac{1}{{2{T^2}}}\frac{{{f_{4,n+i}}}}{{{f_{1,n+i}}}}} \right) \\
 - \left( {1 - {\delta _{i,N - n}}} \right)\frac{{l_{n + i}^2}}{{{k_{n + i}} - {k_{n + i + 1}}}}\left( {\frac{{{f_{4,n+i}} - {f_{4,n+i+1}}}}{{{k_{n + i}} - {k_{n + i + 1}}}} + \frac{1}{{2{T^2}}}\frac{{{f_{4,n+i}}}}{{{f_{1,n+i}}}}} \right)\\
+ {\delta _{i1}}\frac{{l_n^2}}{{{{\left( {{k_n} - {k_{n + 1}}} \right)}^2}}}\left[ { - \frac{{\left( {{f_{1,n}} - {f_{1,n+1}}} \right)\left( {{f_{2,n}} - {f_{2,n+1}}} \right)}}{{{f_{1,n}}{f_{1,n+1}}}}} + \frac{{{f_{2,n+1}}}}{{{f_{1,n+1}}}}\frac{{{{\left( {{f_{1,n}} - {f_{1,n+1}}} \right)}^2}}}{{{f_{1,n}}{f_{1,n+1}}}} \right. \\
\left. { + \frac{{{f_{1,n+1}} - {f_{2,n+1}}}}{{2f_{1,n+1}^2}}\frac{{{{\left( {{f_{3,n}} - {f_{3,n+1}}} \right)}^2}}}{{{f_{3,n}}}}} \right] .
\end{multline}
and
\begin{align}
\left( {{\gamma ^{ - 1}}\beta } \right)_i^{2\left( 2 \right)} &= \frac{{{l_{n + i}}{l_{n + i + 1}}}}{{{k_{n + i}} - {k_{n + i + 1}}}}\left( {\frac{{{f_{4,n+i}} - {f_{4,n+i+2}}}}{{{k_{n + i}} - {k_{n + i + 2}}}} } {- \frac{{{f_{4,n+i+1}} - {f_{4,n+i+2}}}}{{{k_{n + i + 1}} - {k_{n + i + 2}}}}} \right) , \\
\left( {{\gamma ^{ - 1}}\beta } \right)_i^{ - 2\left( 2 \right)} &= \frac{{{l_{n + i}}{l_{n + i + 1}}}}{{{k_{n + i + 1}} - {k_{n + i + 2}}}}\left( {\frac{{{f_{4,n+i}} - {f_{4,n+i+1}}}}{{{k_{n + i}} - {k_{n + i + 1}}}} } {- \frac{{{f_{4,n+i}} - {f_{4,n+i+2}}}}{{{k_{n + i}} - {k_{n + i + 2}}}}} \right) .
\end{align}
This concludes the perturbative calculation of the matrix $\gamma^{-1} \beta$ in the $\varepsilon$ expansion up to second order.

\subsection{The Eigenvalues in Non-Degenerate Perturbation Theory}
\label{subsec:app_hopping_nondeg}

So far, we have calculated the matrix $\gamma^{-1} \beta$, perturbatively in $\varepsilon$. As we have already discussed in section \ref{sec:chain}, a small $\varepsilon$ is sufficient for the perturbative calculation of the matrix, but not of its eigenvalues. For this purpose, it is necessary to know whether the non-diagonal elements of $K$ are larger or smaller than the differences of the diagonal elements of $K$ and not the elements themselves. In the following we present two approaches for the perturbative calculation of the eigenvalues of the matrix $\gamma^{-1} \beta$ and consequently the entanglement entropy and the mutual information. In this subsection, we consider the case where the off-diagonal elements of the matrix $K$ are smaller than the differences of the diagonal ones. We refer to this approach as ``non-degenerate'' perturbation theory.

In this case one can consider $\left(\gamma^{-1} \beta \right)^{\left( 0 \right)}$ as an unperturbed, exactly solvable problem, and treat $\left(\gamma^{-1} \beta \right)^{\left( 1 \right)}$ and $\left(\gamma^{-1} \beta \right)^{\left( 2 \right)}$ as perturbative corrections. Since $\left(\gamma^{-1} \beta \right)^{\left( 0 \right)}$ is diagonal, in an obvious manner the unperturbed eigenvectors are $\left| {v^j} \right\rangle$, where
\begin{equation}
{\left( {{v^j}} \right)_i} = {\delta _{ij}} .
\end{equation}
At zeroth and first order, the eigenvalues of the matrix ${{\gamma ^{ - 1}}\beta }$ are trivially
\begin{align}
\beta _{Di}^{\left( 0 \right)} &= \left( {{\gamma ^{ - 1}}\beta } \right)_i^{0\left( 0 \right)} = {f_{4,n+i}} , \label{eq:app_non_deg_eigenvalue_0}\\
\beta _{Di}^{\left( 1 \right)} &= \left\langle {v^i} \right|{\left( {{\gamma ^{ - 1}}\beta } \right)^{\left( 1 \right)}}\left| {v^i} \right\rangle  = 0 . \label{eq:app_non_deg_eigenvalue_1}
\end{align}
At second order, one has to take account of the second order correction from the first order perturbation, as well as the first order correction from the second order perturbation. It is a matter of algebra to find that
\begin{multline}
\beta _{Di}^{\left( 2 \right)} = \left\langle {v^i} \right|{\left( {{\gamma ^{ - 1}}\beta } \right)^{\left( 2 \right)}}\left| {v^i} \right\rangle  + \sum\limits_{j \ne i} {\frac{{\left\langle {v^i} \right|{{\left( {{\gamma ^{ - 1}}\beta } \right)}^{\left( 1 \right)}}\left| {v^j} \right\rangle \left\langle {v^j} \right|{{\left( {{\gamma ^{ - 1}}\beta } \right)}^{\left( 1 \right)}}\left| {v^i} \right\rangle }}{{\beta _{Di}^{\left( 0 \right)} - \beta _{Dj}^{\left( 0 \right)}}}} \\
 = \left( {{\gamma ^{ - 1}}\beta } \right)_i^{0\left( 2 \right)} + \frac{{{{\left[ {\left( {{\gamma ^{ - 1}}\beta } \right)_i^{1\left( 1 \right)}} \right]}^2}}}{{\left( {{\gamma ^{ - 1}}\beta } \right)_i^{0\left( 0 \right)} - \left( {{\gamma ^{ - 1}}\beta } \right)_{i + 1}^{0\left( 0 \right)}}}\left( {1 - {\delta _{i,N - n}}} \right) + \frac{{{{\left[ {\left( {{\gamma ^{ - 1}}\beta } \right)_{i - 1}^{1\left( 1 \right)}} \right]}^2}}}{{\left( {{\gamma ^{ - 1}}\beta } \right)_i^{0\left( 0 \right)} - \left( {{\gamma ^{ - 1}}\beta } \right)_{i - 1}^{0\left( 0 \right)}}}\left( {1 - {\delta _{i,1}}} \right)\\
 = \frac{1}{{2{T^2}}}\frac{{{f_{4,n+i}}}}{{{f_{1,n+i}}}}\left( {\frac{{l_{n + i - 1}^2}}{{{k_{n + i - 1}} - {k_{n + i}}}} - \frac{{l_{n + i}^2}}{{{k_{n + i}} - {k_{n + i - 1}}}}\left( {1 - {\delta _{i,N - n}}} \right)} \right)\\
 + {\delta _{i1}}\frac{{l_n^2}}{{{{\left( {{k_n} - {k_{n + 1}}} \right)}^2}}}\frac{1}{{{f_{1,n+1}}}}\left[ {{f_{1,n}}\left( {{f_{4,n}} - {f_{4,n+1}}} \right)} { + \left( {1 + {f_{4,n+1}}} \right)\frac{{{{\left( {{f_{3,n}} - {f_{3,n+1}}} \right)}^2}}}{{2{f_{3,n}}}}} \right] . \label{eq:app_non_deg_eigenvalue_2}
\end{multline}

In a similar manner, had we considered the complementary subsystem, we would have found similar expressions for the eigenvalues. We give here the second order correction of those
\begin{multline}
\beta _{Di}^{\left( 2 \right)} = \frac{1}{{2{T^2}}}\frac{{{f_4}\left( {\frac{{{k_i}}}{{{T^2}}}} \right)}}{{{f_1}\left( {\frac{{{k_i}}}{{{T^2}}}} \right)}}\left( {\frac{{l_{i - 1}^2}}{{{k_{i - 1}} - {k_i}}}\left( {1 - {\delta _{i,1}}} \right) - \frac{{l_i^2}}{{{k_i} - {k_{i + 1}}}}} \right)\\
 + {\delta _{in}}\frac{{l_n^2}}{{{{\left( {{k_n} - {k_{n + 1}}} \right)}^2}}}\frac{1}{{{f_{1,n}}}}\left[ - {{f_{1,n+1}}\left( {{f_{4,n}} - {f_{4,n+1}}} \right)} { + \left( {1 + {f_{4,n}}} \right)\frac{{{{\left( {{f_{3,n}} - {f_{3,n+1}}} \right)}^2}}}{{2{f_{3,n+1}}}}} \right] .
\end{multline}

The corresponding calculation of the thermal entropy requires the perturbative calculation of the eigenvalues of the matrix $K$. It is trivial to show that
\begin{align}
k_i^{\left( 0 \right)} &= {k_i} , \\
k_i^{\left( 1 \right)} &= 0 , \\
k_i^{\left( 2 \right)} &=  - \left( {\frac{{l_{i - 1}^2}}{{{k_{i - 1}} - {k_i}}}\left( {1 - {\delta _{i,1}}} \right) - \frac{{l_i^2}}{{{k_i} - {k_{i + 1}}}}\left( {1 - {\delta _{i,N}}} \right)} \right) .
\end{align}

The entanglement and thermal entropies can now be calculated in terms of the quantities $\xi_i = \frac{\beta_{Di}}{1+\sqrt{1-\beta_{Di}^2}}$ and $\zeta_i = e^{-\sqrt{k_i}/T}$, respectively. These quantities give identical contributions to the entanglement and thermal entropy respectively, namely $S_{\mathrm{EE}} = \sum \left( -\ln \left( 1-\xi_i \right)- \frac{\xi_i}{1-\xi_i} \ln \xi_i \right)$ and $S_{\mathrm{th}} = \sum \left( -\ln \left( 1-\zeta_i \right) - \frac{\zeta_i}{1-\zeta_i} \ln \zeta_i \right)$. It is a matter of algebra to show that
\begin{align}
{\xi _i} &= \xi _i^{\left( 0 \right)} + \xi _i^{\left( 2 \right)} + \mathcal{O} \left( l^3 \right) = \xi _i^{\left( 0 \right)} + \xi _i^{\left( 0 \right)}\frac{T}{{\sqrt {{k_i}} }}\frac{{{f_{1,i}}}}{{{f_{4,i}}}}\beta _{Di}^{\left( 2 \right)} + \mathcal{O} \left( l^3 \right) , \\
{\zeta _i} &= \zeta _i^{\left( 0 \right)} + \zeta _i^{\left( 2 \right)} + \mathcal{O} \left( l^3 \right) = \zeta _i^{\left( 0 \right)} - \zeta _i^{\left( 0 \right)}\frac{1}{{2T\sqrt {{k_i}} }}k_i^{\left( 2 \right)} + \mathcal{O} \left( l^3 \right) .
\end{align}
The index $i$ runs from 1 to $N$ for both cases. In the case of the $\xi_i$'s the $i\leq n$ values correspond to the eigenvalues that we get when we trace out the $i>n$ subsystem and vice versa. The formulae obtained above for the expansions of $\beta_{Di}$ and $k_i$ imply
\begin{align}
\xi _i^{\left( 0 \right)} &= \zeta _i^{\left( 0 \right)} = {e^{ - \frac{{\sqrt {{k_i}} }}{T}}} , \\
\xi _i^{\left( 1 \right)} &= \zeta _i^{\left( 1 \right)} = 0 .
\end{align}
The second order corrections are
\begin{multline}
\xi _i^{\left( 2 \right)} = \zeta _i^{\left( 2 \right)} \\
+ \frac{{T l_n^2}}{{{{\left( {{k_n} - {k_{n + 1}}} \right)}^2}}} \Bigg\{ {\delta _{i,n}}\frac{{e^{ - \frac{{\sqrt {{k_n}} }}{T}}}}{{\sqrt {{k_n}} f_{4,n} }} \left[ {{f_{1,n+1}}\left( {{f_{4,n+1}} - {f_{4,n}}} \right)} { + \left( {1 + {f_{4,n}}} \right)\frac{{{{\left( {{f_{3,n}} - {f_{3,n+1}}} \right)}^2}}}{{2{f_{3,n+1}}}}} \right] \\
{\delta _{i,n + 1}}\frac{{e^{ - \frac{{\sqrt {{k_{n + 1}}} }}{T}}}}{{\sqrt {{k_{n + 1}}} f_{4,n+1} }} \left[ {{f_{1,n}}\left( {{f_{4,n}} - {f_{4,n+1}}} \right)} { + \left( {1 + {f_{4,n+1}}} \right)\frac{{{{\left( {{f_{3,n}} - {f_{3,n+1}}} \right)}^2}}}{{2{f_{3,n}}}}} \right] \Bigg\}
\label{eq:non_degenerate_xi2}
\end{multline}
and
\begin{equation}
\zeta _i^{\left( 2 \right)} =  - \frac{1}{{2T\sqrt {{k_i}} }}{e^{ - \frac{{\sqrt {{k_i}} }}{T}}}\left( {\frac{{l_i^2}}{{{k_i} - {k_{i + 1}}}}\left( {1 - {\delta _{iN}}} \right) + \frac{{l_{i - 1}^2}}{{{k_i} - {k_{i - 1}}}}\left( {1 - {\delta _{i1}}} \right)} \right) .
\end{equation}

The expansive expression for the mutual information is
\begin{multline}
I = \sum\limits_{i = 1}^N \left( { - \ln \left( {1 - \xi _i^{\left( 0 \right)}} \right) - \frac{{\xi _i^{\left( 0 \right)}}}{{1 - \xi _i^{\left( 0 \right)}}}\ln \xi _i^{\left( 0 \right)} - \frac{{\ln \xi _i^{\left( 0 \right)}}}{{{{\left( {1 - \xi _i^{\left( 0 \right)}} \right)}^2}}}\xi _i^{\left( 2 \right)} } \right. \\
\left. { { + \ln \left( {1 - \zeta _i^{\left( 0 \right)}} \right) + \frac{{\zeta _i^{\left( 0 \right)}}}{{1 - \zeta _i^{\left( 0 \right)}}}\ln \zeta _i^{\left( 0 \right)}}  + \frac{{\ln \zeta _i^{\left( 0 \right)}}}{{{{\left( {1 - \zeta _i^{\left( 0 \right)}} \right)}^2}}}\zeta _i^{\left( 2 \right)}} \right)
\end{multline}
It follows from the equations above that there are only two contributions to the mutual information, which appear at second order. These originate from the two eigenvalues for whom the corresponding $\xi^{\left( 2 \right)}$ and $\zeta^{\left( 2 \right)}$ are not identical (see equation \eqref{eq:non_degenerate_xi2}), namely the $\left(\xi_n , \zeta_n \right)$ and $\left(\xi_{n+1} , \zeta_{n+1}\right)$ ones,
\begin{equation}
I =  - \frac{{\ln \xi _n^{\left( 0 \right)}}}{{{{\left( {1 - \xi _n^{\left( 0 \right)}} \right)}^2}}}\left( {\xi _n^{\left( 2 \right)} - \zeta _n^{\left( 2 \right)}} \right) - \frac{{\ln \xi _{n + 1}^{\left( 0 \right)}}}{{{{\left( {1 - \xi _{n + 1}^{\left( 0 \right)}} \right)}^2}}}\left( {\xi _{n + 1}^{\left( 2 \right)} - \zeta _{n + 1}^{\left( 2 \right)}} \right) .
\end{equation}
After some algebra, it turns out that the mutual information is equal to
\begin{equation}
I = \frac{{l_n^2}}{{4{T^2}\left( {{k_n} - {k_{n + 1}}} \right)}}\left( {\frac{1}{{{f_{3,n+1}}}} - \frac{1}{{{f_{3,n}}}}} \right) + \mathcal{O}\left( {{l^3}} \right) .
\end{equation}

\subsection{The Eigenvalues in Degenerate Perturbation Theory}
\label{subsec:app_hopping_deg}

When, the off-diagonal elements of the matrix $K$ are larger than the differences of the diagonal ones, a different approach is required. Then, the unperturbed problem is the problem where the diagonal elements are all identical. In such cases, it is clear that even the formulae that we have written down in the previous section for the matrix $\gamma^{-1} \beta$ in the large $m$ expansion need rephrasing, since they are undetermined. The expansion of the powers of the matrix $K$ reads
\begin{align}
\left( {{K^n}} \right)_{ij}^{\left( 0 \right)} &= {k^n}{\delta _{ij}} , \\
\left( {{K^n}} \right)_{ij}^{\left( 1 \right)} &= nl{k^{n - 1}}\left( {{\delta _{i + 1,j}} + {\delta _{i,j + 1}}} \right) , \\
\left( {{K^n}} \right)_{ij}^{\left( 2 \right)} &= \frac{{n\left( {n - 1} \right)}}{2}{l^2}{k^{n - 2}}\left( {\left( {2 - {\delta _{i,1}} - {\delta _{i,N}}} \right){d _{ij}} + {\delta _{i + 2,j}} + {\delta _{i,j + 2}}} \right) ,
\end{align}
The formulae \eqref{eq:app:f1p_Taylor} and \eqref{eq:app:f2p_Taylor} imply that
\begin{multline}
\left( f\left( {\frac{K}{{{T^2}}}} \right) \right)_{ij}= f\left( {\frac{k}{{{T^2}}}} \right){\delta _{ij}} + \frac{l}{{{T^2}}}f'\left( {\frac{k}{{{T^2}}}} \right)\left( {{\delta _{i + 1,j}} + {\delta _{i,j + 1}}} \right)\\
+ \frac{{{l^2}}}{{2{T^4}}}f''\left( {\frac{k}{{{T^2}}}} \right)\left( {\left( {2 - {\delta _{i,1}} - {\delta _{i,N}}} \right){d _{ij}} + {\delta _{i + 2,j}} + {\delta _{i,j + 2}}} \right) .
\end{multline}
At this order
\begin{equation}
{d _{ij}} = \frac{{{l^2}}}{{{T^4}}}\Delta {\delta _{i,1}}{\delta _{j,1}} + O\left( {{l^3}} \right),\quad \Delta  = \frac{{{{\left[ {{f_3}'\left( {\frac{k}{{{T^2}}}} \right)} \right]}^2}}}{{{f_3}\left( {\frac{k}{{{T^2}}}} \right)}} ,
\end{equation}
implying that
\begin{multline}
{\gamma _{ij}} = {f_1}\left( {\frac{k}{{{T^2}}}} \right){\delta _{ij}} + \frac{l}{{{T^2}}}{f_1}'\left( {\frac{k}{{{T^2}}}} \right)\left( {{\delta _{i + 1,j}} + {\delta _{i,j + 1}}} \right)\\
+ \frac{{{l^2}}}{{2{T^4}}}\left[ {{f_1}''\left( {\frac{k}{{{T^2}}}} \right)\left( {\left( {2 - {\delta _{i,1}} - {\delta _{i,N}}} \right){\delta _{ij}} + {\delta _{i + 2,j}} + {\delta _{i,j + 2}}} \right) - \Delta {\delta _{i,1}}{\delta _{j,1}}} \right] ,
\end{multline}
\begin{multline}
{\beta _{ij}} =  - {f_2}\left( {\frac{k}{{{T^2}}}} \right){\delta _{ij}} - \frac{l}{{{T^2}}}{f_2}'\left( {\frac{k}{{{T^2}}}} \right)\left( {{\delta _{i + 1,j}} + {\delta _{i,j + 1}}} \right)\\
 - \frac{{{l^2}}}{{2{T^4}}}\left[ {{f_2}''\left( {\frac{k}{{{T^2}}}} \right)\left( {\left( {2 - {\delta _{i,1}} - {\delta _{i,N}}} \right){\delta _{ij}} + {\delta _{i + 2,j}} + {\delta _{i,j + 2}}} \right) - \Delta {\delta _{i,1}}{\delta _{j,1}}} \right] .
\end{multline}

It is a matter of algebra to show that
\begin{align}
\left( {{\gamma ^{ - 1}}\beta } \right)_i^{0\left( 0 \right)} &= {f_4}\left( {\frac{k}{{{T^2}}}} \right) , \\
\left( {{\gamma ^{ - 1}}\beta } \right)_i^{1\left( 1 \right)} &= \frac{l}{{{T^2}}}{f_4}'\left( {\frac{k}{{{T^2}}}} \right) , \\
\left( {{\gamma ^{ - 1}}\beta } \right)_i^{2\left( 2 \right)} &= \frac{{{l^2}}}{{2{T^4}}}{f_4}''\left( {\frac{k}{{{T^2}}}} \right) , \\
\left( {{\gamma ^{ - 1}}\beta } \right)_i^{0\left( 2 \right)} &= \frac{{{l^2}}}{{2{T^4}}}\left( {{f_4}''\left( {\frac{k}{{{T^2}}}} \right)\left( {2 - {\delta _{i,1}} - {\delta _{i,N - n}}} \right) + {\beta _1}{\delta _{i,1}}} \right) \nonumber \\
&\equiv \frac{{{l^2}}}{{{T^4}}}\left( {{f_4}''\left( {\frac{k}{{{T^2}}}} \right) + {B_1}{\delta _{i,1}} + {B_{N - n}}{\delta _{i,N - n}}} \right) ,
\end{align}
where
\begin{multline}
{\beta _1} = \frac{1}{{{{\left( {{f_1}\left( {\frac{k}{{{T^2}}}} \right)} \right)}^2}}}\left[ {\left( {{f_1}\left( {\frac{k}{{{T^2}}}} \right) - {f_2}\left( {\frac{k}{{{T^2}}}} \right)} \right)\Delta } \right.\\
\left. { - \left( {{f_1}\left( {\frac{k}{{{T^2}}}} \right){f_2}''\left( {\frac{k}{{{T^2}}}} \right) - {f_1}''\left( {\frac{k}{{{T^2}}}} \right){f_2}\left( {\frac{k}{{{T^2}}}} \right)} \right)} \right] .
\end{multline}

The above imply that the eigenvalues at zeroth order are
\begin{equation}
\beta _D^{j\left( 0 \right)} = {f_4}\left( {\frac{k}{{{T^2}}}} \right) ,
\end{equation}
they are all equal and they do not determine the eigenvectors. At first order the matrix $\gamma^{-1} \beta$ is proportional to the matrix ${{\delta _{i + 1,j}} + {\delta _{i,j + 1}}}$. Its normalized eigenvectors $v^j$ are
\begin{equation}
v_i^j = \sqrt {\frac{2}{{N + 1}}} \sin \frac{{ij\pi }}{{N + 1}}
\end{equation}
with corresponding eigenvalues
\begin{equation}
{\lambda ^j} = 2\cos \frac{{j\pi }}{{N + 1}}.
\end{equation}
It follows that the eigenvalues of the matrix $\gamma^{-1} \beta$ at first order equal
\begin{equation}
\beta _D^{j\left( 1 \right)} = \frac{{2l}}{{{T^2}}}{f_4}'\left( {\frac{k}{{{T^2}}}} \right)\cos \frac{{j\pi }}{{N - n + 1}}.
\end{equation}
Now we may proceed with perturbation theory to determine the eigenvalues at second order. They equal
\begin{equation}
\beta _D^{j\left( 2 \right)} = \left\langle {{v^j}} \right|{\left( {{\gamma ^{ - 1}}\beta } \right)^{\left( 2 \right)}}\left| {{v^j}} \right\rangle .
\end{equation}
There are three contributions to the above formula. The first one is trivial and comes from the part of ${\left( {{\gamma ^{ - 1}}\beta } \right)^{\left( 2 \right)}}$ that is proportional to the identity matrix. It equals
\begin{equation}
\beta _{D1}^{j\left( 2 \right)} = \frac{{{l^2}}}{{{T^4}}}{f_4}''\left( {\frac{k}{{{T^2}}}} \right) .
\end{equation}
The second contribution comes from the corrections at the edges of the diagonal part. This equals
\begin{equation}
\begin{split}
\beta _{D2}^{m\left( 2 \right)} &= \frac{{{l^2}}}{{{T^4}}}\sum\limits_{i = 1}^{N - n} {\sum\limits_{j = 1}^{N - n} {v_i^m\left( {{B_1}{\delta _{i,1}}{\delta _{j,1}} + {B_{N - n}}{\delta _{i,N - n}}{\delta _{j,N - n}}} \right)v_j^m} } \\
 &= \frac{{{l^2}}}{{{T^4}}}\frac{{2\left( {{B_1} + {B_{N - n}}} \right)}}{{N - n + 1}}{\sin ^2}\frac{{m\pi }}{{N - n + 1}} .
\end{split}
\end{equation}
Finally, the third contribution comes from the off-diagonal part. It equals
\begin{equation}
\begin{split}
\beta _{D3}^{m\left( 2 \right)} &= \frac{{{l^2}}}{{2{T^4}}}{f_4}''\left( {\frac{k}{{{T^2}}}} \right)\sum\limits_{i = 1}^{N - n} {\sum\limits_{j = 1}^{N - n} {v_i^m\left( {{\delta _{i + 2,j}} + {\delta _{i,j + 2}}} \right)v_j^m} } \\
 &= \frac{{{l^2}}}{{{T^4}}}{f_4}''\left( {\frac{k}{{{T^2}}}} \right)\left( {1 - \frac{4}{{N - n + 1}}{{\sin }^2}\frac{{m\pi }}{{N - n + 1}}} \right) .
\end{split}
\end{equation}
Putting everything together, we find
\begin{equation}
\beta _D^{m\left( 2 \right)} = \frac{{{l^2}}}{{{T^4}}}\left( {2{f_4}''\left( {\frac{k}{{{T^2}}}} \right){{\cos }^2}\frac{{m\pi }}{{N - n + 1}} + \frac{{{\beta _1}}}{{N - n + 1}}{{\sin }^2}\frac{{m\pi }}{{N - n + 1}}} \right) .
\end{equation}

The quantities $\xi^i$ are
\begin{equation}
\xi^i = {\xi ^{i\left( 0 \right)}} + {\xi ^{i\left( 1 \right)}} \varepsilon + {\xi ^{i\left( 2 \right)}} \varepsilon^2 + \mathcal{O} \left( \varepsilon^3 \right) ,
\end{equation}
where
\begin{align}
{\xi ^{i\left( 0 \right)}} &= {e^{ - \frac{{\sqrt k }}{T}}} , \\
{\xi ^{i\left( 1 \right)}} &=  - \frac{{l}}{{T\sqrt k }}{e^{ - \frac{{\sqrt k }}{T}}}\cos \frac{{i\pi }}{{N - n + 1}} , \\
{\xi ^{i\left( 2 \right)}} &= \frac{{{l^2}}}{{2{T^2}k}}{e^{ - \frac{{\sqrt k }}{T}}}\left[ {\left( {1 + \frac{T}{{\sqrt k }}} \right){{\cos }^2}\frac{{i\pi }}{{N - n + 1}}} \right. \nonumber \\
&\left. { + \frac{T}{{2\sqrt k }}\left( {1 + {f_1}\left( {\frac{k}{{{T^2}}}} \right) - {f_2}\left( {\frac{k}{{{T^2}}}} \right) - 1/{f_2}\left( {\frac{k}{{{T^2}}}} \right)} \right)\frac{1}{{N - n + 1}}{{\sin }^2}\frac{{i\pi }}{{N - n + 1}}} \right] .
\end{align}

Similarly, we may calculate the quantities $\zeta^i$ that enter into the calculation of the thermal entropy, perturbatively. This is trivial as they equal ${e^{ - \frac{{\sqrt k_i }}{T}}}$, where $k_i$ are the known eigenvalues of the matrix $K$. They equal
\begin{equation}
\zeta^i  = {e^{ - \frac{{\sqrt k }}{T}}}\left( {1 - \frac{l}{{\sqrt k T}}\cos \frac{{i\pi }}{{N + 1}} + \frac{{{l^2}}}{{2k{T^2}}}\left( {1 + \frac{T}{{\sqrt k }}} \right){{\cos }^2}\frac{{i\pi }}{{N + 1}}} \right) + \mathcal{O}\left( {{l^3}} \right)
\end{equation}

Putting everything together, the entanglement entropy at this order equals
\begin{multline}
{S_A} = \left( {N - n} \right)\left[ {\frac{{\sqrt k }}{T}\frac{{{e^{ - \frac{{\sqrt k }}{T}}}}}{{1 - {e^{ - \frac{{\sqrt k }}{T}}}}} - \ln \left( {1 - {e^{ - \frac{{\sqrt k }}{T}}}} \right)} \right] \\
+ \frac{{{l^2}}}{{32{k^{\frac{3}{2}}}{T^3}}}\left[ {\sqrt k T \csch^2 \frac{{\sqrt k }}{{2T}} + \coth \frac{{\sqrt k }}{{2T}}\left( {2{T^2} + k\left( {2\left( {N - n} \right) - 1} \right) \csch^2 \frac{{\sqrt k }}{{2T}}} \right)} \right] \\
+ \mathcal{O}\left( {{l^3}} \right) ,
\end{multline}
The thermal entropy equals
\begin{multline}
{S_{\textrm{th}}} = N\left[ {\frac{{\sqrt k }}{T}\frac{{{e^{ - \frac{{\sqrt k }}{T}}}}}{{1 - {e^{ - \frac{{\sqrt k }}{T}}}}} - \ln \left( {1 - {e^{ - \frac{{\sqrt k }}{T}}}} \right)} \right]\\
 + \frac{{{l^2}}}{{32\sqrt k {T^3}}}\left( {N - 1} \right)\csch^4\frac{{\sqrt k }}{{2T}}\sinh \frac{{\sqrt k }}{T} + \mathcal{O}\left( {{l^3}} \right)
\end{multline}
and finally, the mutual information equals
\begin{equation}
I = \frac{{{l^2}}}{{16{k^{\frac{3}{2}}}{T^3}}} \csch^2 \frac{{\sqrt k }}{{2T}}\left( {\sqrt k  + T\sinh \frac{{\sqrt k }}{T}} \right) + \mathcal{O}\left( {{l^3}} \right) .
\end{equation}

\setcounter{equation}{0}
\section{Low Temperature Expansion in a Chain of Oscillators}
\label{sec:Small_T_chain}

We have seen in section \ref{sec:lowT} that it is not simple to find a low temperature for the eigenvalues $\beta_D$ for a generic oscillatory system. However, in the case of a chain of oscillators, since we managed to perturbatively calculate these eigenvalues, we can perform this task. As we have already encountered in section \ref{sec:lowT}, the functions that we need to expand are not analytic at $T=0$. However, they can be expanded in a series of exponentials and we expect to find that all deviations from the $T=0$ result should be exponentially suppressed.

First we expand the eigenvalues, which have been calculated up to second order in the $l/k$ expansion in the previous appendix (see equations \eqref{eq:app_non_deg_eigenvalue_0}, \eqref{eq:app_non_deg_eigenvalue_1} and \eqref{eq:app_non_deg_eigenvalue_2}). The generic eigenvalues, i.e. all eigenvalues except $\beta_{Dn}$ and $\beta_{Dn+1}$, can be expanded as
\begin{equation}
{\beta_{Di}} = 2 \exp \left[ { - \frac{{\sqrt {{k_i}} }}{T}\left( {1 + \frac{{k_i^{\left( 2 \right)}}}{{2k_i^{\left( 0 \right)}}}} + \mathcal{O}\left( {{l^3}} \right) \right)} \right] + \dots , \quad i \ne n,n + 1 .
\end{equation}
The two special ones are a little different. Since they do not vanish at zero temperature they can be expanded around this value to yield
\begin{align}
{\beta_{Dn}} &= \beta _n^{\left( 0 \right)} + 2\left( {1 + \frac{{k_n^{\left( 2 \right)}}}{{2T\sqrt {k_n^{\left( 0 \right)}} }}} + \mathcal{O}\left( {{l^3}} \right) \right) \exp\left[ { - \frac{{\sqrt {{k_n}} }}{T}} \right] \nonumber \\
&- \frac{{\left( {{k_n} + {k_{n + 1}}} \right)l_n^2}}{{\sqrt {{k_n}} \sqrt {{k_{n + 1}}} {{\left( {{k_n} - {k_{n + 1}}} \right)}^2}}}\left( { \exp\left[ { - \frac{{\sqrt {{k_n}} }}{T}} \right] -  \exp\left[ { - \frac{{\sqrt {{k_{n + 1}}} }}{T}} \right]} \right) + \dots , \\
{\beta_{D n + 1}} &= \beta _{n + 1}^{\left( 0 \right)} + 2\left( {1 + \frac{{k_{n + 1}^{\left( 2 \right)}}}{{2T\sqrt {k_{n + 1}^{\left( 0 \right)}} }}} + \mathcal{O}\left( {{l^3}} \right) \right) \exp\left[ { - \frac{{\sqrt {{k_{n + 1}}} }}{T}} \right] \nonumber \\
&- \frac{{\left( {{k_n} + {k_{n + 1}}} \right)l_n^2}}{{\sqrt {{k_n}} \sqrt {{k_{n + 1}}} {{\left( {{k_n} - {k_{n + 1}}} \right)}^2}}}\left( { \exp\left[ { - \frac{{\sqrt {{k_{n + 1}}} }}{T}} \right] -  \exp\left[ { - \frac{{\sqrt {{k_n}} }}{T}} \right]} \right) + \dots ,
\end{align}
where $\beta _{Dn}^{\left( 0 \right)} $ and $ \beta _{D n + 1}^{\left( 0 \right)}$ are the zero temperature values of $\beta _{Dn}$ and $ \beta _{D n + 1}$. At second order in the $l/k$ expansion, they are
\begin{equation}
\beta _{Dn}^{\left( 0 \right)} = \beta _{D n + 1}^{\left( 0 \right)} = \frac{{l_n^2}}{{2\sqrt {{k_n}} \sqrt {{k_{n + 1}}} {{\left( {\sqrt {{k_n}}  + \sqrt {{k_{n + 1}}} } \right)}^2}}} .
\end{equation}

One can observe a basic difference between the low temperature expressions of the generic eigenvalue and the two special ones. In the first case, the $l/k$ expansion is performed in the argument of the exponential, whereas this is not the case for the two special eigenvalues. This is due to the fact that the latter do not vanish at zero temperature, which enforces us to make a different expansion around $T=0$. However, as discussed in appendix \ref{sec:lowT}, we expect that the result should be exponentially suppressed, with the eigenfrequencies of the overall system appearing in the exponents. This implies that naturally, the $l/k$ expansion should appear in the exponents of the low temperature expansion terms. This argument strongly suggests that the expressions for the two special eigenvalues should be resummed, so that the first terms read
\begin{align}
{\beta _{Dn}} &= \beta _n^{\left( 0 \right)} + 2 \exp\left[ { - \frac{{\sqrt {{k_n}} }}{T}\left( {1 + \frac{{k_n^{\left( 2 \right)}}}{{2k_n^{\left( 0 \right)}}}} + \mathcal{O}\left( {{l^3}} \right) \right)} \right] \nonumber \\
& - \frac{{\left( {{k_n} + {k_{n + 1}}} \right)l_n^2}}{{\sqrt {{k_n}} \sqrt {{k_{n + 1}}} {{\left( {{k_n} - {k_{n + 1}}} \right)}^2}}}\left( { \exp\left[ { - \frac{{\sqrt {{k_n}} }}{T}} \right] -  \exp\left[ { - \frac{{\sqrt {{k_{n + 1}}} }}{T}} \right]} \right) + \dots , \\
{\beta _{D n + 1}} &= \beta _{n + 1}^{\left( 0 \right)} + 2 \exp\left[ { - \frac{{\sqrt {{k_{n + 1}}} }}{T}\left( {1 + \frac{{k_{n + 1}^{\left( 2 \right)}}}{{2k_{n + 1}^{\left( 0 \right)}}}} + \mathcal{O}\left( {{l^3}} \right) \right)} \right] \nonumber \\
& - \frac{{\left( {{k_n} + {k_{n + 1}}} \right)l_n^2}}{{\sqrt {{k_n}} \sqrt {{k_{n + 1}}} {{\left( {{k_n} - {k_{n + 1}}} \right)}^2}}}\left( { \exp\left[ { - \frac{{\sqrt {{k_{n + 1}}} }}{T}} \right] -  \exp\left[ { - \frac{{\sqrt {{k_n}} }}{T}} \right]} \right) + \dots .
\end{align}

It is now straightforward to show that the contributions to the entanglement entropy from each generic eigenvalue are
\begin{multline}
S_i =  \left[ {1 + \frac{{\sqrt {{k_i}} }}{T}\left( {1 + \frac{{k_i^{\left( 2 \right)}}}{{2k_i^{\left( 0 \right)}}}} + \mathcal{O}\left( {{l^3}} \right) \right)} \right] \\
\times \exp\left[ { - \frac{{\sqrt {{k_i}} }}{T}\left( {1 + \frac{{k_i^{\left( 2 \right)}}}{{2k_i^{\left( 0 \right)}}}} + \mathcal{O}\left( {{l^3}} \right) \right)} \right] + \dots , \quad i \ne n,n + 1 .
\end{multline}
The two special contributions are
\begin{equation}
\begin{split}
S_n + S_{n + 1} &=  - \frac{1}{2}\log \left( {\frac{{\beta _n^{\left( 0 \right)}}}{2}} \right)\left( {1 + 2\beta _n^{\left( 0 \right)}} \right)\left( {\beta _n^{\left( 1 \right)} + \beta _{n + 1}^{\left( 1 \right)}} \right)\\
 &=  - \log \left( {\frac{{\beta _n^{\left( 0 \right)}}}{2}} \right)\left( {1 + 2\beta _n^{\left( 0 \right)}} \right)\left( { \exp\left[ { - \frac{{\sqrt {{k_n}} }}{T}\left( {1 + \frac{{k_n^{\left( 2 \right)}}}{{2k_n^{\left( 0 \right)}}}} \right)} \right] } \right. \\
&\quad\quad\quad\quad\quad\quad\quad\quad\quad\quad \left. { +  \exp\left[ { - \frac{{\sqrt {{k_{n + 1}}} }}{T}\left( {1 + \frac{{k_{n + 1}^{\left( 2 \right)}}}{{2k_{n + 1}^{\left( 0 \right)}}}} \right)} \right]} \right) .
\end{split}
\end{equation}

Finally, we may follow the same procedure to calculate the thermal entropy, and, thus the mutual information. The quantities $\zeta_i$ are
\begin{equation}
{\zeta _i} =  \exp\left[ { - \frac{{\sqrt {{k_i}} }}{T}\left( {1 + \frac{{k_i^{\left( 2 \right)}}}{{2k_i^{\left( 0 \right)}}}} + \mathcal{O}\left( {{l^3}} \right) \right)} \right] + \dots .
\end{equation}
The contribution of each $\zeta$ to the thermal entropy reads
\begin{equation}
\begin{split}
S_{\textrm{th}i} &= {\zeta _i}\left( {1 - \log {\zeta _i}} \right)\\
&= \left[ {1 + \frac{{\sqrt {{k_i}} }}{T}\left( {1 + \frac{{k_i^{\left( 2 \right)}}}{{2k_i^{\left( 0 \right)}}}} + \mathcal{O}\left( {{l^3}} \right) \right)} \right] \exp\left[ { - \frac{{\sqrt {{k_i}} }}{T}\left( {1 + \frac{{k_i^{\left( 2 \right)}}}{{2k_i^{\left( 0 \right)}}}} + \mathcal{O}\left( {{l^3}} \right) \right)} \right] + \dots .
\end{split}
\end{equation}

Putting everything together, it is evident that the mutual information receives non-vanishing contributions only from the two special eigenvalues. It is equal to
\begin{multline}
I =  - \log \left( {\frac{{\beta _n^{\left( 0 \right)}}}{2}} \right)\left( {1 + 2\beta _n^{\left( 0 \right)}} \right) + \left( n \to n + 1 \right) \\
 + \left[ { - \log \left( {\frac{{\beta _n^{\left( 0 \right)}}}{2}} \right)\left( {1 + \beta _n^{\left( 0 \right)}} \right) - \left( {1 + \frac{{\sqrt {{k_n}} }}{T}\left( {1 + \frac{{k_n^{\left( 2 \right)}}}{{2k_n^{\left( 0 \right)}}}} + \mathcal{O}\left( {{l^3}} \right) \right)} \right)} \right] \\
\times \exp\left[ { - \frac{{\sqrt {{k_n}} }}{T}\left( {1 + \frac{{k_n^{\left( 2 \right)}}}{{2k_n^{\left( 0 \right)}}}} + \mathcal{O}\left( {{l^3}} \right) \right)} \right] + \left( n \to n + 1 \right) + \ldots .
\end{multline}

\end{document}